\documentclass[graybox, nosecnum]{svmult}

\usepackage{amsmath}       
\usepackage{amssymb}       
\usepackage{mathptmx}       
\usepackage{helvet}         
\usepackage{courier}        
\usepackage{type1cm}        
\usepackage{makeidx}         
\usepackage{graphicx}        
\usepackage{multicol}        
\usepackage[bottom]{footmisc}
\usepackage{hyperref}        
\usepackage{soul}            
\hypersetup{colorlinks=true,urlcolor=blue}
\usepackage[square,numbers]{natbib}
\makeindex             

\newcommand{\mnras}{MNRAS }
\newcommand{\solphys}{Sol.\ Phys.\ }
\newcommand{\apj}{ApJ }
\newcommand{\apjs}{ApJS}
\newcommand{\apjl}{ApJL}
\newcommand{\aap}{A\&A}
\newcommand{\aapr}{A\&A Rev.\ }
\newcommand{\pasj}{PASJ}
\newcommand{\apss}{Astrophysics and Space Science }
\newcommand{\jgr}{JGR}
\newcommand{\grl}{GRL}

\newcommand{\zap}{Zeitschrift für Astrophysik }
\newcommand{\prl}{PRL}
\newcommand{\nat}{Nature }
\newcommand{\araa}{Annual Reviews of Astronomy \& Astrophysics }
\newcommand{\ssr}{Space Science Reviews }

\newcommand{\siiv}{Si\,{\sc iv}}
\newcommand{\feix}{Fe\,{\sc ix}}

\newcommand{\fexiv}{Fe\,{\sc xiv}}
\newcommand{\fexv}{Fe\,{\sc xv}}
\newcommand{\fexvi}{Fe\,{\sc xvi}}

\newcommand{\fexviii}{Fe\,{\sc xviii}}

\newcommand{\fexxi}{Fe\,{\sc xxi}}
\newcommand{\fexxv}{Fe\,{\sc xxv}}
\newcommand{\caxix}{Ca\,{\sc xix}}
\newcommand{\oviii}{O\,{\sc viii}}
\newcommand{\neix}{Ne\,{\sc ix}}
\newcommand{\mgxi}{Mg\,{\sc xi}}
\newcommand{\sixiii}{Si\,{\sc xiii}}
\newcommand{\sxv}{S\,{\sc xv}}
\newcommand{\kms}{~km~s$^{-1}$}

\begin{document}
\title*{The Solar X-ray Corona}
\author{Paola Testa \thanks{corresponding author} and Fabio Reale}
\institute{Paola Testa \at Harvard-Smithsonian Center for Astrophysics, 60 Garden St, Cambridge, MA 02193, USA, \email{ptesta@cfa.harvard.edu}
\and Fabio Reale \at Dipartimento di Fisica e Chimica, Università di Palermo,  Piazza del Parlamento 1, 90134 Palermo, Italy, and INAF-Osservatorio Astronomico di Palermo, Piazza del Parlamento 1, 90134 Palermo, Italy, \email{fabio.reale@unipa.it}}

%
%
\maketitle
\abstract{The X-ray emission from the Sun reveals a very dynamic hot atmosphere, the corona, which is characterized by a complex morphology and broad range of timescales of variability and spatial structuring. The solar magnetic fields play a fundamental role in the heating and structuring of the solar corona. Increasingly higher quality X-ray solar observations with high spatial (down to subarcsec) and temporal resolution provide fundamental information to refine our understanding of the solar magnetic activity and of the underlying physical processes leading to the heating of the solar outer atmosphere. Here we provide a brief historical overview of X-ray solar observations and we summarize recent progress in our understanding of the solar corona as made possible by state-of-the-art current X-ray observations. 
}
\section{Keywords} 
Solar Physics -- Solar Atmosphere -- Solar Activity -- Solar Flares -- Solar Corona -- High-Energy Astrophysics -- Plasma Astrophysics  -- X-ray Astronomy -- Magnetohydrodynamics (MHD) -- Solar Radiation

\section{\textit{1. Introduction}}
\label{sec:intro}

Observations of solar eclipses, in visible light, allowed occasional glimpses of the solar corona. Spectroscopic observations during eclipses detected a bright green optical line, initially tentatively assigned to a new element, "coronium", but in the 20th century it was finally correctly interpreted as a line emitted by highly ionized iron (\fexiv\ \cite{Grotrian1939,Edlen1943}) and therefore it was recognized that the solar corona is extremely hot ($\gtrsim 10^6$~K).
The hot plasma in the solar atmosphere primarily emits at high energies (X-ray and extreme-ultraviolet [EUV]), which are however very efficiently absorbed by the terrestrial atmosphere therefore rendering ground observations at those short wavelengths infeasible. So it was only with the advent of space astronomy and in  particular X-ray observations, from the 1960s onward, first with rocket experiments and then with satellite-based observatories (e.g., \cite{Blake1963,Dere1974,Hall1970,Vaiana1976,Acton1980}), that the nature of the solar outer atmosphere could be properly investigated (see e.g., \cite{Golub2009}).

\begin{figure*}[!t]
\centering
\includegraphics[width=0.99\textwidth]{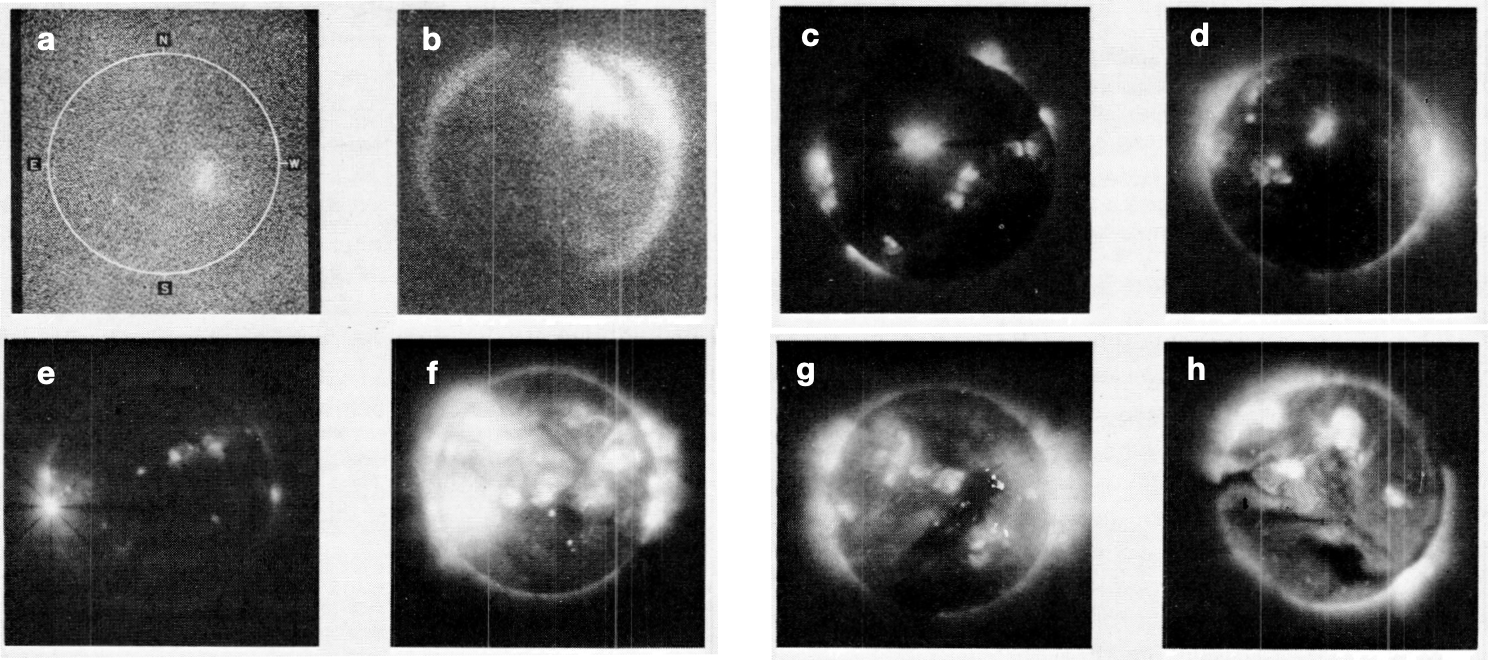}
\caption{Early solar X-ray observations illustrating the rapid progress of X-ray imaging techniques from 1963 to 1973: (a) first image of solar X-ray emission  made with a grazing incidence telescope (October 15 1963); (b) image from March  17 1965, with $\sim 30$" resolution;  (c) image from June 8 1968 (these data are discussed in \cite{Vaiana1973b}); (d) image from April 8 1969, in which X-ray bright points start being resolved; (e) image from November 4 1969, also detecting a flare at east limb; (f) X-ray image shortly after fourth contact of the solar  eclipse of March 7 1970; (g) image from November 24 1970;  (h) image from March 8 1973, obtained from the first successful flight of an X-ray telescope with higher efficiency.  (Image adapted from Figure~1 of \cite{Vaiana1973b})
}\label{fig:vaiana73}
\end{figure*}

The first X-ray imaging observations of the Sun, carried out with pin-hole cameras on-board rocket flights, were characterized by limited spatial resolution (arcmins), and nevertheless immediately uncovered the highly structured spatial distribution of the high-energy emission (e.g., \cite{Blake1963,Blake1965,Russell1965}). The X-ray emission was in fact observed to have significantly enhanced emission in small regions, which were co-spatial with H$_\alpha$ and Ca K plages, and with radio emission (e.g., \cite{Blake1965,Russell1965}).
Higher spatial resolution (down to few arcsecs) and sensitivity in X-ray astronomical observations were achieved with grazing incidence optics, focusing high energy photons using coaxial and confocal surfaces (e.g., \cite{Giacconi1960,Giacconi1962,Giacconi1965};  Figure~\ref{fig:vaiana73}).
The higher resolution provided further convincing evidence of the connection between magnetic active regions (ARs hereafter) and coronal features with brighter X-ray emission
(see e.g., \cite{deJager1965,Goldberg1967} for early reviews).
Furthermore, the hot coronal plasma appeared to be mostly confined in closed-loop features connecting regions of opposite polarity (e.g., \cite{Vaiana1973a,Vaiana1973b}), tracing the three-dimensional configuration of the magnetic field in the corona (\cite{Krieger1971}). These early observations highlighted the fundamental role of the solar magnetic fields in shaping and heating the corona.
Plasma diagnostics based on X-ray spectroscopy (e.g., \cite{Blake1964,Blake1965,Fritz1967,Evans1967,Evans1968,Gabriel1969,Gabriel1973,Dere1974}), or on analysis of X-ray imaging observations (ratios of images in two X-ray passbands, e.g., \cite{Vaiana1973b}) provided early determinations of plasma properties of the coronal plasma, such as temperature and density, necessary to constrain early models of coronal structures (e.g., \cite{RTV}).

\begin{figure*}[!t]
\centering
\includegraphics[width=0.515\textwidth]{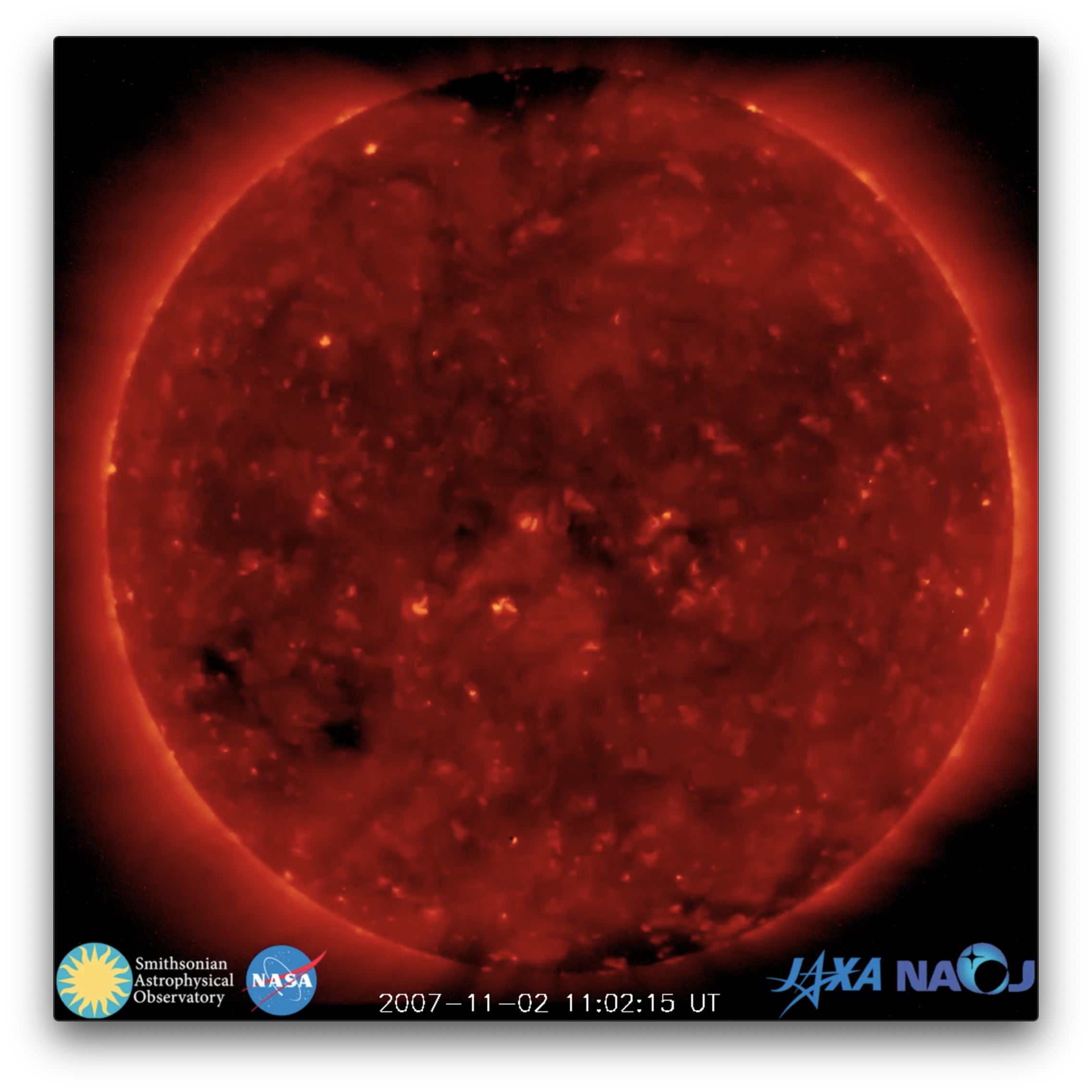}\hspace{-0.5cm}
\includegraphics[width=0.515\textwidth]{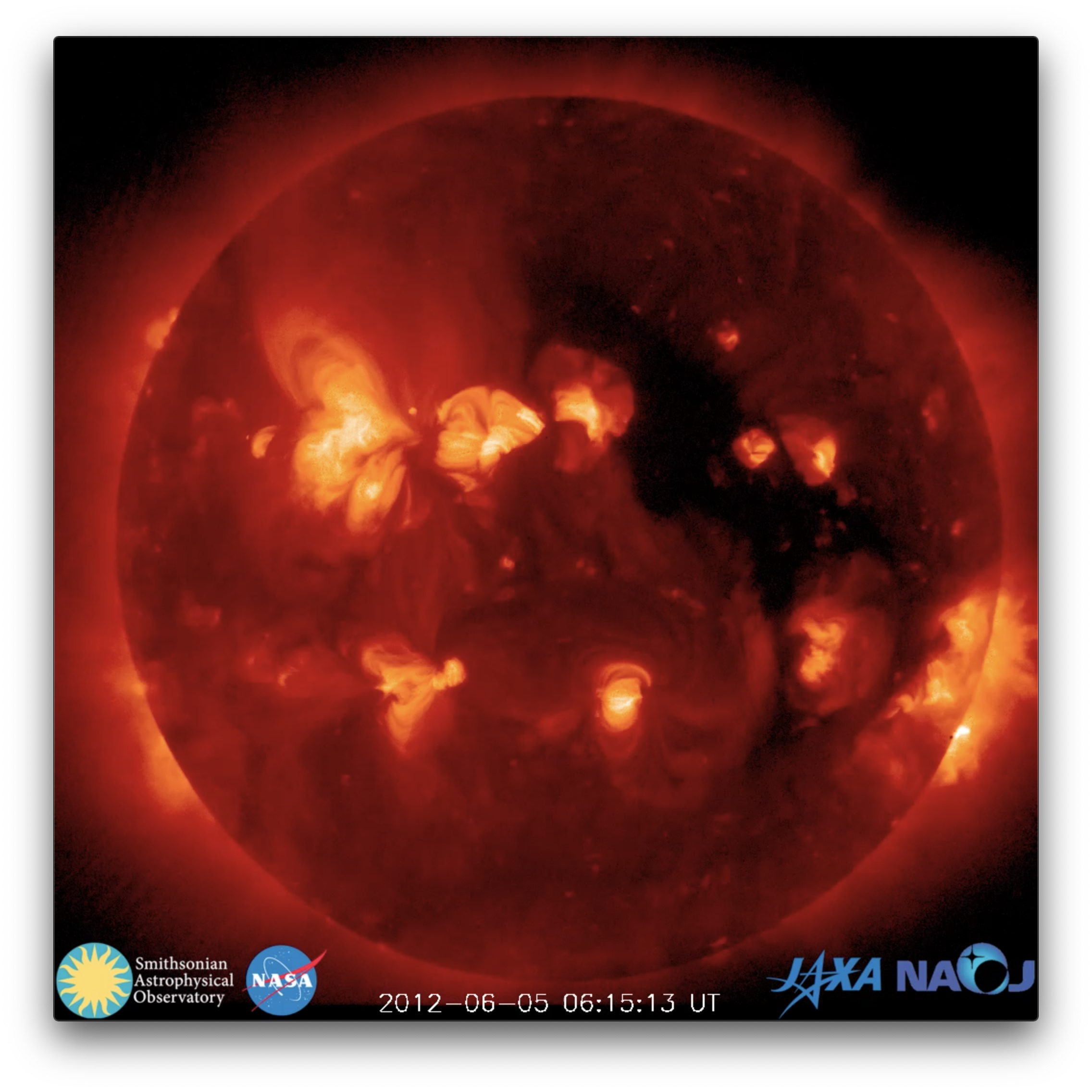}
\caption{Full disk images of the Hinode X-ray Telescope (in the Ti\_poly filter; see \cite{Golub2007} for details), characterized by $\sim 1$" pixels, for two times about 5 years apart and showing very different X-ray emission properties at different phases of the solar activity cycle. The XRT images illustrate the variety of X-ray solar features as well as the significant variation of the X-ray emission during the activity cycle. The Quiet Sun image (November 2007, {\em left}) shows diffuse emission, enhanced emission in small coronal Bright Points (CBP), and dark regions or Coronal Holes (CH, here mostly close to the Poles). The active Sun image (June 2012, {\em right}) shows bright emission in Active Regions (ARs; a dozen AR are present in this image), where most of the X-ray emitting plasma is confined by the magnetic field in loop-like arches (including loops connecting different ARs), CBPs (although partly obscured by the AR emission), and CH.
(Images adapted from XRT Picture Of the Week archive, https://xrt.cfa.harvard.edu/xpow/20120915.html)}\label{fig:XRT_syn}
\end{figure*}

The rich rocket program of the pioneering early age of solar X-ray astronomy produced a wealth of X-ray solar observations that engendered a rapid progress in the understanding of the solar corona. However, rocket experiments observed only for a few minutes, and therefore only with instrumentation onboard satellites, starting with the Skylab mission (\cite{Vaiana1977}), it became possible to obtain long time series of X-ray observations and investigate the temporal variability of the solar coronal features on longer timescales (from hours to days; e.g., \cite{Golub1974}).
These observations unveiled a scenario of a solar corona characterized by highly dynamic and spatially structured coronal features, even in the "quiet Sun", conclusively dispelling the idea of a "quiet homogeneous corona" \cite{Vaiana1973a}. 
The observed coronal structures encompass (see  Figure~\ref{fig:XRT_syn} for an illustration) "quiescent coronal structures", with relatively steady or slowly varying X-ray emission, and "flares", highly dynamic events during which the X-ray emission can increase by orders of magnitude on timescales of seconds, and which are accompanied by significant restructuring of the magnetic field configuration over short timescales, as well as, not uncommonly, by coronal mass ejections (CMEs; see section~4~\ref{sec:fl}). The quiescent (non-flaring) corona includes, quiet Sun (QS; section~2~\ref{sec:qs}), coronal holes (CHs, regions where the solar magnetic field extends into interplanetary space as an open field, and which are characterized by cooler and less dense coronal plasma, and diffuse and weak X-ray emission; section~2~\ref{sec:qs}), coronal bright points (CBP, initially named X-ray bright points, XBP, which are small regions in QS with increased X-ray emission; section~2~\ref{sec:qs}), active regions (ARs; regions with increased coronal temperature and density, and, in turn, X-ray emission, and which are associated with strong and complex photospheric magnetic field; section~3~\ref{sec:ar}), as well as coronal features interconnecting ARs.
In the quiescent corona outside of ARs (e.g., QS, CHs) the typical observed temperatures are of the order of $\sim 10^{6}$~K, and the coronal densities of the order of a few $\sim 10^{8}$~cm$^{-3}$. In coronal emission of ARs the temperatures on average reach values of $\sim 4 \times 10^{6}$~K (\cite{Warren2012}), and the densities are typically $\sim 10^{8}$-$10^{9}$~cm$^{-3}$, although localized heating events lead to short-lived temperature and density increases (up to $\sim 10$~MK and $\sim 10^{10}$~cm$^{-3}$ respectively;  e.g., \cite{Reale2019a,Testa2020}). Large flares can reach temperatures of several tens of MK and high densities of $\sim 10^{12}$~cm$^{-3}$, and have duration ranging from minutes to several hours (see e.g., \cite{Benz2017} for a recent review).

Solar X-ray observations show that the structure and dynamics of the solar corona is dominated by the magnetic field. The brightest high-energy emission mainly originates from plasma which is: (a) confined by the magnetic field in loop-like structures connecting regions of different magnetic field polarity, and (b) characterized by high temperature ($\gtrsim 1$~MK), low density ($\lesssim 10^{9}$~cm$^{-3}$), and therefore highly ionized and an efficient heat conductor.
The importance of the magnetic field for the coronal activity is also highlighted by the variations of the X-ray emission with the phases of the solar 11-yr magnetic cycle (\cite{Hathaway2015,Charbonneau2020}) as shown for example in Figure~\ref{fig:XRT_syn}.
The importance of the magnetic field for coronal activity is clear for the Sun (e.g., \cite{Wiegelmann2012,Petrie2015,Brun2017}), and strong evidence is provided by observations of other stars that X-ray coronae are associated with magnetic fields. In fact, strikingly, a power-law relation is observed between X-ray luminosity and magnetic flux over many orders of magnitude, ranging from solar quiet-regions, through active regions, to dwarf and T-Tauri stars (\cite{Fisher1998,Pevtsov2003}).

These observations prompted the key question of how the outer solar atmosphere is heated to such high temperatures of millions of degrees. The issue of the coronal heating is important not only in the context of solar and stellar physics,  but also because of the fundamental role of the solar high-energy emission on the Earth's atmosphere, and, analogously, to understand the effect of stellar X-ray and EUV emission on exoplanetary atmospheres (e.g., \cite{Lammer2003,Ribas2005,Penz2008,Sanz-Forcada2011}).
The magneto-convective energy at the underlying photosphere is clearly the source powering the corona, but the details of the processes converting it to heat of the corona are still under investigation. The shuffling of the photospheric footpoints of the coronal magnetic field lines, due to the convective motion of the plasma at and below the photosphere, can generate atmospheric heating mainly via two general mechanisms: (1) production of MHD (Alfv\'en) waves, that get transported to and dissipated in the corona (\cite{Alfven1947}), and (2) magnetic stresses building current sheets, dissipated via magnetic reconnection events ("nanoflares"; \cite{Parker1988}, having typical total energies $\sim$ 9 orders of magnitude smaller than large flares). There is observational evidence for the presence of both processes, although their relative importance is not well established as well as whether different coronal features are dominated by different mechanisms. Advances in computational power allow ever more sophisticated simulations to model these processes, including e.g., 3D radiative MHD models of the solar atmosphere from the convection zone up to the corona, which self-consistently produce a hot atmosphere (primarily via dissipation of currents, e.g., \cite{Gudiksen2011,Hansteen2015,Rempel2017}), and e.g., a 3D reduced MHD model for the propagation and dissipation of Alfv\'en waves in a coronal loop (\cite{vanBallegooijen2011,vanBallegooijen2017}). 
Another intriguing mystery of the solar corona is its anomalous chemical composition, at odds with the underlying photospheric abundances, as observed since the earliest spectroscopic coronal studies. The coronal plasma, in fact, is often characterized by higher abundances of elements with low first ionization potential (FIP) compared to the underlying photosphere (e.g., \cite{Feldman1998,Feldman2003}). This chemical fractionation is expected to occur in the chromosphere, where low-FIP and high-FIP elements are differently ionized, and is likely associated with the heating mechanism, therefore providing additional clues and constraints to heating models (see e.g., reviews by \cite{Testa10ssrv,Testa2015,Laming2015} and references therein). We note that also the coronae of other stars are typically characterized by abundance anomalies, compared to their photospheric composition, although their abundance patterns display a wide range of variation from solar-like FIP effect for low to intermediate activity start to inverse FIP effect (i.e., with coronal depletion of low-FIP elements) for high activity stars (see e.g., \cite{Testa10ssrv,Testa2015}, and Drake \& Stelzer in this handbook). 
There are a number of observational obstacles that hamper progress in our understanding of the coronal heating and chemical fractionation processes, including the small (unresolvable) spatial scales characterizing the energy release, and the fact that the whole atmosphere is a complex coupled system with both energy and mass transferred in both directions between the corona and the lower thin atmospheric layer formed by chromosphere and transition region (see e.g., \cite{Klimchuk2006,Parnell2012,Testa2015,DePontieu2021}, for reviews).

\begin{figure*}[!t]
\centering
\includegraphics[width=0.99\textwidth]{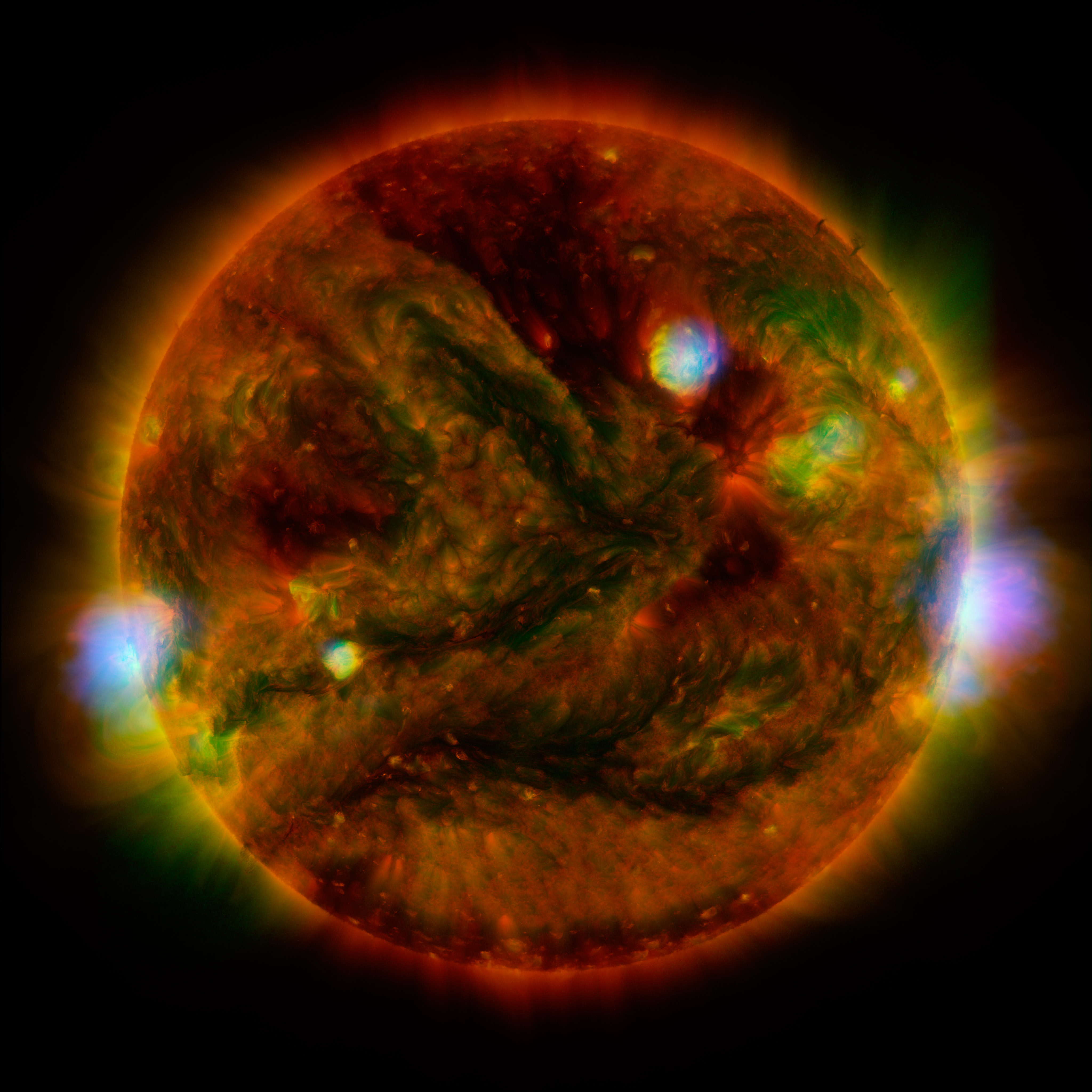}
\caption{This multi-color full disk solar image combines images obtained by several high-energy observatories: the NuSTAR observatory ({\em blue}; this image shows the hard X-ray emission in the $\sim 2$-6~KeV range), the Hinode X-ray Telescope ({\em green}; shows soft X-ray emission between 0.2 and 2.4~keV), and the Atmospheric Imaging Assembly onboard the Solar Dynamics Observatory ({\em yellow} and {\em red}; the AIA passbands shown here, 171\AA\ and  193\AA, detect EUV emission typically emitted by plasma around $\sim 0.8$~MK and  $\sim 1.5$~MK respectively). This composite image nicely shows how the higher energy emission is generally confined to the bright ARs.
(Image from https://www.nasa.gov/jpl/pia19821/nustar-stares-at-the-sun. See also \cite{Grefenstette2016}.)}\label{fig:Xray_composite}
\end{figure*}

The importance of these open issues, in our astronomical backyard, motivated the several satellite based X-ray observatories \footnote{See e.g., \cite{Aschwanden2005book} for a more detailed list of solar coronal instrumentation, also including other wavelengths, such as the extreme-ultraviolet, ultraviolet and visible, which we are mostly not discussing in this overview.} that allowed going beyond the early discoveries, including: 
\begin{itemize}
    \item SkyLab (1973 – 1979), equipped with grazing-incidence X-ray telescopes, EUV/UV spectrographs \citep{Vaiana1976};
    \item the Solar Maximum Mission (SMM, 1980, 1984-1989), which included $\gamma$-ray, hard and soft X-ray spectrometers, as well as an X-ray imager; \citep{Acton1980};
    \item Yohkoh (1991–2001), which included hard and soft X-ray telescopes, as well as spectrometers at $\gamma$-ray, hard and soft X-ray wavelengths \citep{yohkoh};
    \item CORONAS-I (1994; \cite{SobelMan1996}), CORONAS-F (2001-2005; \cite{Zhitnik2003}), CORONAS-PHOTON (2009, \cite{Kuzin2009});
    \item the Hinode X-ray Telescope (2006–present; \cite{Golub2007});
    \item the Atmospheric Imaging Assembly (AIA; \cite{Lemen12}) onboard the Solar Dynamics Observatory (SDO, 2010-present); note that AIA mostly observes in the EUV, and its passbands at shorter wavelengths in soft-X-ray, at 94\AA\ and 131\AA, are mainly sensitive to hot plasma at 5-10~MK.
\end{itemize}  
Although here we mostly focus on soft X-rays, several observatories -- such as e.g., RHESSI (2002-2018;\cite{Lin2002}), the FOXSI sounding rockets (2012, \cite{Krucker2014}; 2014, \cite{Christe2016}), NuSTAR (2012-present; \cite{Harrison2013} \footnote{Note however that NuSTAR is an astrophysical observatory, not designed to observe the Sun, and not suitable for observations of bright solar sources such as flares or even bright ARs}) -- have recently provided solar observations at hard-X-ray energies (above $\sim 5-10$~keV), which are important to study the highest energy phenomena,  including particle acceleration to relativistic speeds, especially prominent in flares.   Figure~\ref{fig:Xray_composite} shows an example of composite of hard X-ray, soft X-ray, and EUV imaging observations obtained with different observatories.  Saint-Hilaire et al.\ (in this handbook) discuss hard X-ray and Gamma-Ray of solar flares.

High-resolution X-ray spectroscopy is of key importance to solve the coronal heating mystery, providing unique diagnostics of plasma temperature, density, dynamic (large scale flows, turbulence), chemical composition (see e.g., \cite{DelZanna2018} for a review) not accessible to imaging observations. Although a wealth of spectral observations of the solar corona exist, the vast majority of them have focused on either relatively low temperature ($\sim 3$~MK) emission, e.g., the Hinode/EIS (Extreme ultraviolet Imaging Spectrometer \cite{Culhane2007}) and the Interface Region Imaging Spectrograph (IRIS; \cite{DePontieu14}), or on very high temperature emission ($\gtrsim 10$~MK) mostly observed in large flares. The 5-10~MK temperature regime, most relevant to study the heating process, is however poorly covered by present and past solar coronal spectrometers (see e.g., \cite{DelZanna2021}), whereas astrophysical observatories such as {\em Chandra} and {\em XMM-Newton} have provided, as discussed at length in the next chapters (Drake \& Stelzer; Schneider et  al.; Sciortino; Kastner \& Principe; all in this handbook), excellent coverage of these temperatures via high-resolution spectroscopy of stellar coronae in the soft X-ray range ($\sim 1$-170\AA).

Indeed, parallel efforts in X-ray astronomy have provided high quality high-energy observations of stars, that have enabled comparing and contrasting the X-ray properties of the Sun with other stars and funded the basis of the solar-stellar connection (refer to other book chapters; and other refs e.g., \cite{Testa2010,Testa2015}). 
Understanding stellar magnetic activity is of crucial importance also due to the key role magnetic fields play in several aspects of stellar astrophysics, ranging from the formation and evolution of stars and planets, controlling angular momentum loss, and orbital decay and evolution in close binary systems.

\section{\textit{2. Quiet Sun, Coronal Bright Points, Coronal Holes}} \label{sec:qs}
The quiet corona ({\em quiet sun}, QS) is significantly less X-ray bright than active regions, and, at modest spatial resolution, appears to be characterized by a more diffuse nature and structures smaller than typical AR loops.  In QS, which is devoid of sunspots, the underlying photospheric fields have mixed polarity and are characterized by a reticular pattern of strong fields ({\em magnetic network}, tracing the downdrafts of flows at the boundaries of convective supergranular cells) and weaker field {\em internetwork} regions in between them (\cite{Bellot2019}). 
Even in these quiet regions the temporal variability of EUV emission suggests that the QS corona is impulsively heated (\cite{Upendran2021}).

Small bipolar regions in QS give rise to {\em coronal bright points} (CBP). CBPs are scaled-down ($\lesssim 60$") and shorter lived ($\lesssim 1$~day) versions of ARs, with intense emission in the X-ray and EUV wavelength ranges, and composed of small-scale loops connecting magnetic flux concentrations of opposite polarities (\cite{Madjarska2019} and references therein). 
The evolution of the X-ray emission of CBPs shows a rapid growth and a slow decay, with a diffuse emission preceding the appearance of the bright core (\cite{Golub1974,Mou2018}). The average growth rate ($\sim 5$" per hour, or $\sim 1$\kms) is similar to the horizontal velocity of supergranulation cells suggesting a strong connection of CBPs with supergranular evolution/flows (\cite{Golub1974}).
Contrary to ARs, which are confined to a belt of active latitudes (within $\sim \pm 40^{\circ} $ from the solar equator), the CBPs are observed at all latitudes. 
Up to several hundreds of CBPs are present on the solar disk in a day (\cite{Alipour2015}), although a subset of them are only visible at EUV wavelengths and not in X-rays due to a broad range of temperature distributions in CBPs (\cite{Madjarska2019}).
Several transient phenomena, such as microflaring (e.g., \cite{Strong1992}), coronal jets (e.g., \cite{Brueckner1983,Raouafi2016})  and mini-CMEs (e.g., \cite{Mou2018}) are observed in CBPs.
The variation of CBPs during the solar cycle was studied with Skylab data and their number was found to anticorrelate with the number of sunspots (\cite{Davis1977,Golub1979}). On the basis of a more recent analysis of data with the Soft X-ray Telescope (SXT) \cite{Tsuneta1991sxt} onboard Yohkoh it was however concluded that this anticorrelation is largely an artifact of the contrast with the soft X-ray intensity of the background corona and the obscuring effect of ARs (\cite{Hara2003}). In fact, Hara et al.\ \cite{Hara2003} find that the number density of observed CBPs is largely independent on the solar activity cycle, but in dark areas (such as CHs) the CBPs number anticorrelates with the sunspot cycle and in the activity belt it correlates with it, suggesting a different mechanism of formation for CBPs in CHs. More recently McIntosh et al.\ \cite{McIntosh2014} suggested that small-scale magnetic flux emergence activity associated to CBPs is related to and could be used to forecast the solar cycle. 

{\em Coronal Holes} (CH) are large-scale coronal features appearing as extended dark areas in X-ray images.  Their underlying photospheric magnetic fields are more uniform and unipolar than other solar regions, and nearly all the open magnetic flux on the Sun -- and the high-speed solar wind flowing outwards into the heliosphere along open magnetic field lines -- originates in CHs (\cite{Krieger1973,Cranmer2009,Harvey2013,Petrie2015}).
The spatial distribution of CHs is observed to vary as a function of the solar activity cycle.  Polar coronal holes, prominent for several years around solar minimum, tend to disappear around solar maximum (see e.g., Figure~\ref{fig:XRT_syn}) -- with a lack of any distinct presence 1-2 yrs at solar maximum, and the gradual reappearance of polar CHs in the declining phase soon after solar maximum, as several high-latitude CHs start collecting at the poles (e.g., \cite{Timothy1975,Harvey2002}). On the other hand, very large meridian coronal holes can appear at medium latitudes during periods of high activity, as shown on the right of Fig.~\ref{fig:XRT_syn}. A very famous CH with a vague shape of a boot was observed to be steady for several solar rotations during the Skylab mission \footnote{http://soi.stanford.edu/results/SolPhys200/Hudson/2000/001020/001020.html}.
Coronal plasma in CHs is characterized by elemental abundances close to photospheric abundances, in contrast to QS and ARs in which the coronal plasma typically shows a significant enhancement in the corona of the abundances of elements with low first ionization potential (FIP) compared to the underlying photosphere.
These abundance anomalies also provide useful diagnostics to trace the flows of solar wind from the solar origin into the heliosphere (e.g., \cite{vonSteiger1995,Zurbuchen2002,Brooks2015}). 

Albeit Yohkoh/SXT provided good observations of the high temperature emission from the quiet corona, its successor, Hinode/XRT has provided a much improved view of several coronal phenomena, especially in the quiet corona, due to its higher spatial resolution, temporal cadence, and broader temperature coverage (extending its sensitivity to lower coronal temperatures, $\sim 1$~MK). A case in point are X-ray jets which XRT has observed to occur frequently in coronal holes (CHs) and quiet Sun (QS) including polar regions, and indicated that these jets are characterized by plasma velocity, temperature and density of the order of $\sim 160$~\kms, $\sim 1-2$~MK, and a few $10^8$~cm$^{-3}$ respectively (e.g., \cite{Savcheva2007,Shimojo2007,Cirtain2007,Paraschiv2015}). XRT has elucidated the details of the magnetic reconnection process producing these jets (see e.g., \cite{hinodereview2019} for a review). The XRT observations of coronal jets also allow to investigate whether they might be significant contributors to the coronal heating of QS and to the solar wind, and suggest their contributions is limited (e.g., \cite{Paraschiv2015,Yu2014,Sako2013}).

Another significant contribution of Hinode/XRT observations is its continuous monitoring of the full disk X-ray solar emission over more than a solar cycle.  In fact, Hinode/XRT has regularly performed synoptic observations in different filters (therefore also allowing temperature diagnostic) providing a very valuable archive of monitoring (over about 15 yrs) of the properties of the full Sun X-ray emission with high temporal cadence (typically higher than daily). These data are perfectly suitable for studies of the solar activity cycle, and for solar-stellar connections (see also e.g., \cite{Testa2015}). Adithya et al.\ (2021) \cite{Adithya2021} used XRT full Sun observations over 13 years (2007-2020) to study the variability of the solar X-ray irradiance, and investigate the contribution of different solar features (AR, CH, QS, XBP) to it.  They find that QS and AR emission has the greatest impact on the observed fluctuations of the solar irradiance.  
They find that the QS emission varies with the cycle, undermining the idea of a constant background X-ray emission (see also, \cite{Kariyappa1999}, and \cite{Zhang2001} on the cyclic variation of QS emission from the low chromosphere; and, e.g., \cite{Zender2017} for a study of irradiance variability in EUV and UV). Detailed investigations of the solar irradiance in high energy (X-ray/EUV) are also very relevant to the understanding of the effects of stellar radiation on escape of exoplanetary atmospheres (e.g., \cite{Hazra2020}).

\section{\textit{3. Active Regions}} \label{sec:ar}

While the coronal emission of QS is often confined to lower temperatures ($\sim 1$~MK), active regions (ARs) typically contain plasma at a variety of temperatures up to several MK (see e.g., \cite{Reale2014} for a review). 
ARs are characterized by a typical evolution showing rapidly increasing high temperature emission and variability in the first few days from first emergence, followed by a longer decay during which the AR spreads out and its core temperature and variability of emission systematically decreases (\cite{vanDriel2015,DelZanna2015}). There is some evidence that this evolution is also accompanied by a significant change of chemical fractionation of the coronal plasma (e.g., \cite{Widing2001,Testa10ssrv,Baker2015,Baker2018}).

\begin{figure*}[!t]
\centering
\includegraphics[width=0.99\textwidth]{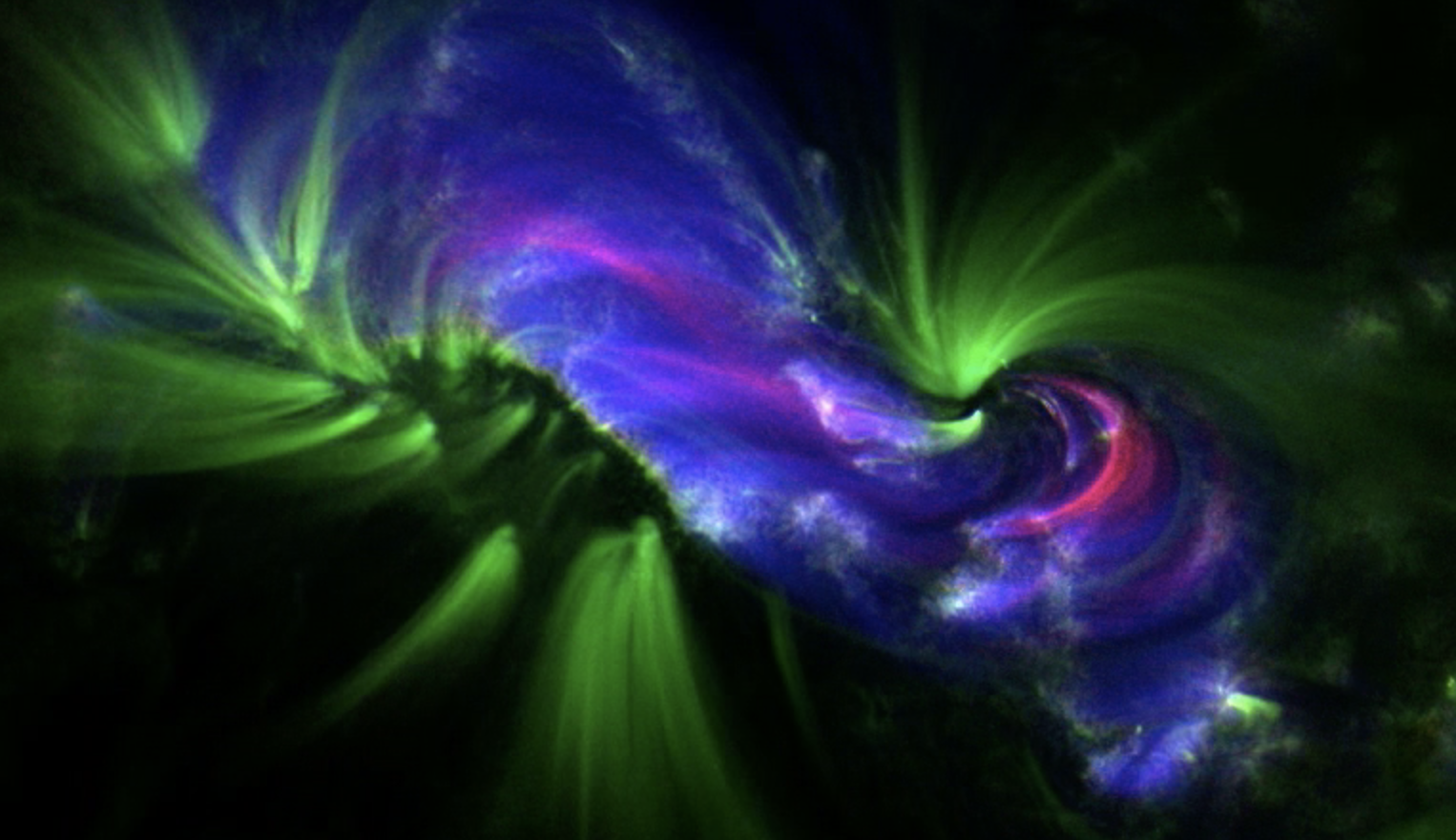}
\caption{This 3-color image combines SDO/AIA observations of AR 11289 (on September 13 2011, around 19UT) in 3 EUV narrow passbands: 171\AA\ (green; dominated by \feix\ emission mostly emitted by $\sim 1$~MK plasma), 335\AA\ (blue; with a strong \fexvi\ line emitted prominently around 2-3~MK), and 94\AA\ (red; with a hot \fexviii\ line emitted at T$\gtrsim 5$~MK; see e.g., \cite{Testa2012b}). The field of view shown here is about $400" \times 250"$ (i.e., roughly $290 \times 180$~Mm). These data clearly show the presence of cooler (green loops) close to the edges of the AR, and hotter material (blue loops and diffuse emission; 2-4~MK) in the AR core, with additional hottest emission (red loops; $\gtrsim 5$~MK) which is highly dynamic (e.g., \cite{UgarteUrra2014,Testa2013,Testa2020b}).
}\label{fig:3col_AR}
\end{figure*}

Warm AR loops ($\sim 1$~MK), typically observed outside the AR core and especially clear in the cool EUV emission (see Figure~\ref{fig:3col_AR}; see also e.g., \cite{hinodereview2019} and references therein), are characterized by a narrow temperature distribution.
High temperature emission is observed in the AR core (see Figure~\ref{fig:3col_AR}): the temperature distribution in the AR cores is often strongly peaked around 4~MK, and the amount of high temperature plasma appears to be correlated with the total unsigned magnetic flux (e.g., \cite{Warren2012}). Emission from even hotter plasma (T $\gtrsim 5$~MK), even in heating events ("nanoflares" to "microflares") smaller than large flares (the next section will discuss flares in details) is observed to be highly transient (e.g., \cite{UgarteUrra2006,UgarteUrra2009,Testa2012b,Testa2013,UgarteUrra2014,Graham2019,UgarteUrra2019,Reale2019a,Testa2020,Testa2020b}), and typically a few of these brightenings per hour are observed in an AR (e.g., \cite{UgarteUrra2014}).

High spatial and temporal resolution coronal observations are necessary to pinpoint the physical mechanisms responsible for coronal heating. High-resolution X-ray observations, with Hinode/XRT and the short wavelength bands of SDO/AIA (94\AA, and 131\AA), over a large AR-sized field of view (and for AIA over the full disk, and continuously at 12s cadence), provide, together with coronal spectral diagnostics provided e.g., by Hinode/EIS, important constraints on theories of coronal heating for the hotter portion of AR coronal plasma. 
Viable coronal heating models predict energy release on small spatial and temporal scales (see e.g., reviews by \cite{Klimchuk2006,Klimchuk2015,Reale2014}), and evidence of impulsive heating is provided by coronal observations (e.g., \cite{Terzo2011,Testa2013,Testa2014,Viall2011,Viall2017,Reale2019a,Testa2020}); it is however not well established whether heating events typically repeat over timescales shorter than the loops cooling times (i.e., heating is effectively steady), or instead with low frequency (i.e., significant cooling occurs in between heating events).
Several coronal observables have been analyzed and modeled to gain insights into the heating of the AR cores, such as e.g., the properties, and in particular the slopes, of the thermal emission measure distribution (e.g., \cite{Klimchuk2001,Cargill2004,Testa2005,Warren2012,Bradshaw2012,Cargill2015,Barnes2019}). The presence of a faint high temperature component is another important heating diagnostic, and one of the main predictions of impulsive nanoflare heating. It has been investigated using imaging observations (e.g., \cite{Reale2009,Schmelz2009a}) also in combination with spectral data (e.g., \cite{McTiernan2009,Ko2009,Sylwester2010,Testa2011,Testa2012b,Miceli2012,petralia14,Brosius2014,Ishikawa2014,Athiray2020}), and the observations are typically compatible with such a low level hot emission (2-4 orders of magnitude lower than the emission measure peak at $2$-4~MK), which however is often close to the detection limits of current instrumentation.
An accurate determination of these subtle properties of the thermal distribution of the AR plasma is in fact somewhat difficult because of the limitations of observations and inversion methods (e.g., \cite{Testa2012c,Winebarger2012}), as well as intrinsic characteristics of the coronal plasma including for instance efficient conduction and non-equilibrium effects (e.g., \cite{Reale2008}) that drastically limit the emission of the hot plasma in the initial phases heating events.

Higher spatio-temporal resolution observations of the solar corona, in particular at sub-arcsec spatial resolution (e.g., with the rocket experiments Normal Incidence X-ray Telescope, NIXT, \cite{Golub1990}, and High-resolution Coronal imager, Hi-C, \cite{Kobayashi2014}), have shown that, although the heating release is expected to happen on much smaller scales, coronal loops typically appear to have a coherence on spatial scales of the order of $\sim 0.3$-0.5" (e.g., \cite{Antolin2012,Brooks2013,Aschwanden2017,Williams2020}). These rocket experiments have also revealed, thanks to their unprecedented resolution, previously unobserved coronal features, such as e.g., the "moss", i.e., the high-temperature transition region (TR) of hot loops (\cite{Peres04,DePontieu1999,Fletcher1999}), or the appearance of braiding of magnetic loops (\cite{Cirtain2013}). High resolution EUV observations with EUI onboard Solar Orbiter are providing interesting new observations of quiescent small-scale dynamic coronal features (e.g.,\cite{Berghmans2021}).
Another notable new result made possible by subarcsecond observations of the solar atmosphere is the detection of fast and  bursty "nanojets",  which are a direct observational signature of reconnection-driven nanoflares \cite{Antolin2021}. These nanojets were observed in the cool (transition region) UV emission with IRIS, and the lack of coronal observations at similar resolution prevents us from investigating these magnetic reconnection features in hotter coronal loops.

\begin{figure*}[!t]
\centering
\includegraphics[width=0.99\textwidth]{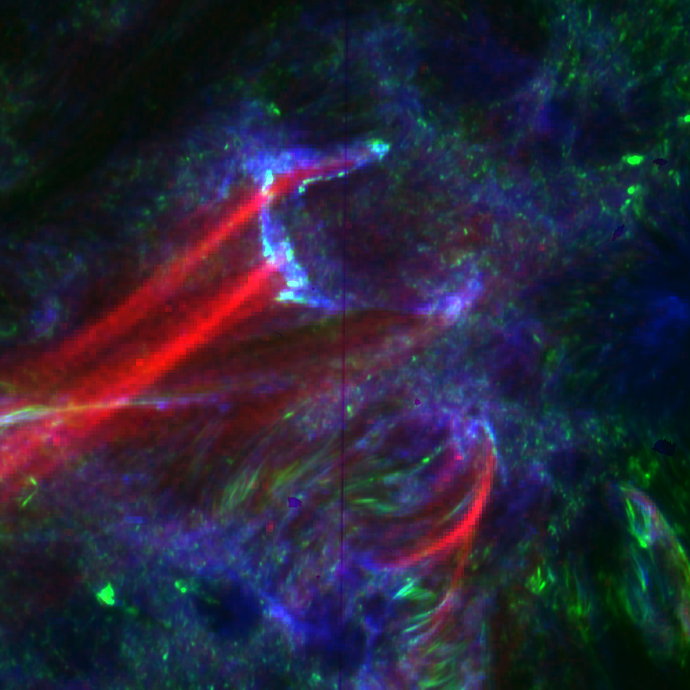}
\caption{3-color image combining SDO/AIA EUV observations and IRIS (Interface Region Imaging Spectrograph; \cite{DePontieu14}) FUV slit-jaw imaging observations of AR 11890 on November 9 2013, around 12:15UT (same data analyzed by \cite{Testa2014}): IRIS 1400\AA\ transition region (TR) emission (green; dominated by \siiv\ emission mostly emitted by $\sim 0.1$~MK plasma; the black vertical line is the IRIS slit where FUV and NUV spectral data are obtained), AIA 171\AA\ (blue; dominated by \feix\ emission mostly emitted by $\sim 1$~MK plasma), and AIA 94\AA\ (red; with a hot \fexviii\ line emitted at T$\gtrsim 5$~MK). The field of view shown here is about $115" \times 115"$ (i.e., roughly $83 \times 83$~Mm). These data clearly show brightenings in the cooler TR emission (green and blue) at the footpoints of dynamic hot ($\sim 10$~MK) loops (red). Both the hot coronal emission and the TR emission are highly transient, with TR emission brightenings typically lasting $\sim 10-60$s (\cite{Testa2013,Testa2014,Testa2020}).
}\label{fig:3col_AR_moss}
\end{figure*}

The study of the moss TR emission provides alternative diagnostics of coronal heating to overcome some of the above described obstacles in detecting heating events directly in the corona. The location of the heating is debated (see e.g., \cite{Reale2014} for a detailed review of observational evidence for footpoint vs.\ looptop heating) -- but, even if energy is released in the corona, the immediate response of the coronal plasma is often too difficult to detect, as discussed earlier in this section, and the hot plasma (e.g., pink/red loops in Fig.~\ref{fig:3col_AR}, and Fig.~\ref{fig:3col_AR_moss}) is typically observed mostly in the cooling phase. The response of the narrow and dense TR layer of the loops to the heating events is instead readily observable as the TR undergoes rapid and intense brightenings that can be easily detected and studied. Initial observations of the moss variability showed typically low temporal variability (e.g., \cite{Antiochos2003,Brooks2009,Tripathi2010}), and they were interpreted as an evidence of steady (high frequency) heating. 
However, the high spatio-temporal cadence of the Hi-C rocket observations have revealed the presence of highly variable moss at the footpoints of transient hot loops \cite{Testa2013}. These results suggest that the insufficient combined spatial and temporal resolution of the observations is one of the reasons for the previous lack of evidence of significant moss variability (see also \cite{Graham2019}). In fact, the IRIS (\cite{DePontieu14}) high spatial ($\lesssim  0.4$") and temporal (down to $\sim 1$~s) resolution has allowed to detect, at cooler TR temperatures, many of this type of footpoint brightenings associated with heating and typically observed simultaneously at both loop footpoints (e.g., \cite{Testa2014,Testa2020}; see example in Fig.~\ref{fig:3col_AR_moss}). IRIS provides also powerful additional diagnostics thanks to its spectral observations that also reveal the TR plasma velocities, which combined with coronal observations (XRT and AIA) and state-of-the-art modeling constrain the properties of the heating and of the energy transport (e.g., \cite{Testa2014,Polito2018,Testa2020}). The coronal observations indicate that these impulsively heated AR core loops typically reach temperatures of $\sim 10$~MK, and their morphology suggests that these hotter loops might arise as a consequence of large-angle magnetic reconnection in the corona (\cite{Reale2019a,Testa2020}; although in a few events a possible alternative scenario of energy release due to flux cancellation has been proposed; \cite{Priest2018,Chitta2018}), i.e., more significant magnetic rearrangements than the typically assumed small-angle reconnection Parker scenario for nanoflares (\cite{Parker1988}). 

The TR observations reveal for several of the observed events signatures of the presence of accelerated (non-thermal) electrons (NTE)), similar to what is observed in large flares (see also next section), and allow to constrain the properties of the NTE distribution. Non-thermal particles, accelerated to relativistic speeds during the magnetic reconnection process, are primarily studied using hard X-ray observations (e.g., with RHESSI; \cite{Lin2002}) which can detect the radiation directly emitted by the NTE. The observed hard X-ray emission from NTE is generally compatible with a single (or double) power-law, implying a power-law for the electron flux distribution as well, with  power-law index $\delta$ and a low-energy cutoff $E_{\rm C}$. Although direct observations of NTE in hard X-rays remain of fundamental importance to study their properties, these new indirect diagnostics of NTE in small events, based on spectral observations of footpoint (TR) emission, are of particular interest because (1) they are sensitive to very small energy events (nanoflares) which are typically inaccessible to hard X-ray observations due to the very low emission level close to or below the sensitivity threshold of hard X-ray observatories; and because (2) they can constrain the low-energy cutoff ($E_{\rm C}$) of the NTE power-law distributions, which is poorly constrained by hard X-ray spectra because of the overlap of thermal and non-thermal spectra.  

The small heating events (nano- to micro-flares) observed in the ARs core in many respects appear like scaled-down versions of large flares with high temperatures (up to $\sim 10$~MK, \cite{Reale2019a,Reale2019b,Glesener2020,Testa2020,Testa2020b,Cooper2020}), abundances closer to photospheric (\cite{Warren2016}, as often observed for larger flares), presence of NTE although often with smaller low-energy cutoff and steeper slopes ($E_{\rm C} \sim 5-15$~keV, $\delta \gtrsim 7$; \cite{Hannah2008,Testa2014,Wright2017,Testa2020,Glesener2020,Cooper2021}) than observed for large flares.
For these small events is often difficult to detect the hard X-ray emission of NTE (\cite{Hannah2016,Hannah2019,Glesener2017,Cooper2020,Vievering2021}), even with the astrophysical observatory NuSTAR (\cite{Harrison2013}) and the FOXSI rocket experiments which have higher sensitivity than RHESSI, because of the presence of a dominant thermal component and, for NuSTAR, instrumental limitations largely preventing it from observing the active Sun.

\section{\textit{4. Solar Flares and Coronal Mass Ejections}} \label{sec:fl}

Solar flares are the brightest X-ray events in the solar system. They are powerful explosions occurring in the solar corona, and they last from a few minutes to several hours. 
Their intensity spans several orders of magnitude, and the largest ones can easily exceed the luminosity of the rest of the solar corona. There is a standard classification of flare magnitude based on the maximum flux measured by the GOES X-ray detector in the 1-8\AA\ band, and the flux decades are marked with letters, A, B, C, M and X -- from low to high (e.g., https://spaceweather.com/glossary/flareclasses.html). Flaring plasma is also much hotter than typical quiescent plasma, easily reaching temperatures much above 10~MK.

\begin{figure*}[!t]
\centering
\includegraphics[width=0.95\textwidth]{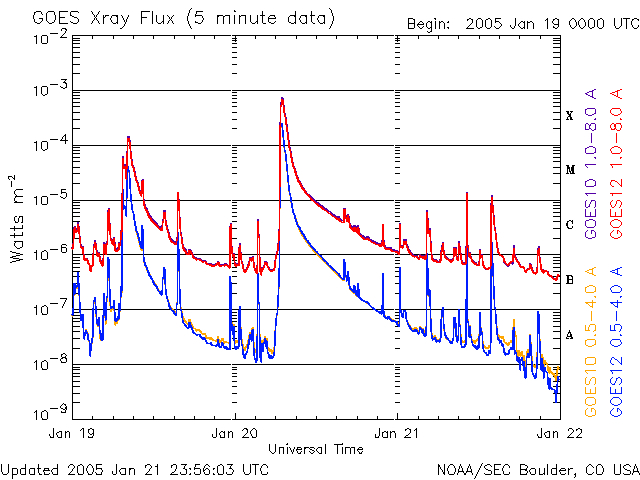}
\caption{X-ray flare light curves as detected by GOES in two different bands, with typical fast rise and slow decay. (from $https://hdrl.gsfc.nasa.gov/LWS_Space_Weather/GOES_Xray_descrip.html$).}\label{fig:goes}
\end{figure*}

\begin{figure*}[!t]
\centering
\includegraphics[width=0.99\textwidth]{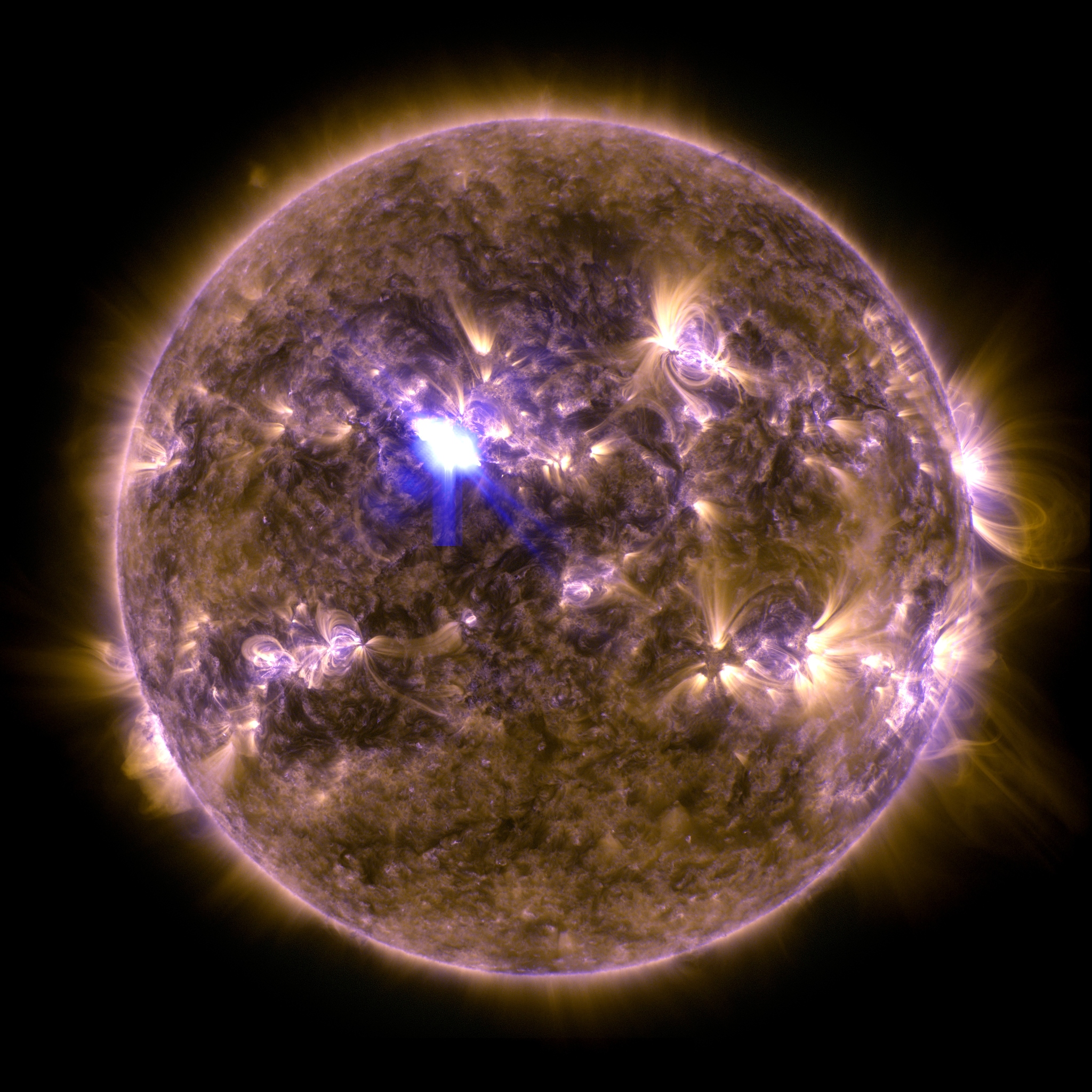}
\caption{False-color composite image in extreme ultraviolet light, from the Atmospheric Imaging Assembly onboard the  Solar Dynamics Observatory, of a solar flare recorded on April 11 at 07:11 UTC. The event, a moderate M6.5 class flare erupting from active region AR 11719, is near the center of the solar disk. Other active regions, areas of intense magnetic fields where sunspot groups are observed in visible light, mottle the surface as the solar maximum approaches. Loops and arcs of glowing plasma trace the active regions' magnetic field lines. (From NASA Astronomy Picture of the Day [APOD] website, 13 April 2013, https://apod.nasa.gov/apod/ap130413.html). }\label{fig:imgflare}
\end{figure*}

Flares are very complex phenomena, and their emission is detected at all wavelengths from radio to $\gamma$-rays \cite{Priest2002b,Benz2017}.
Here we focus on flares as observed in the soft X-rays and EUV. They are generally characterized by a fast/steep rise phase  and a much more gradual decay (Fig.~\ref{fig:goes}), and occur in extremely localized areas, typically inside active regions (Fig.~\ref{fig:imgflare}).
In the following we  present an historical overview of the main flare characteristics derived from X-ray observations.

\subsection{\textit{4.1 Surprising flares: rocket experiments and Skylab}} \label{sec:skylab}

The first X-ray flare observations date back to rocket experiments.
In particular, an early rocket launch was meticulously organized to occur "within a few minutes of the observation of a solar flare by an alarm network of ground-based solar observations" \cite{Vaiana1968}. It was a very important enterprise because, although it was known that flares do not involve the whole solar surface, until then the real extension of flaring regions was unknown. The flare was classified as a "large flare of the parallel ribbon type", i.e., a flare involving an arcade of loops the footpoints of which are visible in the lower atmosphere (photosphere/chromosphere/transition region) as two parallel ribbons (see below). The grazing incidence telescope allowed for high spatial resolution (down to about 2 arcsec) observations that revealed that the flare was extremely localized, and it was brighter than the other X-ray bright regions by at least one order of magnitude. The emitted energy was estimated to be about $10^{31}$~erg.

Another breakthrough in flare X-ray observations was achieved when telescopes could be placed on-board satellite missions. This allowed to study the flare evolution by providing time coverage of flares in their entirety. Particular care was devoted to analyzing the morphology and evolution of a specific flare observed on June 15, 1973 with telescope S-054 on board the Skylab mission \cite{Pallavicini1975}. An ingenious observation technique allowed to capture the broad dynamic range required for a flare observation: photographic images were taken in sequences of frames with different exposure times, from hundredth of seconds to few seconds. This allowed for an accurate morphological analysis and measurement of the length of the flaring structure. Furthermore, the temperature and emission measure, averaged over the volume, could also be derived, assuming an isothermal emitting plasma, by using a technique based on the principle of the optically thin emission:

\begin{equation}
I = \int_V{n^2 G(T) dV} \sim G(<T>) \int_V{n^2 dV}
\label{eq:emission}
\end{equation}
where $I$ is the emission intensity, $V$ is the emitting volume, $T$ is its temperature, $n$ is its particle density, and $G(T)$ is the emissivity per unit volume, as described clearly in \cite{Vaiana1973b}. The ratio of the emission of the same plasma volume in two filters with different passbands is a function of the temperature only. If the temperature is known we can derive the emission measure directly by inverting Eq.~(\ref{eq:emission}). 

A survey of flare observations from Skylab opened the way to the systematic analysis of solar flare properties. A flare classification based on the morphology of the flaring structures was proposed, in particular differentiating flares occurring in single loop structures (compact flares), point-like flares, and flares in loop systems (or "two-ribbon flares") \cite{Pallavicini1977}. With the advent of subsequent flare-dedicated satellite missions, this classification has become much more blurred and somewhat obsolete. Nevertheless, it is useful to point out here that, although flares are in general very complex phenomena, in some cases they can be described as rather simple systems, even single magnetic loops as observed in the X-rays. These early X-ray  observations of flares, although still rather coarse, were sufficient to stimulate the development of several flare models, based on the assumption that flaring plasma moves and transports energy along the field lines of a single coronal loop and can be therefore described by purely one-dimensional hydrodynamics (i.e., no magnetic force is in action; \cite{Nagai1980}).

\begin{figure*}[!t]
\centering
\includegraphics[width=0.97\textwidth]{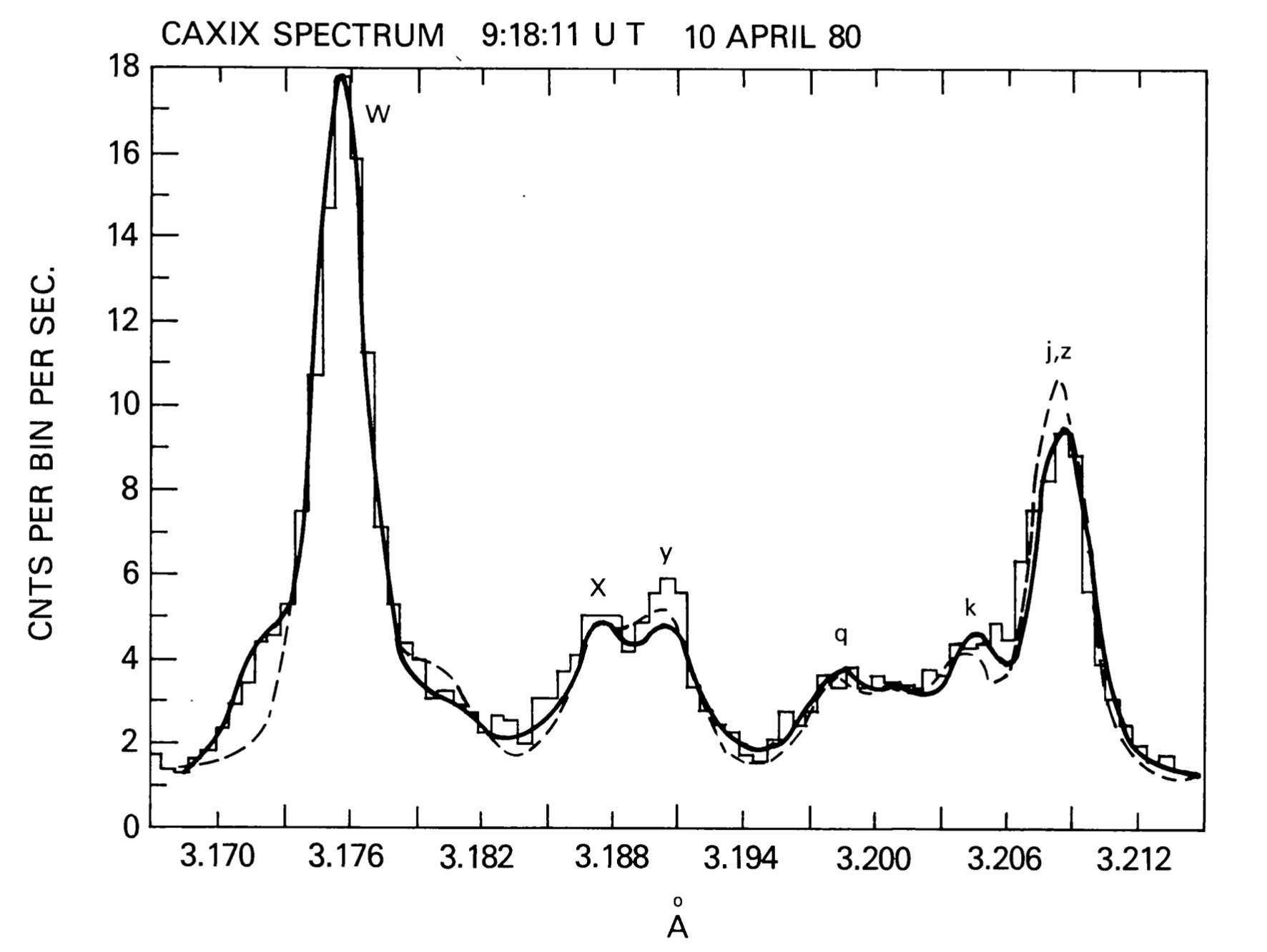}
\caption{\caxix\ SMM/BCS spectrum obtained during the impulsive phase of a flare observed on April 10, 1980 (09:18:11 UT). This spectrum shows a strong blue-shifted secondary component (the bump on the left of the main line). The best fit (thick line) is the superposition of two components, including a blue-shifted one (corresponding to velocities of $>300$\kms) which is a signature of chromospheric evaporation, triggered by the very intense and short lived energy release at the beginning of the flare. (From \cite{Antonucci1982}). }\label{fig:evap}
\end{figure*}

\subsection{\textit{4.2 The power of spectroscopy: The Solar Maximum Mission}}

The Skylab mission stimulated much interest in dynamic events on the Sun, and a new mission was specifically devoted to investigate their physical characteristics. The Solar Maximum Mission (SMM) was launched in 1980, close to the maximum of the 11-year solar activity cycle. SMM carried several instruments for observation of the solar atmosphere, but here we focus our attention on the soft X-ray spectrometers -- the Flat Crystal Spectrometer (FCS) and the Bent Crystal Spectrometer (BCS) -- which were part of the X-Ray Polichromator (XRP). The FCS was able to capture flare X-ray spectra in a relatively broad band (in the range from 5.68\AA\ to 18.97\AA, \cite{Phillips1982}), including several bright H-like or He-like lines from highly-ionized elements (\oviii, \neix, \mgxi, \sixiii, \sxv, \caxix, \fexv). The BCS detected very high resolution spectra (up to $\lambda/\Delta \lambda \sim 15000$, \cite{Sylwester2020}) especially around two specific important high-energy He-like lines, \caxix\ 3.16 \AA\ and \fexxv\ 1.84 \AA. The Ca\,{\sc xviii-xix} and Fe\,{\sc xxiv-xxv} satellite lines could be clearly resolved and used for diagnostics \cite{Culhane81}. One very interesting result was the clear detection of a significant blue-shifted component of the 3.16 \AA\ resonance line of \caxix\ \cite{Antonucci1982} in the initial (so-called impulsive) phase of the flare  (see Figure~\ref{fig:evap}). This was the signature of strong upflows at hundreds of \kms\ at the beginning of the flare. A possible physical origin of these flows can be ascribed to a strong flare heat pulse causing a pressure excess that makes the low loop atmosphere to rapidly expand into the relatively empty overlying magnetic tube. This effect was therefore named "chromospheric evaporation". This scenario was modeled in detail with hydrodynamic loop simulations (\cite{Antonucci1987}; Fig.~\ref{fig:evap}). A very exhaustive review and discussion of flare spectroscopic data from SMM data is provided by \cite{Doschek90}. Multi-line (with SMM/FCS) and multi-band observations provided us with a detailed scenario and plenty of constraints for flare models (\cite{MacNeice1985}) that allowed us to discriminate among different heating mechanisms and to constrain heating parameters \cite{Peres87}.

\begin{figure*}[!t]
\centering
\includegraphics[width=0.95\textwidth]{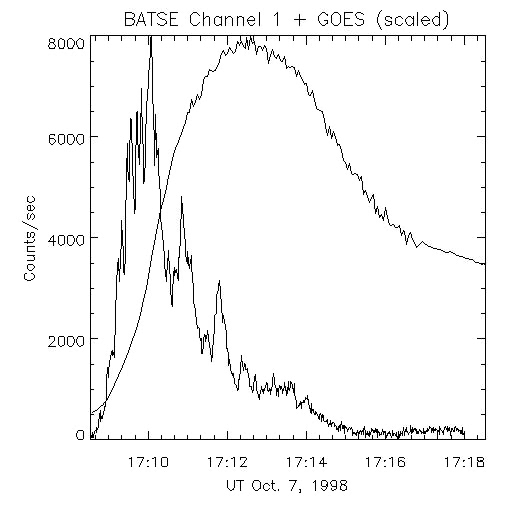}
\caption{The Neupert effect: flare soft X-ray light curve (from the GOES soft channel; the line peaking later [$\sim$17:13UT]), and hard X-ray light curve (left peak [$\sim$17:10UT]) measured with CGRO/BATSE (20-25~keV). Both curves show a fast rise and a slower decay. The soft X-ray light curve is approximately equal to the integral of the hard X-ray light curve (from http://soi.stanford.edu/results/SolPhys200/Hudson/1998/981016/981016.html). 
}\label{fig:neupert}
\end{figure*}

Flare observations have been used as laboratories where to study plasma physics and heating mechanisms. Since many of them are localized and fast events, strong assumptions can be made regarding the relevant physical mechanisms at work. 
During the flare rise phase the plasma dynamics play a very important role. It is generally accepted that the emission in hard X-rays is a tracer of the energy deposition. One important evidence for this is the so-called "Neupert effect": the soft X-ray light curve in the initial rising phase matches the time-integrated emission in hard X-rays (\cite{Neupert1968}; Fig.~\ref{fig:neupert}). This has been observed to occur in the majority of flares, at least those with single and well-defined heat pulses, but there are exceptions that demonstrate that the energy release in flares can be rather complex both in locations and in time \cite{dennis93}.

\begin{figure*}[!t]
\centering
\includegraphics[width=0.97\textwidth]{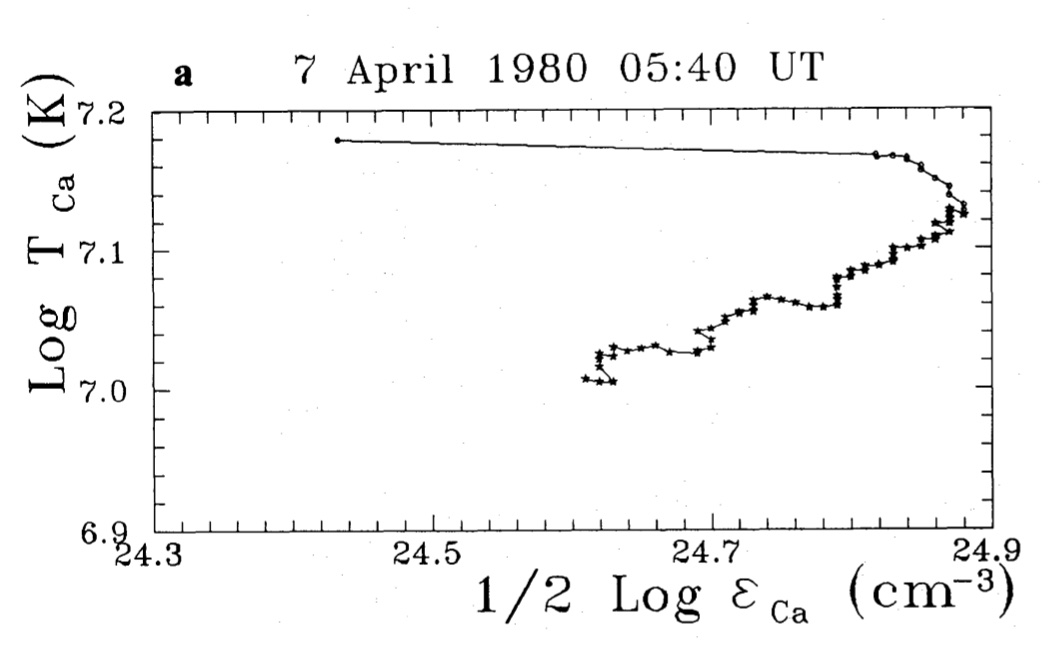}
\caption{Evolution in the emission measure-temperature diagram for a flare observed with SMM on April 7 1980. The dots mark measurements equispaced in time. Time goes clockwise.  (From \cite{Sylwester1993}). }\label{fig:ntdiag}
\end{figure*}

In the flare decay phase, when the plasma slowly drains down, the plasma dynamics can be largely neglected, and we can work under the assumption of an imbalance between the heating, which has ceased, and the cooling, which then determines most of the evolution. In spite of the impulsive rise phase, we can start here from a quasi-equilibrium condition which can be described by Rosner, Tucker and Vaiana (RTV) scaling laws \cite{RTV}. It is then  possible to analytically derive a very simple scaling law for the decay time $\tau_{s}$ of a single flaring loop \cite{Serio91,Reale2007mod}:

\begin{equation}
\tau_{s} = 3.7 \times 10^{-4} \frac{L}{\sqrt{T_0}}
\sim 120 \frac{L_9}{\sqrt{T_{0,7}}}
\label{eq:tserio}
\end{equation}
where $L$ ($L_9$) is the loop half-length (in units of $10^9$ cm), and $T_0$ ($T_{0,7}$) the loop maximum temperature (in units of $10^7$ K). This time scale estimate  is a lower limit of the actual decay time, for various reasons. First, the hypothesis of equilibrium condition when the decay starts is a strong assumption, and it maximizes the initial flare density. Real flares are impulsive and the heat pulse rarely lasts so long as to reach equilibrium density. A lower density implies a slower radiation cooling. Second, some residual heating in the decay phase, which also slows down the cooling, is thought to occur frequently in flares. 
Keeping these caveats in mind, this scaling law can in principle be easily applied to estimate the loop length, in the cases where the flaring loop is not resolved, as is the case for stellar flares (see also Drake \& Stelzer, in this handbook).

However, as such, the equation can provide only an upper limit to the loop length, because the possible presence of heating in the decay can lead to an overestimate of the actual loop length. Nevertheless, X-ray observations provide additional diagnostics of the presence of heating during the decay phase, as explained here below. Spectral fittings to flare data from observations with SMM allowed to derive the evolution of the emission measure and temperature during the flares. The (square root of the) emission measure can be used as a proxy of the density, if the volume of the emitting plasma does not change much during the flare, as it is largely observed during many flares. It was pointed out that flares follow a well-defined path in a density-temperature diagram \cite{Jakimiec1992}, and that the slope of the path in the decay changes if the heating is absent (steeper) or present (shallower). Different slopes were indeed observed in a survey of SMM flares (\cite{Sylwester1993}; Fig.~\ref{fig:ntdiag}), and a new empirical formula for the loop length,  incorporating  these additional diagnostics, was calibrated on hydrodynamic flaring loop models, and tested on solar flares \cite{Reale97}.

\subsection{\textit{4.3 The digital era: from Yohkoh to Hinode and onward}}

\begin{figure*}[!t]
\centering
\includegraphics[width=0.99\textwidth]{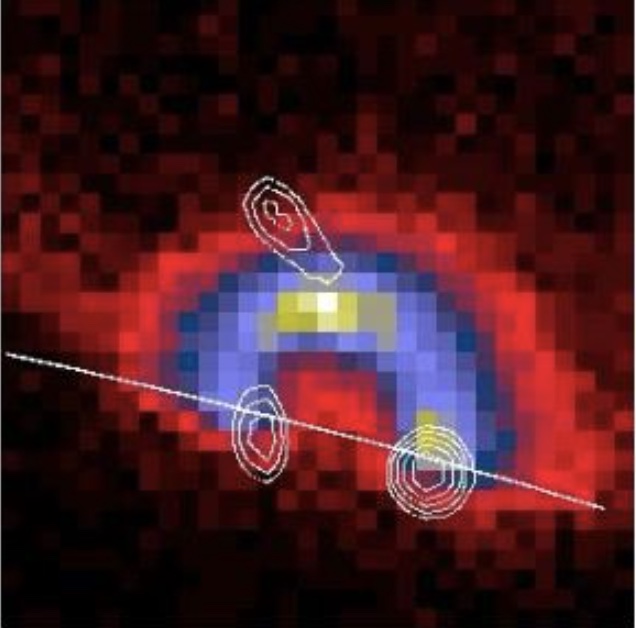}
\caption{False-color image of the flare of 13 January 1992 17:25 UT observed by Yohkoh/SXT. The contours mark the locations of the hard X-ray emission. The straight line marks the solar limb. (From \cite{Masuda1994}). }\label{fig:masuda}
\end{figure*}

\begin{figure*}[!t]
\centering
\includegraphics[width=0.99\textwidth]{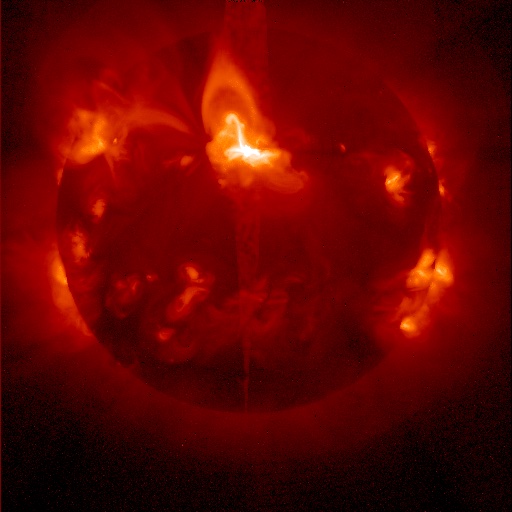}
\caption{Yohkoh/SXT image of the Sun showing cusp-like emission above a flare site.  (From Astronomy Picture Of the Day, APOD, 8 June 2000 https://apod.nasa.gov/apod/ap000608.html)}
\label{fig:cusp}
\end{figure*}

The development of digital detection devices (CCD), which were adopted in next generation missions, and in particular Yohkoh and Hinode, launched in 1991 and 2006 respectively, represented a new important step for flare studies. The Yohkoh/SXT had a flag that allowed, soon after the flare start, to switch to a flare mode, characterized by very short exposure times. This allowed to record an unprecedented set of flares, and to image them with a very good spatial (5~arcsec) and temporal (2~s) resolution. With this high resolution it was possible to better observe the basic processes of magnetic energy conversion. Flares were still largely divided into two main categories, i.e., compact (single loop) flare and the so-called Long Duration Events (LDE), involving many loop structures (similar to the two-ribbon flares). Several events of both classes clearly showed hard X-ray emission at the flaring loop footpoints, but for some of them such emission was detected at the loop apex. One famous flare, observed by Yohkoh on 13 January 1992 17:25 UT \cite{Masuda1994} has been the subject of several investigations. This flare occurred at the solar limb and apparently involved a single semicircular coronal loop (compact flare). Fig.~\ref{fig:masuda} clearly shows the three main sources of hard X-ray emission. These locations are generally interpreted as sites of non-thermal particle acceleration. There electrons are accelerated to relativistic speeds as an effect of magnetic reconnection. The electron beams streaming along the flaring loops are very important because it is believed that in many flares they deposit their energy at the loop footpoints and trigger chromospheric evaporation \cite{Hirayama74,Fisher85}. However, the same effect is caused by thermal conduction fronts moving down along the loop from the reconnection site, and in some flares there is evidence that these prevail \cite{Lopez2022}. The source at the loop apex is believed to be strong evidence of magnetic reconnection at that site \cite{Tsuneta1997} and Yohkoh detected cusp-like emission and related jets on top of a flaring system (\cite{Shibata1995}; Fig.~\ref{fig:cusp}). Probably both mechanisms are at work at the same time, with different efficiency in different flares, and this issue deserves further investigation in the future.   

\begin{figure*}[!t]
\centering
\includegraphics[width=0.99\textwidth]{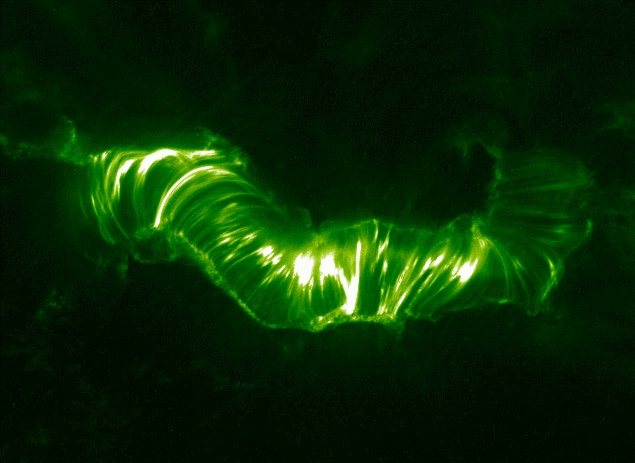}
\caption{Arcade of flaring loops triggered during the 14 July 2000 flare, as imaged by the TRACE mission in the 195\AA\ channel (From astronomy picture of the day, APOD, 20 July 2000 https://apod.nasa.gov/apod/ap000720.html)}\label{fig:bastille}
\end{figure*}

On the side of LDEs we find a flare very well observed by multiple instruments on 14 July 2000, the so-called Bastille-day flare (X5.7; \cite{Aschw01}. This flare has been observed simultaneously in several bands. It started from a very localized region but then extended rapidly to a very long arcade of loops, as imaged impressively by the TRACE mission (\cite{Aschwanden2005book}; Fig.~\ref{fig:bastille}). Also in this case the investigations pointed out strong evidence for on-going magnetic reconnection \cite{Somov2002,Qiu2010}. 

\begin{figure*}[!t]
\centering
\includegraphics[width=0.495\textwidth]{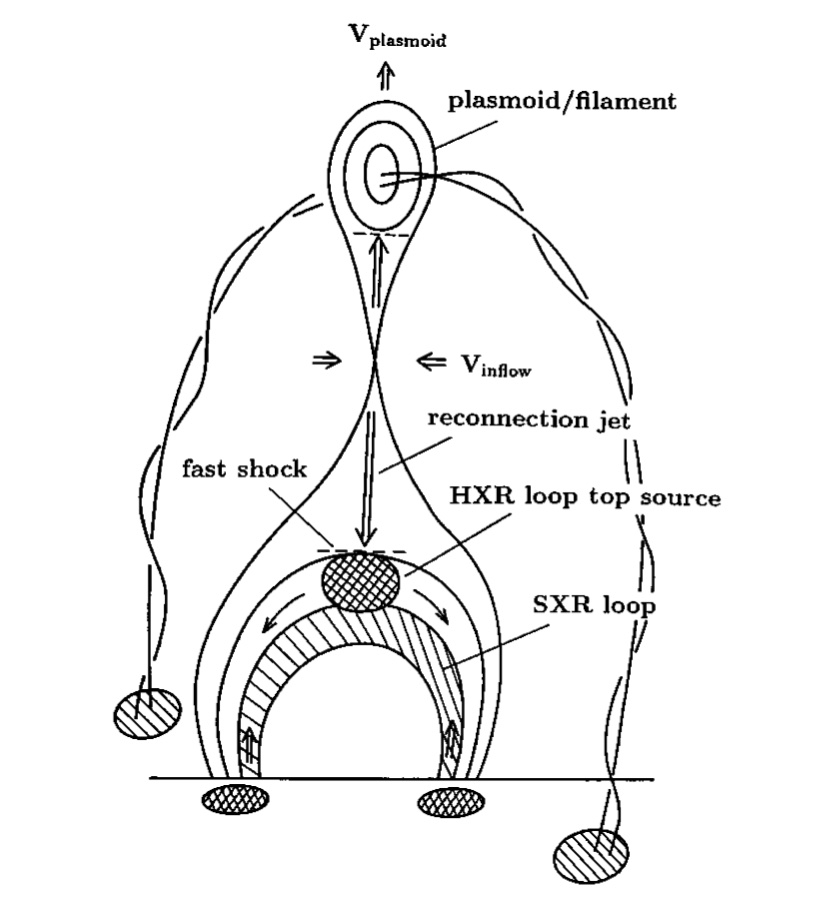}
\includegraphics[width=0.495\textwidth]{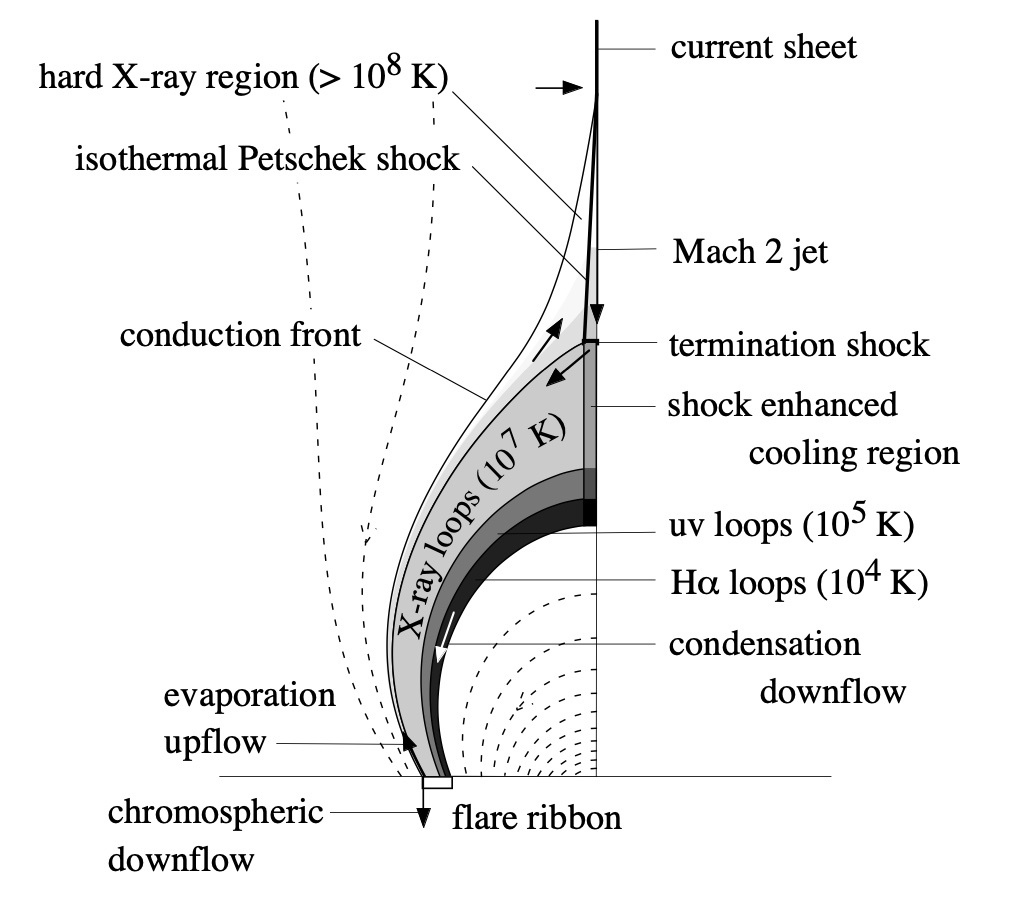}
\caption{Schematic diagrams of a flare loop system driven by magnetic reconnection. (From \cite{Shibata1995} (left), \cite{Priest2002b} (right)). }\label{fig:scheme}
\end{figure*}

The detailed analysis of flares in several energy bands led to build schematic views in which the soft X-ray emission is only one of the manifestations of the energy release from magnetic reconnection (\cite{Shibata1995,Priest2002b}; Fig.~\ref{fig:scheme}).
The extensive database of flares observed in different filterbands by the Yohkoh mission also allowed to extract a collection of samples of the emission measure distribution versus temperature, and of its evolution during the flares \citep{Reale2001}. From such distributions can be used to synthesize the emission that could be detected by non-solar X-ray telescopes. This flaring "Sun-as-a-star" becomes a template to interpret X-ray observations of stellar flares (see also Drake \& Stelzer, in this handbook). 

Current missions that monitor the corona, Hinode, Solar Dynamics Observatory, IRIS, focus more on the details of flare evolution and  the flaring structure.
Although here we will focus mostly on the high energy observations in the soft X-ray band, we first briefly mention some important results obtained with IRIS spectral observations in the ultraviolet band, in particular using the \fexxi\ 1354\AA\ flare line. IRIS has observed hundreds of flares so far, and brought about significant progress in flare studies mainly thanks to its unprecedented spatial ($\sim 0.33$") and temporal (down to $< 1$s) resolution. The unique IRIS flare results include for instance: finally spatially resolving chromospheric evaporation sites at the footpoints of flaring loops (e.g., \cite{Young2015}), which solved a decade-long puzzle and brought agreement with model predictions, although the large symmetric broadenings observed with IRIS are not easily explained by flaring loop models (\cite{Polito2019}); constraints on magnetic reconnection processes, e.g., from \fexxi\ redshifts at looptop, coincident with hard X-ray sources, and interpreted as signatures of termination shocks (e.g., \cite{Polito2018b}); constraints on initiation of CMEs from e.g., \fexxi\ observations in flux ropes prior to eruptions (e.g., \cite{Cheng2016}). We refer the reader to \cite{DePontieu2021} for a recent detailed review of IRIS flare studies (including also a discussion of exciting flare results from the IRIS chromospheric and transition region lines).

Magnetic reconnection remains the main scenario for the interpretation of flare X-ray observations. The high cadence and spatial resolution of the Hinode/XRT captured the shrinking of cusp-shaped flare loops \cite{Reeves2008}. Dark downflows, called supra-arcade downflows \cite{McKenzie1999}, have been observed above flare arcades during the decay phase of some flares \cite{Innes2003a,Innes2003b,Savage2012,Hanneman2014}, probably associated with the reconnection process \cite{Asai2004}. These downflows travel through a very hot medium \cite{Reeves2011}.
Hinode XRT also detected symmetric upflows along the loop legs in several events, most with speeds of 100 km/s but also up to 500 km/s, identified with chromospheric evaporation.
Joint analysis of EUV, soft X-ray emission (GOES) and hard X-ray spectra (RHESSI) showed that flaring structures are multithermal \cite{Warren2013}, and that there is evidence for a superhot ($>30$ MK) component in X-class flares \cite{Caspi2014}.

Multiband observations made with the Hinode mission were devoted to study 
the build-up of flare energy. The free energy can build up as a result of the emergence of sheared magnetic fields above the photosphere \cite{Leka1996,Schrijver2005,Schrijver2007}, shearing motions or rotation of the photospheric footpoints \cite{Gesztelyi1984,Zirin1990,Brown2003}, and the cancellation of flux in the photosphere \cite{Martin1985,Livi1989}. The magnetic stress leads to the formation of highly twisted flux tubes, called sigmoids, which are also highly unstable and can erupt or determine explosive reconnection. Some active regions have been monitored carefully by the Hinode mission, especially at instrument peak performance in the early mission phase.
The emergence of twisted flux tubes and of dark unstable and elongated structures, called filament channels was either correlated to the rotation of the related spots \cite{Min2009} or interpreted as bodily emergence of helical flux ropes \cite{Okamoto2008,Okamoto2009}. The measured local increase of the magnetic helicity might be another magnetic signature  \cite{Park2010}.

Such multiband observations, where the XRT and SOT had a major role, were also supported by advanced modeling, which allowed detailed comparison to data (e.g., \cite{Savcheva2009,Su2009,Savcheva2012}) and emphasized the possible role of flux cancellation in triggering flare eruptions. MHD simulations showed that sunspot rotation can really trigger the formation of sigmoids \cite{Leake2013}. A flare catalog from the Hinode mission is available at https://hinode.isee.nagoya-u.ac.jp/flare\_catalogue/ \cite{Watanabe2012}.

Great attention has been paid quite recently to the generation of quasi-periodic oscillations (QPP) inside flaring coronal loops. Such oscillations have been detected in flare light curves in several spectral bands, from radio to hard X-rays, and they have a number of possible origins and physical drivers (e.g., \cite{Zimovets2021}). 
Flares can trigger substantial wave modes and fronts, such as sausage modes \cite{Shi2019} and magnetosonic modes \cite{Reale2016,Miao2019} which can involve significant density variations and therefore can become visible in the EUV and soft X-ray band. The detection of QPPs can be used as probes of physical conditions inside flaring loops, e.g., through measuring the period and the propagation speed, and provide information about the flare heat pulses \cite{Reale2016}. Evidence for magnetosonic waves has been detected also in EUV observations of microflaring loop systems \cite{Reale2019b}.
Up-to-date much attention has been devoted to the flare and eruption generation and prediction, and their influence on terrestrial activities (space weather) through Coronal Mass Ejections (see next section).

\subsection{\textit{5. Coronal Mass Ejections}}

Flares often determine strong stresses and weakening of the local confining magnetic field. As a consequence, closed magnetic structures can break up and result into large-scale violent eruptions of dense plasma. Relatively hot and dense clouds are then expelled out of the solar atmosphere into the interplanetary medium, the so-called Coronal Mass Ejections (CMEs). 
CMEs are key drivers of space weather in the heliosphere and planetary environments. The study of CMEs is therefore also important to investigate their impact on exoplanetary atmospheres and in particular on exoplanet habitability (e.g., \cite{Yamashiki2019}). This is particularly consequential in light of recent Kepler observations that revealed a high frequency of flares that can be thousands of times more intense than solar ones (the so-called superflares) in solar-like stars (e.g., \cite{Maehara2012}), and detected also in the X-ray band on pre-MS stars \cite{Getman2021}. Direct evidence for a CME has been detected in the X-rays during a stellar flare on an active star \cite{Argiroffi2019}.
However, here we discuss only very briefly this topic because on the Sun CMEs mostly consist of relatively cool plasma ($\leq 1$ MK), and therefore not primarily emitting in the X-ray band. If present, X-ray emission is typically detected in the very initial phases of the ejection, when the ejected plasma is still at coronal temperatures. For instance, hot X-ray plasma ($\sim 3$ MK) was detected from the current sheet associated with the reconnection that generated an eruption \cite{Reeves2018}. Hinode/XRT also discovered that the heating from reconnection continues also during the eruption itself \cite{Gou2020}, and it provided abundant evidence that flux ropes exist in the corona prior to eruption (e.g.,\cite{hinodereview2019}), although the  details of the CME initiation mechanisms are still elusive. Recent studies have been devoted to correlation between flares, CMEs, and SEP (Solar Energetic Particles) events (e.g., \cite{Firoz2019}),  to the details of initial CME acceleration as connected to the energy release during an intense flare \cite{Gou2020}, and to the direct excitation of MHD waves by CMEs \cite{Miao2019}.
We refer to reviews by e.g., \citet{Chen2011,Webb2012,Temmer2021,Nitta2021}, for recent overviews of salient characteristics of CMEs and their importance in the solar physics and heliospheric context.

\section{\textit{6. Conclusions}} \label{sec:concl}
In this chapter we have presented an historical overview of X-ray observations of the solar corona, focusing in particular on recent results obtained with current instrumentation, and we have discussed the status of the resulting understanding of the physical processes at work in the solar outer atmosphere. We note that, to achieve a tractable review of the many aspects of the high energy emission from the Sun, we have discussed a limited selection of topics. We made the choice to focus on the X-ray emission, and soft X-ray emission in particular ($\lesssim 10$ keV, $\gtrsim 1$\AA ), although several results from hard X-ray observations are discussed as well in the sections on active regions and on flares. We note that a discussion of solar gamma-ray emission, and additional hard X-ray results, are included in another chapter of this book (Saint-Hilaire et al.).
We have also included some results from extreme-ultraviolet observations, but especially focusing on the shorter wavelengths (e.g., AIA 94\AA\ and 131\AA\ passbands) more sensitive to the hotter plasma. As a consequence of these choices, our overview is heavily focused on the hottest coronal features, i.e., active region cores and flares.  
The EUV and UV observations of course provide complementary information of fundamental importance to understand some of these phenomena, as briefly referred to here, and thoroughly discussed in ample literature (see e.g., recent reviews of \cite{hinodereview2019,DePontieu2021}). Similarly, another important aspect we have not delved into here is the modeling of the observed high energy phenomena on the Sun, which is necessary to understand the underlying physical processes (see e.g., \cite{Shibata2011,Chen2011,Reale2014,DeMoortel2015,Peter2015,Cargill2015,Velli2015} for some recent reviews).

Our review traced the progress of X-ray observations of the Sun in the last several decades. These X-ray observations at increasing spatial, temporal, and spectral resolution reveal the extremely dynamic nature of the solar hot atmosphere. The solar corona, which is shaped and energized by the solar magnetic field, is highly structured in regions of different activity and X-ray emission, and its emission mirrors the underlying solar magnetic activity cycle.
Solar X-ray observations, thanks to their high resolution, also provide us with a close-up view of several astrophysical phenomena (e.g., magnetic reconnection, jets) that cannot be easily investigated in far-away astrophysical sources.

The issue of which physical processes dominate the conversion of magneto-convective energy to power the corona, i.e., the coronal heating problem, is one of the most important fundamental open issues in astrophysics. 
Despite significant recent progress in understanding the physical phenomena at work in the quiescent solar corona, several open issues remain, including the following, which would benefit the most from X-ray observations:
\begin{itemize}
    \item accurately determine the presence and characteristics (spatial, temporal, thermal properties) of the hot plasma;
    \item determine the characteristic magnitude and frequency of heating events that sustain the quiescent corona;
    \item  determine the extent and limits of similarities between nanoflares and large flares (e.g., are accelerated particles common in the 'quiescent' corona, and do they play a significant role in energizing the solar atmosphere?);
    \item determine the importance (and properties) of different heating mechanisms (e.g., Alfv\'en waves, nanoflares) in different solar regions.
\end{itemize}

Flares remain a central topic of investigation for the future, especially considering the increasing importance of space weather, for its influence on e.g., electromagnetic and telecommunication activities, and human space exploration. 
Similarly to the quiescent solar emission, several aspects of the physical processes driving flares (and CMEs) remain to be fully understood. Important issues that need yet to be addressed include:
\begin{itemize}
    \item capturing the initial impulsive instabilities that lead to flares and CMEs (i.e., the flare/CME triggers);
    \item constraining the components that determine the intensity of the flare;
    \item determining the relative role of particle acceleration and current dissipation;
    \item ultimately reaching an exhaustive flare scenario that allows for forecasting.
\end{itemize}

High spatial, temporal, and spectral resolution observations are necessary to address these issues and, in turn, to pinpoint the physical mechanisms responsible for coronal heating and flare physics. High quality flare observations are still difficult to obtain because of their very dynamic nature, but they might be within reach of next-generation X-ray instruments. Multi-band observations and high spatial/temporal resolution X-ray/EUV spectroscopy are certainly key to constraining advanced modeling. Indeed, increasing computing power has enabled major advances in realism and complexity of numerical modeling. 

A rich ongoing sounding rocket program, and future missions, allow to push the boundaries of current observations and explore novel instrumentation and technologies.
Spectroscopic observations provide unique diagnostics of plasma properties (e.g., plasma flows, density, chemical composition), and are therefore particularly crucial to testing competing models and furthering our understanding of physical processes at work in the corona. However, it is significantly  challenging to design a spectrograph able to obtain spectral measurements at high spatial and temporal cadence, and with large spatial coverage, all of which are necessary to capture the plasma dynamic, at different locations of sizable coronal features. These observational constraints will be achievable with future instrumentation, and it is in particular foreseeable, for instance, with a combination of the Multi-slit Solar Explorer (MUSE; \cite{DePontieu2020,DePontieu2021arXiv,Cheung2021arXiv}) -- a recently selected NASA MIDEX mission, with planned launch in 2026, which will obtain high spatio-temporal spectral and imaging coronal observations, over an active region-size field of view, in EUV narrow spectral bands thanks to its innovative multi-slit design (37 slits) -- and the planned Solar-C/EUVST (\cite{Shimizu2019}) mission -- which, with a traditional single-slit design, will provide spectral observations with very extensive thermal coverage, at matching ($\sim  0.3$") high spatial resolution at different temperatures.  
The few coronal observations available to date at subarcsec resolution, such as for instance the short time series of imaging data from the Hi-C rocket experiments (see sec.~3~\ref{sec:ar}), have provided us with tantalizing glimpses of the significant leap these future high-resolution coronal observatories will certainly produce.

\vspace{2cm}
P.T.\ was supported by NASA contract NNM07AB07C (Hinode/XRT) to the Smithsonian Astrophysical Observatory, by contract 8100002705 (IRIS) from Lockheed-Martin to SAO, by the MUSE contract from Lockheed-Martin to SAO, and by the NASA grants 80NSSC20K0716, 80NSSC21K0737, and 80NSSC20K1272. F.R.\ acknowledges support from the italian Ministero dell’Universit{\`a} e della Ricerca (MUR).

\section{\textit{Cross-references}} \label{sec:crossrefs}
\begin{itemize}
    \item Stellar Coronae (J.\ Drake and B.\ Stelzer)
    \item Star Forming Regions (S.\ Sciortino)
    \item Pre-main sequence: Accretion and Outflows (P.C.\ Schneider, H.M.\ G{\"u}nther, and S. Ustamujic)
    \item X-ray and Gamma-Ray indirect imaging (Saint-Hilaire P., Hurford G., Shih A., and Dennis B.)
\end{itemize}


\begin{thebibliography}{259}
\providecommand{\natexlab}[1]{#1}
\providecommand{\url}[1]{{#1}}
\providecommand{\urlprefix}{URL }
\expandafter\ifx\csname urlstyle\endcsname\relax
  \providecommand{\doi}[1]{DOI~\discretionary{}{}{}#1}\else
  \providecommand{\doi}{DOI~\discretionary{}{}{}\begingroup
  \urlstyle{rm}\Url}\fi
\providecommand{\eprint}[2][]{\url{#2}}

\bibitem[{{Acton} et~al(1980){Acton}, {Culhane}, {Gabriel}, {Bentley},
  {Bowles}, {Firth}, {Finch}, {Gilbreth}, {Guttridge}, {Hayes}, {Joki},
  {Jones}, {Kent}, {Leibacher}, {Nobles}, {Patrick}, {Phillips}, {Rapley},
  {Sheather}, {Sherman}, {Stark}, {Springer}, {Turner}, and
  {Wolfson}}]{Acton1980}
{Acton} LW, {Culhane} JL, {Gabriel} AH, et al.\ (1980) {The soft X-ray polychromator for the Solar Maximum
  Mission.} \solphys 65(1):53--71

\bibitem[{{Adithya} et~al(2021){Adithya}, {Kariyappa}, {Shinsuke}, {Kanya},
  {Zender}, {Dam{\'e}}, {Gabriel}, {DeLuca}, and {Weber}}]{Adithya2021}
{Adithya} HN, {Kariyappa} R, {Shinsuke} I, et al.\ (2021) {Solar Soft X-ray Irradiance
  Variability, I: Segmentation of Hinode/XRT Full-Disk Images and Comparison
  with GOES (1 - 8 A) X-Ray Flux}. \solphys 296(4):71

\bibitem[{{Alfv{\'e}n}(1947)}]{Alfven1947}
{Alfv{\'e}n} H (1947) {Magneto hydrodynamic waves, and the heating of the solar
  corona}. \mnras 107:211

\bibitem[{{Alipour} and {Safari}(2015)}]{Alipour2015}
{Alipour} N, {Safari} H (2015) {Statistical Properties of Solar Coronal Bright
  Points}. \apj 807(2):175

\bibitem[{{Antiochos} et~al(2003){Antiochos}, {Karpen}, {DeLuca}, {Golub}, and
  {Hamilton}}]{Antiochos2003}
{Antiochos} SK, {Karpen} JT, {DeLuca} EE, {Golub} L, {Hamilton} P (2003)
  {Constraints on Active Region Coronal Heating}. \apj 590:547--553

\bibitem[{{Antolin} and {Rouppe van der Voort}(2012)}]{Antolin2012}
{Antolin} P, {Rouppe van der Voort} L (2012) {Observing the Fine Structure of
  Loops through High-resolution Spectroscopic Observations of Coronal Rain with
  the CRISP Instrument at the Swedish Solar Telescope}. \apj 745(2):152

\bibitem[{{Antolin} et~al(2021){Antolin}, {Pagano}, {Testa}, {Petralia}, and
  {Reale}}]{Antolin2021}
{Antolin} P, {Pagano} P, {Testa} P, {Petralia} A, {Reale} F (2021)
  {Reconnection nanojets in the solar corona}. Nature Astronomy 5:54--62

\bibitem[{{Antonucci} et~al(1982){Antonucci}, {Gabriel}, {Acton}, {Culhane},
  {Doyle}, {Leibacher}, {Machado}, {Orwig}, and {Rapley}}]{Antonucci1982}
{Antonucci} E, {Gabriel} AH, {Acton} LW,  et al.\ (1982) {Impulsive Phase of Flares
  in Soft X-Ray Emission}. \solphys 78(1):107--123

\bibitem[{{Antonucci} et~al(1987){Antonucci}, {Dodero}, {Peres}, {Serio}, and
  {Rosner}}]{Antonucci1987}
{Antonucci} E, {Dodero} MA, {Peres} G, {Serio} S, {Rosner} R (1987)
  {Simulations of the CA XIX Spectral Emission from a Flaring Solar Crornal
  Loop. I. Thermal Case}. \apj 322:522

\bibitem[{{Argiroffi} et~al(2019){Argiroffi}, {Reale}, {Drake}, {Ciaravella}, {Testa}, et al.\}}]{Argiroffi2019} {Argiroffi} C, {Reale} F, {Drake} JJ, {Ciaravella} A, {Testa} P, et  al.\ (2019) {A stellar flare-coronal mass ejection event revealed by X-ray plasma motions}. Nature Astronomy 3:742--748

\bibitem[{{Asai} et~al(2004){Asai}, {Yokoyama}, {Shimojo}, and
  {Shibata}}]{Asai2004}
{Asai} A, {Yokoyama} T, {Shimojo} M, {Shibata} K (2004) {Downflow Motions
  Associated with Impulsive Nonthermal Emissions Observed in the 2002 July 23
  Solar Flare}. \apjl 605(1):L77--L80

\bibitem[{{Aschwanden}(2005)}]{Aschwanden2005book}
{Aschwanden} MJ (2005) {Physics of the Solar Corona. An Introduction with
  Problems and Solutions (2nd edition). Praxis Publishing Ltd., Chichester,
  UK; Springer, New York, Berlin, 2005.}

\bibitem[{{Aschwanden} and {Alexander}(2001)}]{Aschw01}
{Aschwanden} MJ, {Alexander} D (2001) {Flare Plasma Cooling from 30 MK down to
  1 MK modeled from Yohkoh, GOES, and TRACE observations during the Bastille
  Day Event (14 July 2000)}. \solphys 204:91--120

\bibitem[{{Aschwanden} and {Peter}(2017)}]{Aschwanden2017}
{Aschwanden} MJ, {Peter} H (2017) {The Width Distribution of Loops and Strands
  in the Solar Corona{\textemdash}Are We Hitting Rock Bottom?} \apj 840(1):4 

\bibitem[{{Athiray} et~al(2020){Athiray}, {Vievering}, {Glesener}, {Ishikawa},
  {Narukage}, {Buitrago-Casas}, {Musset}, {Inglis}, {Christe}, {Krucker}, and
  {Ryan}}]{Athiray2020}
{Athiray} PS, {Vievering} J, {Glesener} L, et al.\ (2020) {FOXSI-2 Solar Microflares. I. Multi-instrument Differential
  Emission Measure Analysis and Thermal Energies}. \apj 891(1):78

\bibitem[{{Baker} et~al(2015){Baker}, {Brooks}, {D{\'e}moulin}, {Yardley}, {van
  Driel-Gesztelyi}, {Long}, and {Green}}]{Baker2015}
{Baker} D, {Brooks} DH, {D{\'e}moulin} P, et al.\ (2015) {FIP Bias Evolution in a Decaying Active
  Region}. \apj 802(2):104

\bibitem[{{Baker} et~al(2018){Baker}, {Brooks}, {van Driel-Gesztelyi}, {James},
  {D{\'e}moulin}, {Long}, {Warren}, and {Williams}}]{Baker2018}
{Baker} D, {Brooks} DH, {van Driel-Gesztelyi} L, et al.\ (2018) {Coronal Elemental Abundances in
  Solar Emerging Flux Regions}. \apj 856(1):71

\bibitem[{{Barnes} et~al(2019){Barnes}, {Bradshaw}, and {Viall}}]{Barnes2019}
{Barnes} WT, {Bradshaw} SJ, {Viall} NM (2019) {Understanding Heating in Active
  Region Cores through Machine Learning. I. Numerical Modeling and Predicted
  Observables}. \apj 880(1):56

\bibitem[{{Bellot Rubio} and {Orozco Su{\'a}rez}(2019)}]{Bellot2019}
{Bellot Rubio} L, {Orozco Su{\'a}rez} D (2019) {Quiet Sun magnetic fields: an
  observational view}. Living Reviews in Solar Physics 16(1):1

\bibitem[{{Benz}(2017)}]{Benz2017}
{Benz} AO (2017) {Flare Observations}. Living Reviews in Solar Physics 14(1):2

\bibitem[{{Berghmans} et~al(2021){Berghmans}, {Auch{\`e}re}, {Long},
  {Soubri{\'e}}, {Zhukov}, {Mierla}, {Sch{\"u}hle}, {Antolin}, {Parenti},
  {Harra}, {Podladchikova}, {Aznar Cuadrado}, {Buchlin}, {Dolla}, {Verbeeck},
  {Gissot}, {Teriaca}, {Haberreiter}, {Katsiyannis}, {Rodriguez}, {Kraaikamp},
  {Smith}, {Stegen}, {Rochus}, {Halain}, {Jacques}, {Thompson}, and
  {Inhester}}]{Berghmans2021}
{Berghmans} D, {Auch{\`e}re} F, {Long} DM, et al.\ (2021) {Extreme UV quiet Sun brightenings
  observed by Solar Orbiter/EUI}. arXiv e-prints arXiv:2104.03382,

\bibitem[{{Blake} et~al(1963){Blake}, {Chubb}, {Friedman}, and
  {Unzicker}}]{Blake1963}
{Blake} RL, {Chubb} TA, {Friedman} H, {Unzicker} AE (1963) {Interpretation of
  X-Ray Photograph of the Sun.} \apj 137:3 

\bibitem[{{Blake} et~al(1964){Blake}, {Chubb}, {Friedman}, and
  {Unzicker}}]{Blake1964}
{Blake} RL, {Chubb} TA, {Friedman} H, {Unzicker} AE (1964) {Solar X-Ray
  Spectrum below 25 Angstroms}. Science 146(3647):1037--1038 

\bibitem[{{Blake} et~al(1965){Blake}, {Chubb}, {Friedman}, and
  {Unzicker}}]{Blake1965}
{Blake} RL, {Chubb} TA, {Friedman} H, {Unzicker} AE (1965) {Spectral and
  Photometric Measurements of Solar X-Ray Emission Below 60 {\r{A}}}. \apj
  142:1

\bibitem[{{Bradshaw} et~al(2012){Bradshaw}, {Klimchuk}, and
  {Reep}}]{Bradshaw2012}
{Bradshaw} SJ, {Klimchuk} JA, {Reep} JW (2012) {Diagnosing the Time-dependence
  of Active Region Core Heating from the Emission Measure. I. Low-frequency
  Nanoflares}. \apj 758(1):53

\bibitem[{{Brooks} et~al(2009){Brooks}, {Warren}, {Williams}, and
  {Watanabe}}]{Brooks2009}
{Brooks} DH, {Warren} HP, {Williams} DR, {Watanabe} T (2009)
  {Hinode/Extreme-Ultraviolet Imaging Spectrometer Observations of the
  Temperature Structure of the Quiet Corona}. \apj 705:1522--1532

\bibitem[{{Brooks} et~al(2013){Brooks}, {Warren}, {Ugarte-Urra}, and
  {Winebarger}}]{Brooks2013}
{Brooks} DH, {Warren} HP, {Ugarte-Urra} I, {Winebarger} AR (2013) {High Spatial
  Resolution Observations of Loops in the Solar Corona}. \apjl 772(2):L19 

\bibitem[{{Brooks} et~al(2015){Brooks}, {Ugarte-Urra}, and
  {Warren}}]{Brooks2015}
{Brooks} DH, {Ugarte-Urra} I, {Warren} HP (2015) {Full-Sun observations for
  identifying the source of the slow solar wind}. Nature Communications 6:5947 

\bibitem[{{Brosius} et~al(2014){Brosius}, {Daw}, and {Rabin}}]{Brosius2014}
{Brosius} JW, {Daw} AN, {Rabin} DM (2014) {Pervasive Faint Fe XIX Emission from
  a Solar Active Region Observed with EUNIS-13: Evidence for Nanoflare
  Heating}. \apj 790(2):112

\bibitem[{{Brown} et~al(2003){Brown}, {Nightingale}, {Alexander}, {Schrijver},
  {Metcalf}, {Shine}, {Title}, and {Wolfson}}]{Brown2003}
{Brown} DS, {Nightingale} RW, {Alexander} D, et al.\ (2003) {Observations of Rotating
  Sunspots from TRACE}. \solphys 216(1):79--108

\bibitem[{{Brueckner} and {Bartoe}(1983)}]{Brueckner1983}
{Brueckner} GE, {Bartoe} JDF (1983) {Observations of high-energy jets in the
  corona above the quiet sun, the heating of the corona, and the acceleration
  of the solar wind}. \apj 272:329--348

\bibitem[{{Brun} and {Browning}(2017)}]{Brun2017}
{Brun} AS, {Browning} MK (2017) {Magnetism, dynamo action and the solar-stellar
  connection}. Living Reviews in Solar Physics 14(1):4

\bibitem[{{Cargill} and {Klimchuk}(2004)}]{Cargill2004}
{Cargill} PJ, {Klimchuk} JA (2004) {Nanoflare Heating of the Corona Revisited}.
  \apj 605:911--920

\bibitem[{{Cargill} et~al(2015){Cargill}, {Warren}, and
  {Bradshaw}}]{Cargill2015}
{Cargill} PJ, {Warren} HP, {Bradshaw} SJ (2015) {Modelling nanoflares in active
  regions and implications for coronal heating mechanisms}. Philosophical
  Transactions of the Royal Society of London Series A
  373(2042):20140260--20140260

\bibitem[{{Caspi} et~al(2014){Caspi}, {Krucker}, and {Lin}}]{Caspi2014}
{Caspi} A, {Krucker} S, {Lin} RP (2014) {Statistical Properties of Super-hot
  Solar Flares}. \apj 781(1):43

\bibitem[{{Charbonneau}(2020)}]{Charbonneau2020}
{Charbonneau} P (2020) {Dynamo models of the solar cycle}. Living Reviews in
  Solar Physics 17(1):4

\bibitem[{{Chen}(2011)}]{Chen2011}
{Chen} PF (2011) {Coronal Mass Ejections: Models and Their Observational
  Basis}. Living Reviews in Solar Physics 8(1):1

\bibitem[{{Cheng} and {Ding}(2016)}]{Cheng2016}
{Cheng} X, {Ding} MD (2016) {Spectroscopic Diagnostics of Solar Magnetic Flux
  Ropes Using Iron Forbidden Line}. \apjl 823:L4

\bibitem[{{Cheung} et~al(2021){Cheung}, {Mart{\'\i}nez-Sykora}, {Testa}, {De
  Pontieu}, {Chintzoglou}, {Rempel}, {Polito}, {Kerr}, {Reeves}, {Fletcher},
  {Jin}, {N{\'o}brega-Siverio}, {Danilovic}, {Antolin}, {Allred}, {Hansteen},
  {Ugarte-Urra}, {DeLuca}, {Longcope}, {Takasao}, {DeRosa}, {Boerner},
  {Jaeggli}, {Nitta}, {Daw}, {Carlsson}, {Golub}, and {the MUSE
  team}}]{Cheung2021arXiv}
{Cheung} MCM, {Mart{\'\i}nez-Sykora} J, {Testa} P, et al.\  (2021) {Probing the Physics of the Solar
  Atmosphere with the Multi-slit Solar Explorer (MUSE): II. Flares and
  Eruptions}. arXiv e-prints arXiv:2106.15591, \eprint{2106.15591}

\bibitem[{{Chitta} et~al(2018){Chitta}, {Peter}, and {Solanki}}]{Chitta2018}
{Chitta} LP, {Peter} H, {Solanki} SK (2018) {Nature of the energy source
  powering solar coronal loops driven by nanoflares}. \aap 615:L9

\bibitem[{{Christe} et~al(2016){Christe}, {Glesener}, {Buitrago-Casas},
  {Ishikawa}, {Ramsey}, {Gubarev}, {Kilaru}, {Kolodziejczak}, {Watanabe},
  {Takahashi}, {Tajima}, {Turin}, {Shourt}, {Foster}, and
  {Krucker}}]{Christe2016}
{Christe} S, {Glesener} L, {Buitrago-Casas} C, et al.\ (2016) {FOXSI-2:
  Upgrades of the Focusing Optics X-ray Solar Imager for its Second Flight}.
  Journal of Astronomical Instrumentation 5(1):1640005-625

\bibitem[{{Cirtain} et~al(2013){Cirtain}, {Golub}, {Winebarger}, and {et
  al.}}]{Cirtain2013}
{Cirtain} JJ, {Golub} L, {Winebarger} A, {et al} (2013) {Energy release in the
  solar corona from spatially resolved magnetic braids}. \nat 493:501--503

\bibitem[{{Cirtain} et~al(2007){Cirtain}, {Golub}, {Lundquist}, {van
  Ballegooijen}, {Savcheva}, {Shimojo}, {DeLuca}, {Tsuneta}, {Sakao}, {Reeves},
  {Weber}, {Kano}, {Narukage}, and {Shibasaki}}]{Cirtain2007}
{Cirtain} JW, {Golub} L, {Lundquist} L, et al.\ (2007) {Evidence for Alfv{\'e}n Waves
  in Solar X-ray Jets}. Science 318(5856):1580

\bibitem[{{Cooper} et~al(2020){Cooper}, {Hannah}, {Grefenstette}, {Glesener},
  {Krucker}, {Hudson}, {White}, and {Smith}}]{Cooper2020}
{Cooper} K, {Hannah} IG, {Grefenstette} BW, et al.\ (2020) {NuSTAR Observation of a Minuscule
  Microflare in a Solar Active Region}. \apjl 893(2):L40

\bibitem[{{Cooper} et~al(2021){Cooper}, {Hannah}, {Grefenstette}, {Glesener},
  {Krucker}, {Hudson}, {White}, {Smith}, and {Duncan}}]{Cooper2021}
{Cooper} K, {Hannah} IG, {Grefenstette} BW, et al.\ (2021) {NuSTAR observations of a
  repeatedly microflaring active region}. \mnras 507(3):3936--3951

\bibitem[{{Cranmer}(2009)}]{Cranmer2009}
{Cranmer} SR (2009) {Coronal Holes}. Living Reviews in Solar Physics 6(1):3

\bibitem[{{Culhane} et~al(1981){Culhane}, {Rapley}, {Bentley}, {Gabriel},
  {Phillips}, {Acton}, {Wolfson}, {Catura}, {Jordan}, and
  {Antonucci}}]{Culhane81}
{Culhane} JL, {Rapley} CG, {Bentley} RD, et al.\ (1981) {X-ray
  spectra of solar flares obtained with a high-resolution bent crystal
  spectrometer}. \apjl 244:L141--L145

\bibitem[{{Culhane} et~al(2007){Culhane}, {Harra}, {James}, {Al-Janabi},
  {Bradley}, {Chaudry}, {Rees}, {Tandy}, {Thomas}, {Whillock}, {Winter},
  {Doschek}, {Korendyke}, {Brown}, {Myers}, {Mariska}, {Seely}, {Lang}, {Kent},
  {Shaughnessy}, {Young}, {Simnett}, {Castelli}, {Mahmoud}, {Mapson-Menard},
  {Probyn}, {Thomas}, {Davila}, {Dere}, {Windt}, {Shea}, {Hagood}, {Moye},
  {Hara}, {Watanabe}, {Matsuzaki}, {Kosugi}, {Hansteen}, and
  {Wikstol}}]{Culhane2007}
{Culhane} JL, {Harra} LK, {James} AM, et al.\ (2007) {The EUV
  Imaging Spectrometer for Hinode}. \solphys 243:19--61

\bibitem[{{Davis} et~al(1977){Davis}, {Golub}, and {Krieger}}]{Davis1977}
{Davis} JM, {Golub} L, {Krieger} AS (1977) {Solar cycle variation of magnetic
  flux emergence.} \apjl 214:L141--L144

\bibitem[{{de Jager}(1965)}]{deJager1965}
{de Jager} C (1965) {Solar X radiation}. Annales d'Astrophysique 28:125

\bibitem[{{De Moortel} and {Browning}(2015)}]{DeMoortel2015}
{De Moortel} I, {Browning} P (2015) {Recent advances in coronal heating}.
  Philosophical Transactions of the Royal Society of London Series A
  373(2042):20140269--20140269

\bibitem[{{De Pontieu} et~al(1999){De Pontieu}, {Berger}, {Schrijver}, and
  {Title}}]{DePontieu1999}
{De Pontieu} B, {Berger} TE, {Schrijver} CJ, {Title} AM (1999) {Dynamics of
  Transition Region `Moss' at high time resolution}. \solphys 190:419--435

\bibitem[{{De Pontieu} et~al(2014){De Pontieu}, {Title}, {Lemen}, {Kushner},
  {Akin}, {Allard}, {Berger}, {Boerner}, {Cheung}, {Chou}, {Drake}, {Duncan},
  {Freeland}, {Heyman}, {Hoffman}, {Hurlburt}, {Lindgren}, {Mathur}, {Rehse},
  {Sabolish}, {Seguin}, {Schrijver}, {Tarbell}, {W{\"u}lser}, {Wolfson},
  {Yanari}, {Mudge}, {Nguyen-Phuc}, {Timmons}, {van Bezooijen}, {Weingrod},
  {Brookner}, {Butcher}, {Dougherty}, {Eder}, {Knagenhjelm}, {Larsen},
  {Mansir}, {Phan}, {Boyle}, {Cheimets}, {DeLuca}, {Golub}, {Gates}, {Hertz},
  {McKillop}, {Park}, {Perry}, {Podgorski}, {Reeves}, {Saar}, {Testa}, {Tian},
  {Weber}, {Dunn}, {Eccles}, {Jaeggli}, {Kankelborg}, {Mashburn}, {Pust},
  {Springer}, {Carvalho}, {Kleint}, {Marmie}, {Mazmanian}, {Pereira}, {Sawyer},
  {Strong}, {Worden}, {Carlsson}, {Hansteen}, {Leenaarts}, {Wiesmann},
  {Aloise}, {Chu}, {Bush}, {Scherrer}, {Brekke}, {Martinez-Sykora}, {Lites},
  {McIntosh}, {Uitenbroek}, {Okamoto}, {Gummin}, {Auker}, {Jerram}, {Pool}, and
  {Waltham}}]{DePontieu14}
{De Pontieu} B, {Title} AM, {Lemen} JR, et al.\ (2014) {The Interface Region Imaging
  Spectrograph (IRIS)}. \solphys 289:2733--2779

\bibitem[{{De Pontieu} et~al(2020){De Pontieu}, {Mart{\'\i}nez-Sykora},
  {Testa}, {Winebarger}, {Daw}, {Hansteen}, {Cheung}, and
  {Antolin}}]{DePontieu2020}
{De Pontieu} B, {Mart{\'\i}nez-Sykora} J, {Testa} P, et al.\ (2020) {The Multi-slit Approach to
  Coronal Spectroscopy with the Multi-slit Solar Explorer (MUSE)}. \apj
  888(1):3

\bibitem[{{De Pontieu} et~al(2021{\natexlab{a}}){De Pontieu}, {Polito},
  {Hansteen}, {Testa}, {Reeves}, {Antolin}, {N{\'o}brega-Siverio}, {Kowalski},
  {Martinez-Sykora}, {Carlsson}, {McIntosh}, {Liu}, {Daw}, and
  {Kankelborg}}]{DePontieu2021}
{De Pontieu} B, {Polito} V, {Hansteen} V, {Testa} P, et al.\ (2021{\natexlab{a}}) {A New
  View of the Solar Interface Region from the Interface Region Imaging
  Spectrograph (IRIS)}. \solphys 296(5):84

\bibitem[{{De Pontieu} et~al(2021{\natexlab{b}}){De Pontieu}, {Testa},
  {Martinez-Sykora}, {Antolin}, {Karampelas}, {Hansteen}, {Rempel}, {Cheung},
  {Reale}, {Danilovic}, {Pagano}, {Polito}, {De Moortel}, {Nobrega-Siverio},
  {Van Doorsselaere}, {Petralia}, {Asgari-Targhi}, {Boerner}, {Carlsson},
  {Chintzoglou}, {Daw}, {DeLuca}, {Golub}, {Matsumoto}, {Ugarte-Urra},
  {McIntosh}, and {the MUSE team}}]{DePontieu2021arXiv}
{De Pontieu} B, {Testa} P, {Martinez-Sykora} J, et al.\ (2021{\natexlab{b}}) {Probing the physics of the solar
  atmosphere with the Multi-slit Solar Explorer (MUSE): I. Coronal Heating}.
  arXiv e-prints arXiv:2106.15584, \eprint{2106.15584}

\bibitem[{{Del Zanna} and {Mason}(2018)}]{DelZanna2018}
{Del Zanna} G, {Mason} HE (2018) {Solar UV and X-ray spectral diagnostics}.
  Living Reviews in Solar Physics 15(1):5

\bibitem[{{Del Zanna} et~al(2015){Del Zanna}, {Tripathi}, {Mason},
  {Subramanian}, and {O'Dwyer}}]{DelZanna2015}
{Del Zanna} G, {Tripathi} D, {Mason} H, {Subramanian} S, {O'Dwyer} B (2015)
  {The evolution of the emission measure distribution in the core of an active
  region}. \aap 573:A104

\bibitem[{{Del Zanna} et~al(2021){Del Zanna}, {Andretta}, {Cargill}, {Corso},
  {Daw}, {Golub}, {Klimchuk}, and {Mason}}]{DelZanna2021}
{Del Zanna} G, {Andretta} V, {Cargill} PJ, et al.\ (2021) {High resolution soft X-ray spectroscopy and
  the quest for the hot (5-10 MK) plasma in solar active regions}. Frontiers in
  Astronomy and Space Sciences 8:33

\bibitem[{{Dennis} and {Zarro}(1993)}]{dennis93}
{Dennis} BR, {Zarro} DM (1993) {The Neupert effect - What can it tell us about
  the impulsive and gradual phases of solar flares?} \solphys 146:177--190

\bibitem[{{Dere} et~al(1974){Dere}, {Horan}, and {Kreplin}}]{Dere1974}
{Dere} KP, {Horan} DM, {Kreplin} RW (1974) {The spectral dependence of solar
  soft X-ray flux values obtained by SOLRAD 9.} Journal of Atmospheric and
  Terrestrial Physics 36:989--994

\bibitem[{{Doschek}(1990)}]{Doschek90}
{Doschek} GA (1990) {Soft X-ray spectroscopy of solar flares - an overview}.
  \apjs 73:117--130

\bibitem[{{Edl{\'e}n}(1943)}]{Edlen1943}
{Edl{\'e}n} B (1943) {Die Deutung der Emissionslinien im Spektrum der
  Sonnenkorona. Mit 6 Abbildungen.} \zap 22:30

\bibitem[{{Evans} and {Pounds}(1967)}]{Evans1967}
{Evans} K, {Pounds} KA (1967) {X-Ray Emission Line Spectrum of a Coronal Active
  Region}. \nat 214(5083):41--42

\bibitem[{{Evans} and {Pounds}(1968)}]{Evans1968}
{Evans} K, {Pounds} KA (1968) {The X-Ray Emission Spectrum of a Solar Active
  Region}. \apj 152:319--+

\bibitem[{{Feldman}(1998)}]{Feldman1998}
{Feldman} U (1998) {FIP Effect in the Solar Upper Atmosphere: Spectroscopic
  Results}. \ssr 85:227--240

\bibitem[{{Feldman} and {Widing}(2003)}]{Feldman2003}
{Feldman} U, {Widing} KG (2003) {Elemental Abundances in the Solar Upper
  Atmosphere Derived by Spectroscopic Means}. \ssr 107(3):665--720

\bibitem[{{Firoz} et~al(2019){Firoz}, {Gan}, {Moon}, {Rodr{\'\i}guez-Pacheco},
  and {Li}}]{Firoz2019}
{Firoz} KA, {Gan} WQ, {Moon} YJ, {Rodr{\'\i}guez-Pacheco} J, {Li} YP (2019) {On
  the Relation between Flare and CME during GLE-SEP and Non-GLE-SEP Events}.
  \apj 883(1):91

\bibitem[{{Fisher} et~al(1985){Fisher}, {Canfield}, and {McClymont}}]{Fisher85}
{Fisher} GH, {Canfield} RC, {McClymont} AN (1985) {Flare loop radiative
  hydrodynamics. V - Response to thick-target heating. VI - Chromospheric
  evaporation due to heating by nonthermal electrons. VII - Dynamics of the
  thick-target heated chromosphere}. \apj 289:414--441

\bibitem[{{Fisher} et~al(1998){Fisher}, {Longcope}, {Metcalf}, and
  {Pevtsov}}]{Fisher1998}
{Fisher} GH, {Longcope} DW, {Metcalf} TR, {Pevtsov} AA (1998) {Coronal Heating
  in Active Regions as a Function of Global Magnetic Variables}. \apj
  508:885--898

\bibitem[{{Fletcher} and {De Pontieu}(1999)}]{Fletcher1999}
{Fletcher} L, {De Pontieu} B (1999) {Plasma Diagnostics of Transition Region
  ``Moss'' using SOHO/CDS and TRACE}. \apjl 520:L135--L138

\bibitem[{{Fritz} et~al(1967){Fritz}, {Kreplin}, {Meekins}, {Unzicker}, and
  {Friedman}}]{Fritz1967}
{Fritz} G, {Kreplin} RW, {Meekins} JF, {Unzicker} AE, {Friedman} H (1967)
  {Solar X-Ray Spectrum from 1.9 to 25 {\r{A}}}. \apjl 148:L133

\bibitem[{{Gabriel} and {Jordan}(1969)}]{Gabriel1969}
{Gabriel} AH, {Jordan} C (1969) {Interpretation of solar helium-like ion line
  intensities}. \mnras 145:241

\bibitem[{{Gabriel} and {Jordan}(1973)}]{Gabriel1973}
{Gabriel} AH, {Jordan} C (1973) {The Temperature Dependence of Line Ratios of
  Helium-Like Ions}. \apj 186:327--334

\bibitem[{{Gesztelyi}(1984)}]{Gesztelyi1984}
{Gesztelyi} L (1984) {Consecutive homologous flares and their relation to
  sunspot motions}. Advances in Space Research 4(7):19--22

\bibitem[{{Getman} and {Feigelson}(2021)}]{Getman2021}
{Getman} KV, {Feigelson} ED (2021) {X-Ray Superflares from Pre-main-sequence
  Stars: Flare Energetics and Frequency}. \apj 916(1):32  

\bibitem[{{Giacconi} and {Rossi}(1960)}]{Giacconi1960}
{Giacconi} R, {Rossi} B (1960) {A `Telescope' for Soft X-Ray Astronomy}. \jgr
  65:773

\bibitem[{{Giacconi} et~al(1962){Giacconi}, {Gursky}, {Paolini}, and
  {Rossi}}]{Giacconi1962}
{Giacconi} R, {Gursky} H, {Paolini} FR, {Rossi} BB (1962) {Evidence for x Rays
  From Sources Outside the Solar System}. \prl 9(11):439--443

\bibitem[{{Giacconi} et~al(1965){Giacconi}, {Reidy}, {Zehnpfennig}, {Lindsay},
  and {Muney}}]{Giacconi1965}
{Giacconi} R, {Reidy} WP, {Zehnpfennig} T, {Lindsay} JC, {Muney} WS (1965)
  {Solar X-Ray Image Obtained Using Grazing-Incidence Optics.} \apj
  142:1274--1278

\bibitem[{{Glesener} et~al(2017){Glesener}, {Krucker}, {Hannah}, {Hudson},
  {Grefenstette}, {White}, {Smith}, and {Marsh}}]{Glesener2017}
{Glesener} L, {Krucker} S, {Hannah} IG, et  al.\ (2017) {NuSTAR Hard X-Ray Observation of a Sub-A
  Class Solar Flare}. \apj 845(2):122

\bibitem[{{Glesener} et~al(2020){Glesener}, {Krucker}, {Duncan}, {Hannah},
  {Grefenstette}, {Chen}, {Smith}, {White}, and {Hudson}}]{Glesener2020}
{Glesener} L, {Krucker} S, {Duncan} J, et  al.\ (2020) {Accelerated Electrons Observed
  Down to \&lt;7 keV in a NuSTAR Solar Microflare}. \apjl 891(2):L34

\bibitem[{{Goldberg}(1967)}]{Goldberg1967}
{Goldberg} L (1967) {Ultraviolet and X Rays from the Sun}. \araa 5:279

\bibitem[{{Golub} and {Pasachoff}(2009)}]{Golub2009}
{Golub} L, {Pasachoff} JM (2009) {The Solar Corona}. Cambridge University Press, 2009. 

\bibitem[{{Golub} et~al(1974){Golub}, {Krieger}, {Silk}, {Timothy}, and
  {Vaiana}}]{Golub1974}
{Golub} L, {Krieger} AS, {Silk} JK, {Timothy} AF, {Vaiana} GS (1974) {Solar
  X-Ray Bright Points}. \apjl 189:L93

\bibitem[{{Golub} et~al(1979){Golub}, {Davis}, and {Krieger}}]{Golub1979}
{Golub} L, {Davis} JM, {Krieger} AS (1979) {Anticorrelation of X-ray bright
  points with sunspot number, 1970 - 1978.} \apjl 229:L145--L150

\bibitem[{{Golub} et~al(1990){Golub}, {Herant}, {Kalata}, {Lovas}, {Nystrom},
  {Pardo}, {Spiller}, and {Wilczynski}}]{Golub1990}
{Golub} L, {Herant} M, {Kalata} K, et al.\ (1990) {Sub-arcsecond observations of the solar X-ray
  corona}. \nat 344(6269):842--844

\bibitem[{{Golub} et~al(2007){Golub}, {Deluca}, {Austin}, {Bookbinder},
  {Caldwell}, {Cheimets}, {Cirtain}, {Cosmo}, {Reid}, {Sette}, {Weber},
  {Sakao}, {Kano}, {Shibasaki}, {Hara}, {Tsuneta}, {Kumagai}, {Tamura},
  {Shimojo}, {McCracken}, {Carpenter}, {Haight}, {Siler}, {Wright}, {Tucker},
  {Rutledge}, {Barbera}, {Peres}, and {Varisco}}]{Golub2007}
{Golub} L, {Deluca} E, {Austin} G, et al.\ (2007) {The X-Ray Telescope
  (XRT) for the Hinode Mission}. \solphys 243:63--86

\bibitem[{{Gou} et~al(2020){Gou}, {Veronig}, {Liu}, {Zhuang}, {Dumbovi{\'c}},
  {Podladchikova}, {Reid}, {Temmer}, {Dissauer}, {Vr{\v{s}}nak}, and
  {Wang}}]{Gou2020}
{Gou} T, {Veronig} AM, {Liu} R, et al.\ (2020)
  {Solar Flare-CME Coupling throughout Two Acceleration Phases of a Fast CME}.
  \apjl 897(2):L36

\bibitem[{{Graham} et~al(2019){Graham}, {De Pontieu}, and {Testa}}]{Graham2019}
{Graham} DR, {De Pontieu} B, {Testa} P (2019) {Automated Detection of Rapid
  Variability of Moss Using SDO/AIA and Its Connection to the Solar Corona}.
  \apjl 880(1):L12

\bibitem[{{Grefenstette} et~al(2016){Grefenstette}, {Glesener}, {Krucker},
  {Hudson}, {Hannah}, {Smith}, {Vogel}, {White}, {Madsen}, {Marsh}, {Caspi},
  {Chen}, {Shih}, {Kuhar}, {Boggs}, {Christensen}, {Craig}, {Forster},
  {Hailey}, {Harrison}, {Miyasaka}, {Stern}, and {Zhang}}]{Grefenstette2016}
{Grefenstette} BW, {Glesener} L, {Krucker} S, et al.\ (2016) {The
  First Focused Hard X-ray Images of the Sun with NuSTAR}. \apj 826(1):20

\bibitem[{{Grotrian}(1939)}]{Grotrian1939}
{Grotrian} W (1939) {Zur Frage der Deutung der Linien im Spektrum der
  Sonnenkorona}. Naturwissenschaften 27(13):214--214

\bibitem[{{Gudiksen} et~al(2011){Gudiksen}, {Carlsson}, {Hansteen}, {Hayek},
  {Leenaarts}, and {Mart{\'{\i}}nez-Sykora}}]{Gudiksen2011}
{Gudiksen} BV, {Carlsson} M, {Hansteen} VH, {Hayek} W, {Leenaarts} J,
  {Mart{\'{\i}}nez-Sykora} J (2011) {The stellar atmosphere simulation code
  Bifrost. Code description and validation}. \aap 531:A154

\bibitem[{{Hall} and {Hinteregger}(1970)}]{Hall1970}
{Hall} LA, {Hinteregger} HE (1970) {Solar radiation in the extreme ultraviolet
  and its variation with solar rotation}. \jgr 75(34):6959

\bibitem[{{Hannah} et~al(2008){Hannah}, {Christe}, {Krucker}, {Hurford},
  {Hudson}, and {Lin}}]{Hannah2008}
{Hannah} IG, {Christe} S, {Krucker} S, {Hurford} GJ, {Hudson} HS, {Lin} RP
  (2008) {RHESSI Microflare Statistics. II. X-Ray Imaging, Spectroscopy, and
  Energy Distributions}. \apj 677(1):704--718

\bibitem[{{Hannah} et~al(2016){Hannah}, {Grefenstette}, {Smith}, {Glesener},
  {Krucker}, {Hudson}, {Madsen}, {Marsh}, {White}, {Caspi}, {Shih}, {Harrison},
  {Stern}, {Boggs}, {Christensen}, {Craig}, {Hailey}, and {Zhang}}]{Hannah2016}
{Hannah} IG, {Grefenstette} BW, {Smith} DM, et al.\
  (2016) {The First X-Ray Imaging Spectroscopy of Quiescent Solar Active
  Regions with NuSTAR}. \apjl 820(1):L14

\bibitem[{{Hannah} et~al(2019){Hannah}, {Kleint}, {Krucker}, {Grefenstette},
  {Glesener}, {Hudson}, {White}, and {Smith}}]{Hannah2019}
{Hannah} IG, {Kleint} L, {Krucker} S, et al.\ (2019) {Joint X-Ray, EUV, and UV Observations of a
  Small Microflare}. \apj 881(2):109

\bibitem[{{Hanneman} and {Reeves}(2014)}]{Hanneman2014}
{Hanneman} WJ, {Reeves} KK (2014) {Thermal Structure of Current Sheets and
  Supra-arcade Downflows in the Solar Corona}. \apj 786(2):95

\bibitem[{{Hansteen} et~al(2015){Hansteen}, {Guerreiro}, {De Pontieu}, and
  {Carlsson}}]{Hansteen2015}
{Hansteen} V, {Guerreiro} N, {De Pontieu} B, {Carlsson} M (2015) {Numerical
  Simulations of Coronal Heating through Footpoint Braiding}. \apj 811:106

\bibitem[{{Hara} and {Nakakubo-Morimoto}(2003)}]{Hara2003}
{Hara} H, {Nakakubo-Morimoto} K (2003) {Variation of the X-Ray Bright Point
  Number over the Solar Activity Cycle}. \apj 589(2):1062--1074

\bibitem[{{Harrison} et~al(2013){Harrison}, {Craig}, {Christensen}, {Hailey},
  {Zhang}, {Boggs}, {Stern}, {Cook}, {Forster}, {Giommi}, {Grefenstette},
  {Kim}, {Kitaguchi}, {Koglin}, {Madsen}, {Mao}, {Miyasaka}, {Mori}, {Perri},
  {Pivovaroff}, {Puccetti}, {Rana}, {Westergaard}, {Willis}, {Zoglauer}, {An},
  {Bachetti}, {Barri{\`e}re}, {Bellm}, {Bhalerao}, {Brejnholt}, {Fuerst},
  {Liebe}, {Markwardt}, {Nynka}, {Vogel}, {Walton}, {Wik}, {Alexander},
  {Cominsky}, {Hornschemeier}, {Hornstrup}, {Kaspi}, {Madejski}, {Matt},
  {Molendi}, {Smith}, {Tomsick}, {Ajello}, {Ballantyne}, {Balokovi{\'c}},
  {Barret}, {Bauer}, {Blandford}, {Brandt}, {Brenneman}, {Chiang},
  {Chakrabarty}, {Chenevez}, {Comastri}, {Dufour}, {Elvis}, {Fabian}, {Farrah},
  {Fryer}, {Gotthelf}, {Grindlay}, {Helfand}, {Krivonos}, {Meier}, {Miller},
  {Natalucci}, {Ogle}, {Ofek}, {Ptak}, {Reynolds}, {Rigby}, {Tagliaferri},
  {Thorsett}, {Treister}, and {Urry}}]{Harrison2013}
{Harrison} FA, {Craig} WW, {Christensen} FE, et al.\ (2013) {The Nuclear
  Spectroscopic Telescope Array (NuSTAR) High-energy X-Ray Mission}. \apj
  770(2):103

\bibitem[{{Harvey}(2013)}]{Harvey2013}
{Harvey} JW (2013) {The Sun in Time}. \ssr 176(1-4):47--58

\bibitem[{{Harvey} and {Recely}(2002)}]{Harvey2002}
{Harvey} KL, {Recely} F (2002) {Polar Coronal Holes During Cycles 22 and 23}.
  \solphys 211(1):31--52

\bibitem[{{Hathaway}(2015)}]{Hathaway2015}
{Hathaway} DH (2015) {The Solar Cycle}. Living Reviews in Solar Physics
  12(1):4

\bibitem[{{Hazra} et~al(2020){Hazra}, {Vidotto}, and {D'Angelo}}]{Hazra2020}
{Hazra} G, {Vidotto} AA, {D'Angelo} CV (2020) {Influence of the Sun-like
  magnetic cycle on exoplanetary atmospheric escape}. \mnras 496(3):4017--4031

\bibitem[{{Hinode Review Team} et~al(2019){Hinode Review Team}, {Al-Janabi},
  {Antolin}, {Baker}, {Bellot Rubio}, {Bradley}, {Brooks}, {Centeno},
  {Culhane}, {Del Zanna}, {Doschek}, {Fletcher}, {Hara}, {Harra}, {Hillier},
  {Imada}, {Klimchuk}, {Mariska}, {Pereira}, {Reeves}, {Sakao}, {Sakurai},
  {Shimizu}, {Shimojo}, {Shiota}, {Solanki}, {Sterling}, {Su}, {Suematsu},
  {Tarbell}, {Tiwari}, {Toriumi}, {Ugarte-Urra}, {Warren}, {Watanabe}, and
  {Young}}]{hinodereview2019}
{Hinode Review Team}, et al.\ (2019) {Achievements of Hinode in the
  first eleven years}. \pasj 71(5):R1

\bibitem[{{Hirayama}(1974)}]{Hirayama74}
{Hirayama} T (1974) {Theoretical Model of Flares and Prominences. I:
  Evaporating Flare Model}. \solphys 34:323--338

\bibitem[{{Innes} et~al(2003{\natexlab{a}}){Innes}, {McKenzie}, and
  {Wang}}]{Innes2003b}
{Innes} DE, {McKenzie} DE, {Wang} T (2003{\natexlab{a}}) {Observations of 1000
  km s$^{-1}$ Doppler shifts in {}10$^{7}$ K solar flare supra-arcade}.
  \solphys 217(2):267--279

\bibitem[{{Innes} et~al(2003{\natexlab{b}}){Innes}, {McKenzie}, and
  {Wang}}]{Innes2003a}
{Innes} DE, {McKenzie} DE, {Wang} T (2003{\natexlab{b}}) {SUMER spectral
  observations of post-flare supra-arcade inflows}. \solphys 217(2):247--265

\bibitem[{{Ishikawa} et~al(2014){Ishikawa}, {Glesener}, {Christe}, {Ishibashi},
  {Brooks}, {Williams}, {Shimojo}, {Sako}, and {Krucker}}]{Ishikawa2014}
{Ishikawa} S, {Glesener} L, {Christe} S, et al.\ (2014) {Constraining hot
  plasma in a non-flaring solar active region with FOXSI hard X-ray
  observations}. \pasj 66:S15

\bibitem[{{Jakimiec} et~al(1992){Jakimiec}, {Sylwester}, {Sylwester}, {Serio},
  {Peres}, and {Reale}}]{Jakimiec1992}
{Jakimiec} J, {Sylwester} B, {Sylwester} J, {Serio} S, {Peres} G, {Reale} F
  (1992) {Dynamics of flaring loops. II - Flare evolution in the
  density-temperature diagram}. \aap 253(1):269--276

\bibitem[{{Kariyappa}(1999)}]{Kariyappa1999}
{Kariyappa} R (1999) {Quiet-Sun Variability with the Solar Cycle}. In:
  {Rimmele} TR, {Balasubramaniam} KS, {Radick} RR (eds) High Resolution Solar
  Physics: Theory, Observations, and Techniques, Astronomical Society of the
  Pacific Conference Series, vol 183, p 501

\bibitem[{{Klimchuk}(2006)}]{Klimchuk2006}
{Klimchuk} JA (2006) {On Solving the Coronal Heating Problem}. \solphys
  234:41--77

\bibitem[{{Klimchuk}(2015)}]{Klimchuk2015}
{Klimchuk} JA (2015) {Key aspects of coronal heating}. Philosophical
  Transactions of the Royal Society of London Series A
  373(2042):20140256--20140256

\bibitem[{{Klimchuk} and {Cargill}(2001)}]{Klimchuk2001}
{Klimchuk} JA, {Cargill} PJ (2001) {Spectroscopic Diagnostics of
  Nanoflare-heated Loops}. \apj 553:440--448

\bibitem[{{Ko} et~al(2009){Ko}, {Doschek}, {Warren}, and {Young}}]{Ko2009}
{Ko} Y, {Doschek} GA, {Warren} HP, {Young} PR (2009) {Hot Plasma in Nonflaring
  Active Regions Observed by the Extreme-Ultraviolet Imaging Spectrometer on
  Hinode}. \apj 697:1956--1970

\bibitem[{{Kobayashi} et~al(2014){Kobayashi}, {Cirtain}, {Winebarger},
  {Korreck}, {Golub}, {Walsh}, {De Pontieu}, {DeForest}, {Title}, {Kuzin},
  {Savage}, {Beabout}, {Beabout}, {Podgorski}, {Caldwell}, {McCracken},
  {Ordway}, {Bergner}, {Gates}, {McKillop}, {Cheimets}, {Platt}, {Mitchell},
  and {Windt}}]{Kobayashi2014}
{Kobayashi} K, {Cirtain} J, {Winebarger} AR, et al.\ (2014) {The High-Resolution Coronal Imager (Hi-C)}. \solphys
  289(11):4393--4412

\bibitem[{{Krieger} et~al(1971){Krieger}, {Vaiana}, and {van
  Speybroeck}}]{Krieger1971}
{Krieger} AS, {Vaiana} GS, {van Speybroeck} LP (1971) {The X-Ray Corona and the
  Photospheric Magnetic Field}. In: {Howard} R (ed) Solar Magnetic Fields,
  vol~43, p 397

\bibitem[{{Krieger} et~al(1973){Krieger}, {Timothy}, and
  {Roelof}}]{Krieger1973}
{Krieger} AS, {Timothy} AF, {Roelof} EC (1973) {A Coronal Hole and Its
  Identification as the Source of a High Velocity Solar Wind Stream}. \solphys
  29(2):505--525

\bibitem[{{Krucker} et~al(2014){Krucker}, {Christe}, {Glesener}, {Ishikawa},
  {Ramsey}, {Takahashi}, {Watanabe}, {Saito}, {Gubarev}, {Kilaru}, {Tajima},
  {Tanaka}, {Turin}, {McBride}, {Glaser}, {Fermin}, {White}, and
  {Lin}}]{Krucker2014}
{Krucker} S, {Christe} S, {Glesener} L, et al.\ (2014)
  {First Images from the Focusing Optics X-Ray Solar Imager}. \apjl 793(2):L32

\bibitem[{{Kuzin} et~al(2009){Kuzin}, {Bogachev}, {Zhitnik}, {Pertsov},
  {Ignatiev}, {Mitrofanov}, {Slemzin}, {Shestov}, {Sukhodrev}, and
  {Bugaenko}}]{Kuzin2009}
{Kuzin} SV, {Bogachev} SA, {Zhitnik} IA, et al.\
  (2009) {TESIS experiment on EUV imaging spectroscopy of the Sun}. Advances in
  Space Research 43(6):1001--1006

\bibitem[{{Laming}(2015)}]{Laming2015}
{Laming} JM (2015) {The FIP and Inverse FIP Effects in Solar and Stellar
  Coronae}. Living Reviews in Solar Physics 12(1):2

\bibitem[{{Lammer} et~al(2003){Lammer}, {Selsis}, {Ribas}, {Guinan}, {Bauer},
  and {Weiss}}]{Lammer2003}
{Lammer} H, {Selsis} F, {Ribas} I, {Guinan} EF, {Bauer} SJ, {Weiss} WW (2003)
  {Atmospheric Loss of Exoplanets Resulting from Stellar X-Ray and
  Extreme-Ultraviolet Heating}. \apjl 598(2):L121--L124

\bibitem[{{Leake} et~al(2013){Leake}, {Linton}, and
  {T{\"o}r{\"o}k}}]{Leake2013}
{Leake} JE, {Linton} MG, {T{\"o}r{\"o}k} T (2013) {Simulations of Emerging
  Magnetic Flux. I. The Formation of Stable Coronal Flux Ropes}. \apj
  778(2):99

\bibitem[{{Leka} et~al(1996){Leka}, {Canfield}, {McClymont}, and {van
  Driel-Gesztelyi}}]{Leka1996}
{Leka} KD, {Canfield} RC, {McClymont} AN, {van Driel-Gesztelyi} L (1996)
  {Evidence for Current-carrying Emerging Flux}. \apj 462:547

\bibitem[{{Lemen} et~al(2012){Lemen}, {Title}, {Akin}, {Boerner}, {Chou},
  {Drake}, {Duncan}, {Edwards}, {Friedlaender}, {Heyman}, {Hurlburt}, {Katz},
  {Kushner}, {Levay}, {Lindgren}, {Mathur}, {McFeaters}, {Mitchell}, {Rehse},
  {Schrijver}, {Springer}, {Stern}, {Tarbell}, {Wuelser}, {Wolfson}, {Yanari},
  {Bookbinder}, {Cheimets}, {Caldwell}, {Deluca}, {Gates}, {Golub}, {Park},
  {Podgorski}, {Bush}, {Scherrer}, {Gummin}, {Smith}, {Auker}, {Jerram},
  {Pool}, {Soufli}, {Windt}, {Beardsley}, {Clapp}, {Lang}, and
  {Waltham}}]{Lemen12}
{Lemen} JR, {Title} AM, {Akin} DJ, et al.\ (2012) {The Atmospheric Imaging Assembly (AIA) on the
  Solar Dynamics Observatory (SDO)}. \solphys 275:17--40

\bibitem[{{Lin} et~al(2002){Lin}, {Dennis}, {Hurford}, {Smith}, {Zehnder},
  {Harvey}, {Curtis}, {Pankow}, {Turin}, {Bester}, {Csillaghy}, {Lewis},
  {Madden}, {van Beek}, {Appleby}, {Raudorf}, {McTiernan}, {Ramaty}, {Schmahl},
  {Schwartz}, {Krucker}, {Abiad}, {Quinn}, {Berg}, {Hashii}, {Sterling},
  {Jackson}, {Pratt}, {Campbell}, {Malone}, {Landis}, {Barrington-Leigh},
  {Slassi-Sennou}, {Cork}, {Clark}, {Amato}, {Orwig}, {Boyle}, {Banks},
  {Shirey}, {Tolbert}, {Zarro}, {Snow}, {Thomsen}, {Henneck}, {Mchedlishvili},
  {Ming}, {Fivian}, {Jordan}, {Wanner}, {Crubb}, {Preble}, {Matranga}, {Benz},
  {Hudson}, {Canfield}, {Holman}, {Crannell}, {Kosugi}, {Emslie}, {Vilmer},
  {Brown}, {Johns-Krull}, {Aschwanden}, {Metcalf}, and {Conway}}]{Lin2002}
{Lin} RP, {Dennis} BR, {Hurford} GJ, et al.\ (2002) {The Reuven Ramaty High-Energy Solar
  Spectroscopic Imager (RHESSI)}. \solphys 210:3--32

\bibitem[{{Livi} et~al(1989){Livi}, {Martin}, {Wang}, and {Ai}}]{Livi1989}
{Livi} SHB, {Martin} S, {Wang} H, {Ai} G (1989) {The Association of Flares to
  Cancelling Magnetic Features on the Sun}. \solphys 121(1-2):197--214

\bibitem[{{L{\'o}pez} et~al(2022){L{\'o}pez}, {Gim{\'e}nez de Castro},
  {Mandrini}, {Sim{\~o}es}, {Cristiani}, {Gary}, {Francile}, and
  {D{\'e}moulin}}]{Lopez2022}
{L{\'o}pez} FM, {Gim{\'e}nez de Castro} CG, {Mandrini} CH, et al.\ (2022) {A solar
  flare driven by thermal conduction observed in mid-infrared}. \aap 657:A51

\bibitem[{{MacNeice} et~al(1985){MacNeice}, {Pallavicini}, {Mason}, {Simnett},
  {Antonucci}, {Shine}, {Rust}, {Jordan}, and {Dennis}}]{MacNeice1985}
{MacNeice} P, {Pallavicini} R, {Mason} HE, et al.\ (1985) {Multiwavelength Analysis of a
  Well Observed Flare from Solar Maximum Mission}. \solphys 99(1-2):167--188

\bibitem[{{Madjarska}(2019)}]{Madjarska2019}
{Madjarska} MS (2019) {Coronal bright points}. Living Reviews in Solar Physics
  16(1):2

\bibitem[{{Maehara} et~al(2012){Maehara}, {Shibayama}, {Notsu}, {Notsu},
  {Nagao}, {Kusaba}, {Honda}, {Nogami}, and {Shibata}}]{Maehara2012}
{Maehara} H, {Shibayama} T, {Notsu} S, et al.\ (2012) {Superflares on solar-type stars}.
  \nat 485(7399):478--481

\bibitem[{{Martin} et~al(1985){Martin}, {Livi}, and {Wang}}]{Martin1985}
{Martin} SF, {Livi} SHB, {Wang} J (1985) {The cancellation of magnetic flux. II
  - In a decaying active region}. Australian Journal of Physics 38:929--959

\bibitem[{{Masuda} et~al(1994){Masuda}, {Kosugi}, {Hara}, {Tsuneta}, and
  {Ogawara}}]{Masuda1994}
{Masuda} S, {Kosugi} T, {Hara} H, {Tsuneta} S, {Ogawara} Y (1994) {A loop-top
  hard X-ray source in a compact solar flare as evidence for magnetic
  reconnection}. \nat 371(6497):495--497

\bibitem[{{McIntosh} et~al(2014){McIntosh}, {Wang}, {Leamon}, {Davey}, {Howe},
  {Krista}, {Malanushenko}, {Markel}, {Cirtain}, {Gurman}, {Pesnell}, and
  {Thompson}}]{McIntosh2014}
{McIntosh} SW, {Wang} X, {Leamon} RJ, et al.\ (2014) {Deciphering Solar Magnetic Activity. I. On the
  Relationship between the Sunspot Cycle and the Evolution of Small Magnetic
  Features}. \apj 792(1):12

\bibitem[{{McKenzie} and {Hudson}(1999)}]{McKenzie1999}
{McKenzie} DE, {Hudson} HS (1999) {X-Ray Observations of Motions and Structure
  above a Solar Flare Arcade}. \apjl 519(1):L93--L96

\bibitem[{{McTiernan}(2009)}]{McTiernan2009}
{McTiernan} JM (2009) {RHESSI/GOES Observations of the Nonflaring Sun from 2002
  to 2006}. \apj 697:94--99

\bibitem[{{Miao} et~al(2019){Miao}, {Liu}, {Shen}, {Li}, {Abidin}, {Elmhamdi},
  and {Kordi}}]{Miao2019}
{Miao} YH, {Liu} Y, {Shen} YD, {Li} HB, {Abidin} ZZ, {Elmhamdi} A, {Kordi} AS
  (2019) {A Quasi-periodic Propagating Wave and Extreme-ultraviolet Waves
  Excited Simultaneously in a Solar Eruption Event}. \apjl 871(1):L2

\bibitem[{{Miceli} et~al(2012){Miceli}, {Reale}, {Gburek}, {Terzo}, {Barbera},
  {Collura}, {Sylwester}, {Kowalinski}, {Podgorski}, and
  {Gryciuk}}]{Miceli2012}
{Miceli} M, {Reale} F, {Gburek} S, et al.\ (2012) {X-ray
  emitting hot plasma in solar active regions observed by the SphinX
  spectrometer}. \aap 544:A139

\bibitem[{{Min} and {Chae}(2009)}]{Min2009}
{Min} S, {Chae} J (2009) {The Rotating Sunspot in AR 10930}. \solphys
  258(2):203--217

\bibitem[{{Mou} et~al(2018){Mou}, {Madjarska}, {Galsgaard}, and
  {Xia}}]{Mou2018}
{Mou} C, {Madjarska} MS, {Galsgaard} K, {Xia} L (2018) {Eruptions from quiet
  Sun coronal bright points. I. Observations}. \aap 619:A55

\bibitem[{{Nagai}(1980)}]{Nagai1980}
{Nagai} F (1980) {A Model of Hot Loops Associated with Solar Flares - Part One
  - Gasdynamics in the Loops}. \solphys 68(2):351--379

\bibitem[{{Neupert}(1968)}]{Neupert1968}
{Neupert} WM (1968) {Comparison of Solar X-Ray Line Emission with Microwave
  Emission during Flares}. \apjl 153:L59

\bibitem[{{Nitta} et~al(2021){Nitta}, {Mulligan}, {Kilpua}, {Lynch}, {Mierla},
  {O'Kane}, {Pagano}, {Palmerio}, {Pomoell}, {Richardson}, {Rodriguez},
  {Rouillard}, {Sinha}, {Srivastava}, {Talpeanu}, {Yardley}, and
  {Zhukov}}]{Nitta2021}
{Nitta} NV, {Mulligan} T, {Kilpua} EKJ, et al.\ (2021) {Understanding the Origins of Problem Geomagnetic Storms
  Associated with ``Stealth'' Coronal Mass Ejections}. \ssr 217(8):82

\bibitem[{{Ogawara} et~al(1992){Ogawara}, {Acton}, {Bentley}, {Bruner},
  {Culhane}, {Hiei}, {Hirayama}, {Hudson}, {Kosugi}, {Lemen}, {Strong},
  {Tsuneta}, {Uchida}, {Watanabe}, and {Yoshimori}}]{yohkoh}
{Ogawara} Y, {Acton} LW, {Bentley} RD, et al.\ (1992) {The status of YOHKOH in orbit
  - an introduction to the initial scientific results}. \pasj 44:L41--L44

\bibitem[{{Okamoto} et~al(2008){Okamoto}, {Tsuneta}, {Lites}, {Kubo},
  {Yokoyama}, {Berger}, {Ichimoto}, {Katsukawa}, {Nagata}, {Shibata},
  {Shimizu}, {Shine}, {Suematsu}, {Tarbell}, and {Title}}]{Okamoto2008}
{Okamoto} TJ, {Tsuneta} S, {Lites} BW, et al.\ (2008) {Emergence of a Helical
  Flux Rope under an Active Region Prominence}. \apjl 673(2):L215

\bibitem[{{Okamoto} et~al(2009){Okamoto}, {Tsuneta}, {Lites}, {Kubo},
  {Yokoyama}, {Berger}, {Ichimoto}, {Katsukawa}, {Nagata}, {Shibata},
  {Shimizu}, {Shine}, {Suematsu}, {Tarbell}, and {Title}}]{Okamoto2009}
{Okamoto} TJ, {Tsuneta} S, {Lites} BW, et al.\ (2009) {Prominence Formation
  Associated with an Emerging Helical Flux Rope}. \apj 697(1):913--922

\bibitem[{{Pallavicini} et~al(1975){Pallavicini}, {Vaiana}, {Kahler}, and
  {Krieger}}]{Pallavicini1975}
{Pallavicini} R, {Vaiana} GS, {Kahler} SW, {Krieger} AS (1975) {Spatial
  Structure and Temporal Development of a Solar X-Ray Flare Observed from
  Skylab on June 15, 1973}. \solphys 45(2):411--433

\bibitem[{{Pallavicini} et~al(1977){Pallavicini}, {Serio}, and
  {Vaiana}}]{Pallavicini1977}
{Pallavicini} R, {Serio} S, {Vaiana} GS (1977) {A survey of soft X-ray limb
  flare images: the relation between their structure in the corona and other
  physical parameters.} \apj 216:108--122

\bibitem[{{Paraschiv} et~al(2015){Paraschiv}, {Bemporad}, and
  {Sterling}}]{Paraschiv2015}
{Paraschiv} AR, {Bemporad} A, {Sterling} AC (2015) {Physical properties of
  solar polar jets. A statistical study with Hinode XRT data}. \aap 579:A96

\bibitem[{{Park} et~al(2010){Park}, {Chae}, and {Wang}}]{Park2010}
{Park} Sh, {Chae} J, {Wang} H (2010) {Productivity of Solar Flares and Magnetic
  Helicity Injection in Active Regions}. \apj 718(1):43--51

\bibitem[{{Parker}(1988)}]{Parker1988}
{Parker} EN (1988) {Nanoflares and the solar X-ray corona}. \apj 330:474--479

\bibitem[{{Parnell} and {De Moortel}(2012)}]{Parnell2012}
{Parnell} CE, {De Moortel} I (2012) {A contemporary view of coronal heating}.
  Philosophical Transactions of the Royal Society of London Series A
  370(1970):3217--3240

\bibitem[{{Penz} et~al(2008){Penz}, {Micela}, and {Lammer}}]{Penz2008}
{Penz} T, {Micela} G, {Lammer} H (2008) {Influence of the evolving stellar
  X-ray luminosity distribution on exoplanetary mass loss}. \aap
  477(1):309--314

\bibitem[{{Peres} et~al(1987){Peres}, {Reale}, {Serio}, and
  {Pallavicini}}]{Peres87}
{Peres} G, {Reale} F, {Serio} S, {Pallavicini} R (1987) {Hydrodynamic flare
  modeling - Comparison of numerical calculations with SMM observations of the
  1980 November 12 17:00 UT flare}. \apj 312:895--908

\bibitem[{{Peres} et~al(2004){Peres}, {Orlando}, and {Reale}}]{Peres04}
{Peres} G, {Orlando} S, {Reale} F (2004) {Are Coronae of Late Type Stars Made
  of Solar-Like Structures? The $F_{X}$-HR Diagram and the Pressure-Temperature
  Correlation}. \apj 612:472--480

\bibitem[{{Peter}(2015)}]{Peter2015}
{Peter} H (2015) {What can large-scale magnetohydrodynamic numerical
  experiments tell us about coronal heating?} Philosophical Transactions of the
  Royal Society of London Series A 373(2042):20150055--20150055

\bibitem[{{Petralia} et~al(2014){Petralia}, {Reale}, {Testa}, and {Del
  Zanna}}]{petralia14}
{Petralia} A, {Reale} F, {Testa} P, {Del Zanna} G (2014) {Thermal structure of
  a hot non-flaring corona from Hinode/EIS}. \aap 564:A3

\bibitem[{{Petrie}(2015)}]{Petrie2015}
{Petrie} GJD (2015) {Solar Magnetism in the Polar Regions}. Living Reviews in
  Solar Physics 12(1):5,

\bibitem[{{Pevtsov} et~al(2003){Pevtsov}, {Fisher}, {Acton}, {Longcope},
  {Johns-Krull}, {Kankelborg}, and {Metcalf}}]{Pevtsov2003}
{Pevtsov} AA, {Fisher} GH, {Acton} LW, et al.\ (2003) {The Relationship Between X-Ray Radiance
  and Magnetic Flux}. \apj 598:1387--1391

\bibitem[{{Phillips} et~al(1982){Phillips}, {Fawcett}, {Kent}, {Gabriel},
  {Leibacher}, {Wolfson}, {Acton}, {Parkinson}, {Culhane}, and
  {Mason}}]{Phillips1982}
{Phillips} KJH, {Fawcett} BC, {Kent} BJ, et al.\ (1982)
  {Solar flare X-ray spectra from the Solar Maximum Mission Flat Crystal
  Spectrometer}. \apj 256:774--787

\bibitem[{{Polito} et~al(2018{\natexlab{a}}){Polito}, {Galan}, {Reeves}, and
  {Musset}}]{Polito2018b}
{Polito} V, {Galan} G, {Reeves} KK, {Musset} S (2018{\natexlab{a}}) {Possible
  Signatures of a Termination Shock in the 2014 March 29 X-class Flare Observed
  by IRIS}. \apj 865(2):161

\bibitem[{{Polito} et~al(2018{\natexlab{b}}){Polito}, {Testa}, {Allred}, {De
  Pontieu}, {Carlsson}, {Pereira}, {Go{\v{s}}i{\'c}}, and {Reale}}]{Polito2018}
{Polito} V, {Testa} P, {Allred} J, et  al.\ (2018{\natexlab{b}}) {Investigating the
  Response of Loop Plasma to Nanoflare Heating Using RADYN Simulations}. \apj
  856(2):178

\bibitem[{{Polito} et~al(2019){Polito}, {Testa}, and {De Pontieu}}]{Polito2019}
{Polito} V, {Testa} P, {De Pontieu} B (2019) {Can the Superposition of
  Evaporative Flows Explain Broad Fe XXIÊProfiles during Solar Flares?} \apjl
  879(2):L17

\bibitem[{{Priest} and {Forbes}(2002)}]{Priest2002b}
{Priest} ER, {Forbes} TG (2002) {The magnetic nature of solar flares}. \aapr
  10(4):313--377

\bibitem[{{Priest} et~al(2018){Priest}, {Chitta}, and {Syntelis}}]{Priest2018}
{Priest} ER, {Chitta} LP, {Syntelis} P (2018) {A Cancellation Nanoflare Model
  for Solar Chromospheric and Coronal Heating}. \apjl 862(2):L24

\bibitem[{{Qiu} et~al(2010){Qiu}, {Liu}, {Hill}, and {Kazachenko}}]{Qiu2010}
{Qiu} J, {Liu} W, {Hill} N, {Kazachenko} M (2010) {Reconnection and Energetics
  in Two-ribbon Flares: A Revisit of the Bastille-day Flare}. \apj
  725(1):319--330

\bibitem[{{Raouafi} et~al(2016){Raouafi}, {Patsourakos}, {Pariat}, {Young},
  {Sterling}, {Savcheva}, {Shimojo}, {Moreno-Insertis}, {DeVore}, {Archontis},
  {T{\"o}r{\"o}k}, {Mason}, {Curdt}, {Meyer}, {Dalmasse}, and
  {Matsui}}]{Raouafi2016}
{Raouafi} NE, {Patsourakos} S, {Pariat} E, et al.\
  (2016) {Solar Coronal Jets: Observations, Theory, and Modeling}. \ssr
  201(1-4):1--53

\bibitem[{{Reale}(2007)}]{Reale2007mod}
{Reale} F (2007) {Diagnostics of stellar flares from X-ray observations: from
  the decay to the rise phase}. \aap 471:271--279

\bibitem[{{Reale}(2014)}]{Reale2014}
{Reale} F (2014) {Coronal Loops: Observations and Modeling of Confined Plasma}.
  Living Reviews in Solar Physics 11(1):4

\bibitem[{{Reale}(2016)}]{Reale2016}
{Reale} F (2016) {Plasma Sloshing in Pulse-heated Solar and Stellar Coronal
  Loops}. \apjl 826(2):L20

\bibitem[{{Reale} and {Orlando}(2008)}]{Reale2008}
{Reale} F, {Orlando} S (2008) {Nonequilibrium of Ionization and the Detection
  of Hot Plasma in Nanoflare-heated Coronal Loops}. \apj 684:715--724

\bibitem[{{Reale} et~al(1997){Reale}, {Betta}, {Peres}, {Serio}, and
  {McTiernan}}]{Reale97}
{Reale} F, {Betta} R, {Peres} G, {Serio} S, {McTiernan} J (1997) {Determination
  of the length of coronal loops from the decay of X-ray flares I. Solar flares
  observed with YOHKOH SXT.} \aap 325:782--790

\bibitem[{{Reale} et~al(2001){Reale}, {Peres}, and {Orlando}}]{Reale2001}
{Reale} F, {Peres} G, {Orlando} S (2001) {The Sun as an X-Ray Star. III.
  Flares}. \apj 557(2):906--920, \doi{10.1086/321598}

\bibitem[{{Reale} et~al(2009){Reale}, {Testa}, {Klimchuk}, and
  {Parenti}}]{Reale2009}
{Reale} F, {Testa} P, {Klimchuk} JA, {Parenti} S (2009) {Evidence of Widespread
  Hot Plasma in a Nonflaring Coronal Active Region from Hinode/X-Ray
  Telescope}. \apj 698:756--765

\bibitem[{{Reale} et~al(2019{\natexlab{a}}){Reale}, {Testa}, {Petralia}, and
  {Graham}}]{Reale2019a}
{Reale} F, {Testa} P, {Petralia} A, {Graham} DR (2019{\natexlab{a}}) {Impulsive
  Coronal Heating from Large-scale Magnetic Rearrangements: From IRIS to
  SDO/AIA}. \apj 882(1):7

\bibitem[{{Reale} et~al(2019{\natexlab{b}}){Reale}, {Testa}, {Petralia}, and
  {Kolotkov}}]{Reale2019b}
{Reale} F, {Testa} P, {Petralia} A, {Kolotkov} DY (2019{\natexlab{b}})
  {Large-amplitude Quasiperiodic Pulsations as Evidence of Impulsive Heating in
  Hot Transient Loop Systems Detected in the EUV with SDO/AIA}. \apj
  884(2):131

\bibitem[{{Reeves}(2018)}]{Reeves2018}
{Reeves} KK (2018) {Hinode Observations of Flows and Heating Associated with
  Magnetic Reconnection During Solar Flares}. In: {Shimizu} T, {Imada} S,
  {Kubo} M (eds) First Ten Years of Hinode Solar On-Orbit Observatory,
  Astrophysics and Space Science Library, vol 449, p 135

\bibitem[{{Reeves} and {Golub}(2011)}]{Reeves2011}
{Reeves} KK, {Golub} L (2011) {Atmospheric Imaging Assembly Observations of Hot
  Flare Plasma}. \apjl 727(2):L52

\bibitem[{{Reeves} et~al(2008){Reeves}, {Seaton}, and {Forbes}}]{Reeves2008}
{Reeves} KK, {Seaton} DB, {Forbes} TG (2008) {Field Line Shrinkage in Flares
  Observed by the X-Ray Telescope on Hinode}. \apj 675(1):868--874

\bibitem[{{Rempel}(2017)}]{Rempel2017}
{Rempel} M (2017) {Extension of the MURaM Radiative MHD Code for Coronal
  Simulations}. \apj 834:10

\bibitem[{{Ribas} et~al(2005){Ribas}, {Guinan}, {G{\"u}del}, and
  {Audard}}]{Ribas2005}
{Ribas} I, {Guinan} EF, {G{\"u}del} M, {Audard} M (2005) {Evolution of the
  Solar Activity over Time and Effects on Planetary Atmospheres. I. High-Energy
  Irradiances (1-1700 {\r{A}})}. \apj 622(1):680--694

\bibitem[{{Rosner} et~al(1978){Rosner}, {Tucker}, and {Vaiana}}]{RTV}
{Rosner} R, {Tucker} WH, {Vaiana} GS (1978) {Dynamics of the quiescent solar
  corona}. \apj 220:643--645

\bibitem[{{Russell}(1965)}]{Russell1965}
{Russell} PC (1965) {Soft X-ray Image of the Sun}. \nat 205(4972):684--685

\bibitem[{{Sako} et~al(2013){Sako}, {Shimojo}, {Watanabe}, and
  {Sekii}}]{Sako2013}
{Sako} N, {Shimojo} M, {Watanabe} T, {Sekii} T (2013) {A Statistical Study of
  Coronal Active Events in the North Polar Region}. \apj 775(1):22

\bibitem[{{Sanz-Forcada} et~al(2011){Sanz-Forcada}, {Micela}, {Ribas},
  {Pollock}, {Eiroa}, {Velasco}, {Solano}, and
  {Garc{\'\i}a-{\'A}lvarez}}]{Sanz-Forcada2011}
{Sanz-Forcada} J, {Micela} G, {Ribas} I, et al.\ (2011) {Estimation of the XUV
  radiation onto close planets and their evaporation}. \aap 532:A6

\bibitem[{{Savage} et~al(2012){Savage}, {McKenzie}, and {Reeves}}]{Savage2012}
{Savage} SL, {McKenzie} DE, {Reeves} KK (2012) {Re-interpretation of
  Supra-arcade Downflows in Solar Flares}. \apjl 747(2):L40

\bibitem[{{Savcheva} and {van Ballegooijen}(2009)}]{Savcheva2009}
{Savcheva} A, {van Ballegooijen} A (2009) {Nonlinear Force-free Modeling of a
  Long-lasting Coronal Sigmoid}. \apj 703(2):1766--1777

\bibitem[{{Savcheva} et~al(2007){Savcheva}, {Cirtain}, {Deluca}, {Lundquist},
  {Golub}, {Weber}, {Shimojo}, {Shibasaki}, {Sakao}, {Narukage}, {Tsuneta}, and
  {Kano}}]{Savcheva2007}
{Savcheva} A, {Cirtain} J, {Deluca} EE, et al.\  (2007) {A Study of Polar Jet Parameters Based on Hinode XRT Observations}.
  \pasj 59:S771

\bibitem[{{Savcheva} et~al(2012){Savcheva}, {Pariat}, {van Ballegooijen},
  {Aulanier}, and {DeLuca}}]{Savcheva2012}
{Savcheva} A, {Pariat} E, {van Ballegooijen} A, {Aulanier} G, {DeLuca} E (2012)
  {Sigmoidal Active Region on the Sun: Comparison of a Magnetohydrodynamical
  Simulation and a Nonlinear Force-free Field Model}. \apj 750(1):15

\bibitem[{{Schmelz} et~al(2009){Schmelz}, {Saar}, {DeLuca}, {Golub}, {Kashyap},
  {Weber}, and {Klimchuk}}]{Schmelz2009a}
{Schmelz} JT, {Saar} SH, {DeLuca} EE, et al.\  (2009) {Hinode X-Ray Telescope Detection of Hot Emission from
  Quiescent Active Regions: A Nanoflare Signature?} \apjl 693:L131--L135

\bibitem[{{Schrijver}(2007)}]{Schrijver2007}
{Schrijver} CJ (2007) {A Characteristic Magnetic Field Pattern Associated with
  All Major Solar Flares and Its Use in Flare Forecasting}. \apjl
  655(2):L117--L1205, \doi{10.1086/511857}

\bibitem[{{Schrijver} et~al(2005){Schrijver}, {De Rosa}, {Title}, and
  {Metcalf}}]{Schrijver2005}
{Schrijver} CJ, {De Rosa} ML, {Title} AM, {Metcalf} TR (2005) {The
  Nonpotentiality of Active-Region Coronae and the Dynamics of the Photospheric
  Magnetic Field}. \apj 628(1):501--513

\bibitem[{{Serio} et~al(1991){Serio}, {Reale}, {Jakimiec}, {Sylwester}, and
  {Sylwester}}]{Serio91}
{Serio} S, {Reale} F, {Jakimiec} J, {Sylwester} B, {Sylwester} J (1991)
  {Dynamics of flaring loops. I - Thermodynamic decay scaling laws}. \aap
  241:197--202

\bibitem[{{Shi} et~al(2019){Shi}, {Li}, {Huang}, and {Chen}}]{Shi2019}
{Shi} M, {Li} B, {Huang} Z, {Chen} SX (2019) {Synthetic Emissions of the Fe XXI
  1354 {\r{A}} Line from Flare Loops Experiencing Fundamental Fast Sausage
  Oscillations}. \apj 874(1):87

\bibitem[{{Shibata} and {Magara}(2011)}]{Shibata2011}
{Shibata} K, {Magara} T (2011) {Solar Flares: Magnetohydrodynamic Processes}.
  Living Reviews in Solar Physics 8(1):6

\bibitem[{{Shibata} et~al(1995){Shibata}, {Masuda}, {Shimojo}, {Hara},
  {Yokoyama}, {Tsuneta}, {Kosugi}, and {Ogawara}}]{Shibata1995}
{Shibata} K, {Masuda} S, {Shimojo} M, et al.\ (1995) {Hot-Plasma Ejections Associated with
  Compact-Loop Solar Flares}. \apjl 451:L83

\bibitem[{{Shimizu} et~al(2019){Shimizu}, {Imada}, {Kawate}, {Ichimoto},
  {Suematsu}, {Hara}, {Katsukawa}, {Kubo}, {Toriumi}, {Watanabe}, {Yokoyama},
  {Korendyke}, {Warren}, {Tarbell}, {De Pontieu}, {Teriaca}, {Sch{\"u}hle},
  {Solanki}, {Harra}, {Matthews}, {Fludra}, {Auch{\`e}re}, {Andretta},
  {Naletto}, and {Zhukov}}]{Shimizu2019}
{Shimizu} T, {Imada} S, {Kawate} T, et al.\ (2019) {The Solar-C\_EUVST mission}.
  In: UV, X-Ray, and Gamma-Ray Space Instrumentation for Astronomy XXI, Society
  of Photo-Optical Instrumentation Engineers (SPIE) Conference Series, vol
  11118, p 1111807

\bibitem[{{Shimojo} et~al(2007){Shimojo}, {Narukage}, {Kano}, {Sakao},
  {Tsuneta}, {Shibasaki}, {Cirtain}, {Lundquist}, {Reeves}, and
  {Savcheva}}]{Shimojo2007}
{Shimojo} M, {Narukage} N, {Kano} R, et al.\ (2007) {Fine
  Structures of Solar X-Ray Jets Observed with the X-Ray Telescope aboard
  Hinode}. \pasj 59:S745

\bibitem[{{Sobel'Man} et~al(1996){Sobel'Man}, {Zhitnik}, {Ignat'ev}, {Korneev},
  {Klepikov}, {Krutov}, {Kuzin}, {Mitrofanov}, {Oparin}, {Pertsov},
  {Salashchenko}, {Slemzin}, {Stepanov}, {Tindo}, {Avetisyan}, {Lomkova},
  {Sukhanov}, and {Forin}}]{SobelMan1996}
{Sobel'Man} II, {Zhitnik} IA, {Ignat'ev} AP, et al.\ (1996) {X-ray spectroscopy of the Sun
  in the 0.84-30.4 nm band in the TEREK-K and RES-K experiments on the
  KORONAS-I satellite}. Astronomy Letters 22(4):539--554

\bibitem[{{Somov} et~al(2002){Somov}, {Kosugi}, {Hudson}, {Sakao}, and
  {Masuda}}]{Somov2002}
{Somov} BV, {Kosugi} T, {Hudson} HS, {Sakao} T, {Masuda} S (2002) {Magnetic
  Reconnection Scenario of the Bastille Day 2000 Flare}. \apj 579(2):863--873

\bibitem[{{Strong} et~al(1992){Strong}, {Harvey}, {Hirayama}, {Nitta},
  {Shimizu}, and {Tsuneta}}]{Strong1992}
{Strong} KT, {Harvey} K, {Hirayama} T, {Nitta} N, {Shimizu} T, {Tsuneta} S
  (1992) {Observations of the Variability of Coronal Bright Points by the Soft
  X-Ray Telescope on YOHKOH}. \pasj 44:L161--L166

\bibitem[{{Su} et~al(2009){Su}, {van Ballegooijen}, {Lites}, {Deluca}, {Golub},
  {Grigis}, {Huang}, and {Ji}}]{Su2009}
{Su} Y, {van Ballegooijen} A, {Lites} BW, et al.\ (2009) {Observations and Nonlinear Force-Free Field
  Modeling of Active Region 10953}. \apj 691(1):105--114

\bibitem[{{Sylwester} et~al(1993){Sylwester}, {Sylwester}, {Serio}, {Reale},
  {Bentley}, and {Fludra}}]{Sylwester1993}
{Sylwester} B, {Sylwester} J, {Serio} S, {Reale} F, {Bentley} RD, {Fludra} A
  (1993) {Dynamics of flaring loops. III - Interpretation of flare evolution in
  the emission measure-temperature diagram}. \aap 267(2):586--594

\bibitem[{{Sylwester} et~al(2010){Sylwester}, {Sylwester}, and
  {Phillips}}]{Sylwester2010}
{Sylwester} B, {Sylwester} J, {Phillips} KJH (2010) {Soft X-ray coronal spectra
  at low activity levels observed by RESIK}. \aap 514:A82

\bibitem[{{Sylwester} et~al(2020){Sylwester}, {Sylwester}, {Phillips}, {Kepa},
  and {Rapley}}]{Sylwester2020}
{Sylwester} J, {Sylwester} B, {Phillips} KJH, {Kepa} A, {Rapley} CG (2020) {A
  Unique Resource for Solar Flare Diagnostic Studies: The SMM Bent Crystal
  Spectrometer}. \apj 894(2):137

\bibitem[{{Temmer}(2021)}]{Temmer2021}
{Temmer} M (2021) {Space weather: the solar perspective}. Living Reviews in
  Solar Physics 18(1):4

\bibitem[{{Terzo} et~al(2011){Terzo}, {Reale}, {Miceli}, {Klimchuk}, {Kano},
  and {Tsuneta}}]{Terzo2011}
{Terzo} S, {Reale} F, {Miceli} M, {Klimchuk} JA, {Kano} R, {Tsuneta} S (2011)
  {Widespread Nanoflare Variability Detected with Hinode/X-Ray Telescope in a
  Solar Active Region}. \apj 736(2):111

\bibitem[{{Testa}(2010{\natexlab{a}})}]{Testa10ssrv}
{Testa} P (2010{\natexlab{a}}) {Element Abundances in X-ray Emitting Plasmas in
  Stars}. \ssr pp 130--+

\bibitem[{{Testa}(2010{\natexlab{b}})}]{Testa2010}
{Testa} P (2010{\natexlab{b}}) {X-ray emission processes in stars and their
  immediate environment}. Proceedings of the National Academy of Science
  107:7158--7163

\bibitem[{{Testa} and {Reale}(2012)}]{Testa2012b}
{Testa} P, {Reale} F (2012) {Hinode/EIS Spectroscopic Validation of Very Hot
  Plasma Imaged with the Solar Dynamics Observatory in Non-flaring Active
  Region Cores}. \apjl 750(1):L105, \doi{10.1088/2041-8205/750/1/L10},

\bibitem[{{Testa} and {Reale}(2020)}]{Testa2020b}
{Testa} P, {Reale} F (2020) {On the Coronal Temperature in Solar Microflares}.
  \apj 902(1):31

\bibitem[{{Testa} et~al(2005){Testa}, {Peres}, and {Reale}}]{Testa2005}
{Testa} P, {Peres} G, {Reale} F (2005) {Emission Measure Distribution in Loops
  Impulsively Heated at the Footpoints}. \apj 622:695--703

\bibitem[{{Testa} et~al(2011){Testa}, {Reale}, {Landi}, {DeLuca}, and
  {Kashyap}}]{Testa2011}
{Testa} P, {Reale} F, {Landi} E, {DeLuca} EE, {Kashyap} V (2011) {Temperature
  Distribution of a Non-flaring Active Region from Simultaneous Hinode XRT and
  EIS Observations}. \apj 728:30--+

\bibitem[{{Testa} et~al(2012){Testa}, {De Pontieu}, {Mart{\'{\i}}nez-Sykora},
  {Hansteen}, and {Carlsson}}]{Testa2012c}
{Testa} P, {De Pontieu} B, {Mart{\'{\i}}nez-Sykora} J, {Hansteen} V, {Carlsson}
  M (2012) {Investigating the Reliability of Coronal Emission Measure
  Distribution Diagnostics using Three-dimensional Radiative
  Magnetohydrodynamic Simulations}. \apj 758:54

\bibitem[{{Testa} et~al(2013){Testa}, {De Pontieu}, {Mart{\'{\i}}nez-Sykora},
  {DeLuca}, {Hansteen}, {Cirtain}, {Winebarger}, {Golub}, {Kobayashi},
  {Korreck}, {Kuzin}, {Walsh}, {DeForest}, {Title}, and {Weber}}]{Testa2013}
{Testa} P, {De Pontieu} B, {Mart{\'{\i}}nez-Sykora} J, et al.\ (2013) {Observing
  Coronal Nanoflares in Active Region Moss}. \apjl 770:L1

\bibitem[{{Testa} et~al(2014){Testa}, {De Pontieu}, {Allred}, {Carlsson},
  {Reale}, {Daw}, {Hansteen}, {Martinez-Sykora}, {Liu}, {DeLuca}, {Golub},
  {McKillop}, {Reeves}, {Saar}, {Tian}, {Lemen}, {Title}, {Boerner},
  {Hurlburt}, {Tarbell}, {Wuelser}, {Kleint}, {Kankelborg}, and
  {Jaeggli}}]{Testa2014}
{Testa} P, {De Pontieu} B, {Allred} J, et al.\ (2014) {Evidence of nonthermal particles in coronal loops heated
  impulsively by nanoflares}. Science 346(6207):1255724

\bibitem[{{Testa} et~al(2015){Testa}, {Saar}, and {Drake}}]{Testa2015}
{Testa} P, {Saar} SH, {Drake} JJ (2015) {Stellar activity and coronal heating:
  an overview of recent results}. Philosophical Transactions of the Royal
  Society of London Series A 373:20140259--20140259

\bibitem[{{Testa} et~al(2020){Testa}, {Polito}, and {De Pontieu}}]{Testa2020}
{Testa} P, {Polito} V, {De Pontieu} B (2020) {IRIS Observations of Short-term
  Variability in Moss Associated with Transient Hot Coronal Loops}. \apj
  889(2):124

\bibitem[{{Timothy} et~al(1975){Timothy}, {Krieger}, and
  {Vaiana}}]{Timothy1975}
{Timothy} AF, {Krieger} AS, {Vaiana} GS (1975) {The Structure and Evolution of
  Coronal Holes}. \solphys 42(1):135--156

\bibitem[{{Tripathi} et~al(2010){Tripathi}, {Mason}, {Del Zanna}, and
  {Young}}]{Tripathi2010}
{Tripathi} D, {Mason} HE, {Del Zanna} G, {Young} PR (2010) {Active region moss.
  Basic physical parameters and their temporal variation}. \aap 518:A42

\bibitem[{{Tsuneta} et~al(1991){Tsuneta}, {Acton}, {Bruner}, {Lemen}, {Brown},
  {Caravalho}, {Catura}, {Freeland}, {Jurcevich}, {Morrison}, {Ogawara},
  {Hirayama}, and {Owens}}]{Tsuneta1991sxt}
{Tsuneta} S, {Acton} L, {Bruner} M, et al.\ (1991) {The Soft X-ray Telescope for the SOLAR-A
  mission}. \solphys 136(1):37--67

\bibitem[{{Tsuneta} et~al(1997){Tsuneta}, {Masuda}, {Kosugi}, and
  {Sato}}]{Tsuneta1997}
{Tsuneta} S, {Masuda} S, {Kosugi} T, {Sato} J (1997) {Hot and Superhot Plasmas
  above an Impulsive Flare Loop}. \apj 478(2):787--798

\bibitem[{{Ugarte-Urra} and {Warren}(2014)}]{UgarteUrra2014}
{Ugarte-Urra} I, {Warren} HP (2014) {Determining Heating Timescales in Solar
  Active Region Cores from AIA/SDO Fe XVIII Images}. \apj 783(1):12

\bibitem[{{Ugarte-Urra} et~al(2006){Ugarte-Urra}, {Winebarger}, and
  {Warren}}]{UgarteUrra2006}
{Ugarte-Urra} I, {Winebarger} AR, {Warren} HP (2006) {An Investigation into the
  Variability of Heating in a Solar Active Region}. \apj 643(2):1245--1257

\bibitem[{{Ugarte-Urra} et~al(2009){Ugarte-Urra}, {Warren}, and
  {Brooks}}]{UgarteUrra2009}
{Ugarte-Urra} I, {Warren} HP, {Brooks} DH (2009) {Active Region Transition
  Region Loop Populations and Their Relationship to the Corona}. \apj
  695(1):642--651

\bibitem[{{Ugarte-Urra} et~al(2019){Ugarte-Urra}, {Crump}, {Warren}, and
  {Wiegelmann}}]{UgarteUrra2019}
{Ugarte-Urra} I, {Crump} NA, {Warren} HP, {Wiegelmann} T (2019) {The Magnetic
  Properties of Heating Events on High-temperature Active-region Loops}. \apj
  877(2):129

\bibitem[{{Upendran} and {Tripathi}(2021)}]{Upendran2021}
{Upendran} V, {Tripathi} D (2021) {On the Impulsive Heating of Quiet Solar
  Corona}. \apj 916(1):59

\bibitem[{{Vaiana} et~al(1968){Vaiana}, {Reidy}, {Zehnpfennig}, {van
  Speybroeck}, and {Giacconi}}]{Vaiana1968}
{Vaiana} GS, {Reidy} WP, {Zehnpfennig} T, {van Speybroeck} L, {Giacconi} R
  (1968) {X-ray Structures of the Sun during the Importance 1N Flare of 8 June
  1968}. Science 161:564--567

\bibitem[{{Vaiana} et~al(1973{\natexlab{a}}){Vaiana}, {Davis}, {Giacconi},
  {Krieger}, {Silk}, {Timothy}, and {Zombeck}}]{Vaiana1973a}
{Vaiana} GS, {Davis} JM, {Giacconi} R, {Krieger} AS, {Silk} JK, {Timothy} AF,
  {Zombeck} M (1973{\natexlab{a}}) {X-Ray Observations of Characteristic
  Structures and Time Variations from the Solar Corona: Preliminary Results
  from SKYLAB}. \apjl 185:L47

\bibitem[{{Vaiana} et~al(1973{\natexlab{b}}){Vaiana}, {Krieger}, and
  {Timothy}}]{Vaiana1973b}
{Vaiana} GS, {Krieger} AS, {Timothy} AF (1973{\natexlab{b}}) {Identification
  and Analysis of Structures in the Corona from X-Ray Photography}. \solphys
  32(1):81--116

\bibitem[{{Vaiana} et~al(1976){Vaiana}, {Krieger}, {Timothy}, and
  {Zombeck}}]{Vaiana1976}
{Vaiana} GS, {Krieger} AS, {Timothy} AF, {Zombeck} M (1976) {ATM Observations,
  X-Ray Results}. \apss 39(1):75--101

\bibitem[{{Vaiana} et~al(1977){Vaiana}, {van Speybroeck}, {Zombeck}, {Krieger},
  {Silk}, and {Timothy}}]{Vaiana1977}
{Vaiana} GS, {van Speybroeck} L, {Zombeck} MV, {Krieger} AS, {Silk} JK,
  {Timothy} A (1977) {The S-054 X-ray telescope experiment on Skylab.} Space
  Science Instrumentation 3:19--76

\bibitem[{{van Ballegooijen} et~al(2011){van Ballegooijen}, {Asgari-Targhi},
  {Cranmer}, and {DeLuca}}]{vanBallegooijen2011}
{van Ballegooijen} AA, {Asgari-Targhi} M, {Cranmer} SR, {DeLuca} EE (2011)
  {Heating of the Solar Chromosphere and Corona by Alfv{\'e}n Wave Turbulence}.
  \apj 736:3

\bibitem[{{van Ballegooijen} et~al(2017){van Ballegooijen}, {Asgari-Targhi},
  and {Voss}}]{vanBallegooijen2017}
{van Ballegooijen} AA, {Asgari-Targhi} M, {Voss} A (2017) {The Heating of Solar
  Coronal Loops by Alfv{\'e}n Wave Turbulence}. \apj 849(1):46

\bibitem[{{van Driel-Gesztelyi} and {Green}(2015)}]{vanDriel2015}
{van Driel-Gesztelyi} L, {Green} LM (2015) {Evolution of Active Regions}.
  Living Reviews in Solar Physics 12(1):1

\bibitem[{{Velli} et~al(2015){Velli}, {Pucci}, {Rappazzo}, and
  {Tenerani}}]{Velli2015}
{Velli} M, {Pucci} F, {Rappazzo} F, {Tenerani} A (2015) {Models of coronal
  heating, turbulence and fast reconnection}. Philosophical Transactions of the
  Royal Society of London Series A 373(2042):20140262--20140262

\bibitem[{{Viall} and {Klimchuk}(2011)}]{Viall2011}
{Viall} NM, {Klimchuk} JA (2011) {Patterns of Nanoflare Storm Heating Exhibited
  by an Active Region Observed with Solar Dynamics Observatory/Atmospheric
  Imaging Assembly}. \apj 738(1):24

\bibitem[{{Viall} and {Klimchuk}(2017)}]{Viall2017}
{Viall} NM, {Klimchuk} JA (2017) {A Survey of Nanoflare Properties in Active
  Regions Observed with the Solar Dynamics Observatory}. \apj 842(2):108

\bibitem[{{Vievering} et~al(2021){Vievering}, {Glesener}, {Athiray},
  {Buitrago-Casas}, {Musset}, {Ryan}, {Ishikawa}, {Duncan}, {Christe}, and
  {Krucker}}]{Vievering2021}
{Vievering} JT, {Glesener} L, {Athiray} PS, et al.\ (2021)
  {FOXSI-2 Solar Microflares. II. Hard X-ray Imaging Spectroscopy and Flare
  Energetics}. \apj 913(1):15

\bibitem[{{von Steiger} et~al(1995){von Steiger}, {Schweingruber}, {Geiss}, and
  {Gloeckler}}]{vonSteiger1995}
{von Steiger} R, {Schweingruber} RFW, {Geiss} J, {Gloeckler} G (1995)
  {Abundance variations in the solar wind}. Advances in Space Research
  15(7):3--12

\bibitem[{{Warren} et~al(2012){Warren}, {Winebarger}, and
  {Brooks}}]{Warren2012}
{Warren} HP, {Winebarger} AR, {Brooks} DH (2012) {A Systematic Survey of
  High-temperature Emission in Solar Active Regions}. \apj 759:141

\bibitem[{{Warren} et~al(2013){Warren}, {Mariska}, and {Doschek}}]{Warren2013}
{Warren} HP, {Mariska} JT, {Doschek} GA (2013) {Observations of Thermal Flare
  Plasma with the EUV Variability Experiment}. \apj 770(2):116

\bibitem[{{Warren} et~al(2016){Warren}, {Brooks}, {Doschek}, and
  {Feldman}}]{Warren2016}
{Warren} HP, {Brooks} DH, {Doschek} GA, {Feldman} U (2016) {Transition Region
  Abundance Measurements During Impulsive Heating Events}. \apj 824(1):56

\bibitem[{{Watanabe} et~al(2012){Watanabe}, {Masuda}, and
  {Segawa}}]{Watanabe2012}
{Watanabe} K, {Masuda} S, {Segawa} T (2012) {Hinode Flare Catalogue}. \solphys
  279(1):317--322

\bibitem[{{Webb} and {Howard}(2012)}]{Webb2012}
{Webb} DF, {Howard} TA (2012) {Coronal Mass Ejections: Observations}. Living
  Reviews in Solar Physics 9(1):3

\bibitem[{{Widing} and {Feldman}(2001)}]{Widing2001}
{Widing} KG, {Feldman} U (2001) {On the Rate of Abundance Modifications versus
  Time in Active Region Plasmas}. \apj 555:426--434

\bibitem[{{Wiegelmann} and {Sakurai}(2012)}]{Wiegelmann2012}
{Wiegelmann} T, {Sakurai} T (2012) {Solar Force-free Magnetic Fields}. Living
  Reviews in Solar Physics 9(1):5

\bibitem[{{Williams} et~al(2020){Williams}, {Walsh}, {Winebarger}, {Brooks},
  {Cirtain}, {De Pontieu}, {Golub}, {Kobayashi}, {McKenzie}, {Morton}, {Peter},
  {Rachmeler}, {Savage}, {Testa}, {Tiwari}, {Warren}, and
  {Watkinson}}]{Williams2020}
{Williams} T, {Walsh} RW, {Winebarger} AR, et al.\ (2020) {Is the High-Resolution Coronal Imager Resolving Coronal Strands?
  Results from AR 12712}. \apj 892(2):134

\bibitem[{{Winebarger} et~al(2012){Winebarger}, {Warren}, {Schmelz}, {Cirtain},
  {Mulu-Moore}, {Golub}, and {Kobayashi}}]{Winebarger2012}
{Winebarger} AR, {Warren} HP, {Schmelz} JT, et  al.\ (2012) {Defining the ``Blind Spot'' of Hinode EIS
  and XRT Temperature Measurements}. \apjl 746(2):L17

\bibitem[{{Wright} et~al(2017){Wright}, {Hannah}, {Grefenstette}, {Glesener},
  {Krucker}, {Hudson}, {Smith}, {Marsh}, {White}, and {Kuhar}}]{Wright2017}
{Wright} PJ, {Hannah} IG, {Grefenstette} BW, et al.\ (2017) {Microflare
  Heating of a Solar Active Region Observed with NuSTAR, Hinode/XRT, and
  SDO/AIA}. \apj 844(2):132

\bibitem[{{Yamashiki} et~al(2019){Yamashiki}, {Maehara}, {Airapetian}, {Notsu},
  {Sato}, {Notsu}, {Kuroki}, {Murashima}, {Sato}, {Namekata}, {Sasaki},
  {Scott}, {Bando}, {Nashimoto}, {Takagi}, {Ling}, {Nogami}, and
  {Shibata}}]{Yamashiki2019}
{Yamashiki} YA, {Maehara} H, {Airapetian} V, et al.\  (2019) {Impact of Stellar Superflares on Planetary Habitability}. \apj
  881(2):114

\bibitem[{{Young} et~al(2015){Young}, {Tian}, and {Jaeggli}}]{Young2015}
{Young} PR, {Tian} H, {Jaeggli} S (2015) {The 2014 March 29 X-flare:
  Subarcsecond Resolution Observations of Fe XXI {$\lambda$}1354.1}. \apj
  799:218

\bibitem[{{Yu} et~al(2014){Yu}, {Jackson}, {Buffington}, {Hick}, {Shimojo}, and
  {Sako}}]{Yu2014}
{Yu} HS, {Jackson} BV, {Buffington} A, {Hick} PP, {Shimojo} M, {Sako} N (2014)
  {The Three-dimensional Analysis of Hinode Polar Jets using Images from LASCO
  C2, the Stereo COR2 Coronagraphs, and SMEI}. \apj 784(2):166

\bibitem[{{Zender} et~al(2017){Zender}, {Kariyappa}, {Giono}, {Bergmann},
  {Delouille}, {Dam{\'e}}, {Hochedez}, and {Kumara}}]{Zender2017}
{Zender} JJ, {Kariyappa} R, {Giono} G, et al.\  (2017) {Segmentation of photospheric magnetic
  elements corresponding to coronal features to understand the EUV and UV
  irradiance variability}. \aap 605:A41

\bibitem[{{Zhang} et~al(2001){Zhang}, {Fang}, {Ding}, and
  {Livingston}}]{Zhang2001}
{Zhang} YX, {Fang} C, {Ding} MD, {Livingston} WC (2001) {Quiet-Sun Variability
  In a Temperature Minimum Region}. \apjl 547(2):L179--L182

\bibitem[{{Zhitnik} et~al(2003){Zhitnik}, {Kuzin}, {Afanas'ev}, {Bugaenko},
  {Ignat'ev}, {Krutov}, {Mitrofanov}, {Oparin}, {Pertsov}, {Slemzin},
  {Sukhodrev}, and {Umov}}]{Zhitnik2003}
{Zhitnik} I, {Kuzin} S, {Afanas'ev} A, et al.\ (2003) {XUV observations of solar corona in the spirit experiment on board
  the coronas-F satellite}. Advances in Space Research 32(4):473--477

\bibitem[{{Zimovets} et~al(2021){Zimovets}, {McLaughlin}, {Srivastava},
  {Kolotkov}, {Kuznetsov}, {Kupriyanova}, {Cho}, {Inglis}, {Reale}, {Pascoe},
  {Tian}, {Yuan}, {Li}, and {Zhang}}]{Zimovets2021}
{Zimovets} IV, {McLaughlin} JA, {Srivastava} AK, et al.\ (2021) {Quasi-Periodic Pulsations in Solar and
  Stellar Flares: A Review of Underpinning Physical Mechanisms and Their
  Predicted Observational Signatures}. \ssr 217(5):66

\bibitem[{{Zirin} and {Wang}(1990)}]{Zirin1990}
{Zirin} H, {Wang} H (1990) {Flows, flares, and formation of umbrae and light
  bridges in BBSO region No. 1167}. \solphys 125(1):45--60

\bibitem[{{Zurbuchen} et~al(2002){Zurbuchen}, {Fisk}, {Gloeckler}, and {von
  Steiger}}]{Zurbuchen2002}
{Zurbuchen} TH, {Fisk} LA, {Gloeckler} G, {von Steiger} R (2002) {The solar
  wind composition throughout the solar cycle: A continuum of dynamic states}.
  \grl 29(9):1352

\end{thebibliography}

\end{document}